\documentclass[aps,preprint,showpacs,superscriptaddress]{revtex4}
\usepackage{graphicx}
\usepackage{bm}
\newcommand{\beq}{\begin{equation}}
\newcommand{\eeq}{\end{equation}}

\begin{document}

\title{\boldmath Universal Cause of High-T$_c$ Superconductivity and
Anomalous Behavior of Heavy Fermion Metals}

\author{V.R.  Shaginyan \footnote{E--mail:
vrshag@thd.pnpi.spb.ru}} \affiliation{Petersburg Nuclear Physics
Institute, Russian Academy of Sciences,  Gatchina, 188300, Russia}
\affiliation{ CTSPS, Clark Atlanta University, Atlanta, Georgia
30314, USA}\affiliation{ Racah Institute of Physics,
the Hebrew University, Jerusalem 91904, Israel}
\author{M.Ya. Amusia}\affiliation{ Racah Institute of Physics,
the Hebrew University, Jerusalem 91904, Israel} \affiliation{ A.F.
Ioffe Physical-Technical Institute, Russian Academy of Sciences,
St. Petersburg, 194021, Russia}
\author{A.Z. Msezane}\affiliation{ CTSPS, Clark Atlanta University,
Atlanta, Georgia 30314, USA}
\author{K.G. Popov}
\affiliation{Komi Science Center,
Ural Division, Russian Academy of Sciences,
Syktyvkar, 167982, Russia}


\begin{abstract}

Unusual properties of strongly correlated liquid observed
in the high-$T_c$ superconductors and heavy-fermion (HF)
metals are determined by quantum phase transitions
taking place at their critical points. Therefore,
direct experimental studies of these 
transitions and critical points are of crucial importance for
understanding the physics of high-$T_c$ superconductors and
HF metals. In case of  high-$T_c$ superconductors such direct
experimental studies are absent since at low temperatures 
corresponding critical points are occupied by the
superconductivity. Recent experimental data on the behavior of
HF metals illuminate both the nature of these critical points
and the nature of the phase transitions.  We show that it is of
crucial importance to simultaneously carry out studies of both
the high-$T_c$ superconductivity and the anomalous behavior of HF
metals. The understanding of this fact has been problematic
largely because of the absence of theoretical guidance.
The main features of the fermion condensation quantum phase
transition (FCQPT), which are distinctive in several aspects
from that of conventional quantum phase transition (CQPT), are
considered. Our paper deals with these fundamental problems 
through studies of the behavior of quasiparticles, leading to
good quantitative agreement with experimental facts.
We show that in contrast to CQPT, whose physics in
the critical region is dominated by thermal and quantum
fluctuations and characterized by the absence of quasiparticles,
the physics of a Fermi system near FCQPT or undergone FCQPT is
controlled by the system of Landau-type 
quasiparticles.  However, contrary to the conventional 
Landau quasiparticles, the
effective mass of these strongly depends on the
temperature $T$, magnetic fields $B$, the number
density $x$, etc. Our general
consideration suggests that FCQPT and the emergence of novel
quasiparticles near and behind FCQPT 
are distinctive features of strongly correlated
substances such as the high-$T_c$ superconductors and HF metals.
We show that the main properties and universal behavior of the
high-$T_c$ superconductors and HF metals
can be understood within the framework of presented here 
theory based on FCQPT. 
A large number of the experimental evidences in favor of
the existence of FCQPT in high-$T_c$ superconductors
and HF metals is presented.
We demonstrate that the essence of strongly correlated
electron liquids can be controlled by both magnetic field B and
temperature T. Thus, the main  properties of heavy-fermion
metal such as magnetoresistance, resistivity, specific
heat, magnetization, volume thermal expansion, etc, are
determined by its position on the $B-T$ phase diagram.
The obtained results are in good agreement with recent
facts and observations.

\end{abstract}


\maketitle

\section{Introduction}
\setcounter{equation}{0}

In the last two decades a new class of materials such as heavy-fermion (HF) 
metals and high-$T_c$ superconductors has been found, which 
display a dazzling variety of physical phenomena, see e.g. \cite{ste,varma}. 
The high-$T_c$ superconductors are strongly correlated metals with normal state 
properties that are not at all those of a normal Landau Fermi liquid (LFL). 
In the case of HF metals, the electronic strong correlations result in 
a renormalization of the effective mass of the quasiparticles, 
which can exceed the bare mass by a factor up to 1000 or even diverge. 
These non-Fermi-liquid (NFL) systems demonstrate anomalous  pure power-law 
temperature dependences in their low temperature properties over broad 
temperature ranges. 
It is believed that in this class 
the basic assumption of the 
LFL theory that at low energies the electrons in a 
metal should behave as 
weakly interacting quasiparticles is violated. 
Therefore, it is generally accepted that 
the fundamental physics that gives rise
to the high-$T_c$ superconductivity and NFL behavior
with a recovery of the LFL behavior under
the application of magnetic fields observed in HF 
metals and high-$T_c$ compounds is controlled by quantum phase
transitions. This has made quantum phase transitions a subject of
intense current interest, see e.g. \cite{sac,varma,voj,uzum}.

A quantum phase transition is
driven by control parameters such as the composition, number density $x$
or magnetic fields $B$, and takes place at a quantum critical point 
(QCP) when temperature $T=0$. QCP separates an ordered phase generated 
by quantum phase transition from a disordered phase.  It is expected 
that the universal behavior is only observable if the system in 
question is very near  QCP, for example, when the correlation length is 
much larger than microscopic length scales.  Quantum phase transitions 
of this sort are quite common, and we shall call 
them as conventional quantum phase transitions (CQPT). In the 
case of CQPT, the physics is dominated by thermal and quantum 
fluctuations of the critical state, which is characterized by the 
absence of quasiparticles.  It is believed that the absence of 
quasiparticle-like excitations is the main cause of the NFL behavior 
and other types of critical behavior in the quantum
critical region. On the base of scaling
related to the divergence of the correlation length, one can
construct the critical contribution to the free energy and evaluate
the corresponding properties such as critical exponents, the NFL
behavior, etc. \cite{sac,varma,voj,uzum}. However along this way one 
may expect difficulties. For example, having the only critical 
contribution, one has to describe different types of the behavior 
exhibited by different HF metals, see e.g.  \cite{geg1,alp,shag4}. Note 
that HF metals are three-dimensional structures, (see e.g. 
\cite{geg1,kadw,geg}) and, thus, the type of behavior cannot be related 
to the dimension. The critical behavior observed in measurements on HF 
metals takes place up to rather high temperatures comparable with the 
effective Fermi temperature $T_k$. For example, the thermal expansion 
coefficient $\alpha(T)$ measured on CeNi$_2$Ge$_2$ shows a $1/\sqrt{T}$ 
divergence over more than two orders of magnitude in temperature 
drop from 6 K down to 
at least 50 mK \cite{geg1}.  It is hardly possible to understand such a 
behavior on the base of the assumption of scaling when the correlation
length has to be much larger than microscopic length scales.
Obviously, such a situation can take place only at $T\to0$. At 
some temperature $T\sim T_k$, 
this macroscopically large correlation length must be destroyed
by thermal fluctuations.  

The next problem is related to explanations
of the recovery of the LFL behavior
under applied magnetic fields $B$, 
observed in HF metals and the high-$T_c$ compounds, see e.g. 
\cite{geg,ste,cyr}.  At $T\to0$, the magnetic field dependence of the 
coefficient $A(B)$, causing an electrical resistivity contribution 
$\Delta\rho=A(B)T^2$, the Sommerfeld coefficient $\gamma(B)$ and 
$\chi(B)$ in specific heat, $C/T=\gamma(B)$, and magnetic
susceptibility, $\chi(B)$, shows that $A(B)\sim\gamma^2(B)$ and
$A(B)\sim\chi^2(B)$, so that the Kadowaki-Woods ratio,
$K=A(B)/\gamma^2(B)$ \cite{kadw}, is $B$-independent and 
conserved \cite{geg}. Such a 
universal behavior is hardly possible to explain within the picture
assuming the absence of quasiparticles which takes place near QCP of
the corresponding CQPT. As a consequence, for example, these facts are
in variance to the spin-density-wave scenario \cite{geg}
and the renormalization group treatment of quantum criticality
\cite{mill}. Moreover, striking recent measurements
of the specific heat, charge and heat transport and the resistivity used
to study the nature of magnetic field-induced QCP in heavy-fermion
metal CeCoIn$_5$ \cite{bi,pag1} certainly seem to disagree with
descriptions based on CQPT.

In a system of interacting bosons at temperatures lower than the
temperature of Bose-Einstein condensation \cite{bel1,bel2},
a finite number of
particles is concentrated in the lowest level. In the case of a
noninteracting Bose gas at zero temperature, $T=0$, this
number is simply equal to the total number of particles in the
system. In a homogeneous system of noninteracting Bosons, the
lowest level is the state with zero momentum, and the ground state
energy is equal to zero. For a noninteracting Fermi system such a
state is impossible, and its ground state energy $E_{gs}$
reduces to the kinetic energy and is proportional to the
total number of particles. Imagine an interacting system of
fermions with a pure repulsive interaction. Let us increase its
interaction strength. As soon as it becomes sufficiently large and the
potential energy starts to prevail over the kinetic energy, we can
expect the system to undergo a phase transition at $T=0$. This
takes place when
the density $x$ tends to the critical quantum point $x\to x_{FC}$
and the kinetic energy $E_k$ becomes
frustrated, while the effective inter-electron interaction, or the
Landau amplitude, being sufficiently large,
starts to determine the occupation numbers of quasiparticles $n({\bf
p})$ which deliver the minimum value to the ground state energy $E[n(p)]$.
As a result, at $x<x_{FC}$, the
function $n({\bf p})$ is given by the standard equation
that determines the minimum of functional $E[n({\bf p})]$ 
\cite{ks,ksn,ksk,dkss,vsl} \beq \frac{\delta E[n({\bf p})]}{\delta 
n({\bf p})}=\mu. \eeq Here we deal with 
three dimensional (3D) case and assume that the phase transition takes 
place at $x<x_{FC}$.  At $T=0$ Eq. (1.1) determines the quasiparticle 
distribution function $n_0({\bf p})$, which delivers the minimum  value 
to the ground state energy $E$. The function $n_0({\bf p})$ being the 
signature of the new state of quantum liquids \cite{vol} does not 
coincide with the quasiparticle distribution function of the LFL theory 
in the region $(p_f-p_i)$, so that $0<n_0({\bf p})<1$ and 
$p_i<p_F<p_f$, with $p_F=(3\pi^2x)^{1/3}$ being the Fermi momentum. Such a 
state was called the state with fermion condensate (FC) because 
quasiparticles located in the region $(p_f-p_i)$ of momentum space are 
pinned to the chemical potential $\mu$ \cite{ksk,ks,vol}. We note that 
the behavior obtained for the single-particle spectrum and 
quasiparticle distribution functions is observed within exactly 
solvable models \cite{dzyal,lid,irk}. Lowering the potential energy, FC 
decreases the total energy.  Unlike the Bose-Einstein condensation, 
which occurs  even in a system of noninteracting bosons, FC
can take place if the coupling constant of
the interaction is large, or the corresponding Landau amplitudes are
large and repulsive.

We note the remarkable peculiarity of FCQPT at $T=0$: this transition
is related to the
spontaneous breaking of gauge symmetry, when the superconducting
order parameter $\kappa({\bf p})=\sqrt{n_0({\bf p})(1-n_0({\bf p}))}$
has a nonzero value over the
region occupied by FC,
with the entropy $S=0$ \cite{vsl,shag2}. At small values of the
pairing coupling constant $\lambda_0$, the gap $\Delta({\bf p})$
is linear in $\lambda_0$ and
vanishes provided that $\lambda_0\to0$,  while $\kappa({\bf p})$
remains finite \cite{dkss,vsl}. As we shall see, this peculiarity 
allows to construct the theory of high-$T_c$ superconductivity based on 
FCQPT.  Thus, the state with FC cannot exist at any finite temperatures 
and is driven by the parameter $x$: at $x>x_{FC}$ the system is on the 
disordered side of FCQPT; at $x=x_{FC}$, Eq. (1.1) possesses the 
non-trivial solutions $n_0({\bf p})$ with $p_i=p_F=p_f$; while at $x<x_{FC}$, 
the system is on the ordered side \cite{shag2}.

One of the most challenging problems of modern condensed matter physics is the
structure and properties of Fermi systems with large coupling constants.
The first solution to this problem was offered by the Landau
theory of Fermi liquids, later called "normal", by introducing the
notion of quasiparticles and so called amplitudes, which characterize the
effective interaction among them \cite{lanl1}.
The Landau theory can be viewed as the low energy
effective theory in which high energy
degrees of freedom are removed
by introducing the effective amplitudes
instead of strong inter-particle interaction. Usually, it is
assumed that the stability of the ground state of Landau liquid is
determined by the Pomeranchuk stability conditions:
the stability is violated when even one of the Landau effective
interaction parameters
is negative and reaches a critical value \cite{lanl1,pom}. 
Note that the new phase, at which the stability
conditions are restored, can in principle be again described within the
framework of the same theory.
However, it has been demonstrated rather recently \cite{ks} that
the Pomeranchuk stability conditions cover not 
all possible instabilities: 
one of them is missed. It corresponds to the situation when, at
the temperature $T=0$, the effective mass, the most important
characteristic of Landau quasiparticles, can become infinitely
large. Such a situation, leading to profound consequences, can
take place when the corresponding Landau amplitude
being repulsive reaches some critical value.
This leads to a completely new class of strongly correlated Fermi
liquids with FC \cite{ks,vol,ksk}, which is separated
from that of a normal Fermi liquid by the fermion condensation
quantum phase transition (FCQPT) \cite{ms,shb}.

In the FCQPT case we are dealing with the strong coupling limit
where an absolutely reliable answer cannot be given on the bases
of pure theoretical first principle foundation. Therefore, the
only way to verify that FC occurs is to consider both
exactly solvable models and experimental
facts, which can be interpreted as confirming the existence of
such a state.
The exactly solvable models unambiguously demonstrate that
Fermi liquids with FC do exist, see e.g. \cite{dzyal,lid,irk}.
On the other hand, these facts are seen in
features of those two-dimensional (2D) systems with interacting
electrons or holes, which can be represented by doped quantum wells
and high-$T_c$ superconductors. Considering the HF metals and
the 2D systems of $^3$He,
we will show that FC exist also in these systems.

The goal of our review is to describe the behavior of Fermi systems
with FC and to show that the existing data on strongly correlated
liquids represented by the electronic (or hole) systems of high-$T_c$
superconductors and HF metals
can be well understood within the theory of Fermi liquids
based on FCQPT. In Section 2, we review the general features of Fermi
liquids with FC in their normal state. Section 3 is devoted to
consideration of the superconductivity in the presence of FC. We
show that the superconducting state is totally transformed by the
presence of FC. For instance, the maximum value $\Delta_1$ of the
superconducting gap can be as large as $\Delta_1\sim 0.1
\varepsilon_F$, while for normal superconductors one has
$\Delta_1\sim 10^{-3} \varepsilon_F$, 
where $\varepsilon_F$ is the Fermi level energy. 
In Section 4 we describe the
quasiparticle's dispersion and its lineshape and show that they
strongly deviate from the case of normal Landau liquids. In Section 5
we consider the field-induced LFL in the heavy electron liquid
with FC. In Section 6 we apply our theory to explain the main properties of
magnetic-field induced Landau Fermi liquid in the high-$T_c$ metals.
Section 7 is devoted to the appearance of FCQPT in different Fermi liquids.
In Section 8 we analyze the main properties of HF metals whose electronic
system is placed on the disordered side of FCQPT. HF metals with
the electronic system located on the ordered side of FCQPT are considered
in Section 9. In Section 10 we show that FC manifests itself in the 
dissymmetry of tunnelling conductivity which can be observed in 
measurements on the high-$T_c$ compounds and HF metals.  Finally, in 
Section 11, we summarize our main results.

\section {Fermi Liquids with Fermion Condensate}
\setcounter{equation}{0}

To study the universal behavior of the high-$T_c$ superconductors and
HF metals at low temperatures, 
we use the heavy electron liquid model in order to ignore the complications 
of the anisotropy of the lattice of solids and its microscopic inhomogeneity.
It is possible since we consider the universal behavior demonstrated
by these materials and processes related to the power-low
divergences of observables 
such as the effective mass, specific heat, thermal expansion, etc. 
These divergences are determined by small momenta
transferred as compared to momenta of the same order of magnitude as 
those of the reciprocal lattice cell, and  
contributions coming from them can be safely ignored. On
the other side, we can simply use the common concept of the
applicability of the LFL theory when describing electronic
properties of metals \cite{lanl1}.
Thus, we may usefully ignore the complications due to 
lattice and its anisotropy.
As a result, we regard the medium as homogeneous heavy
electron isotropic liquid.

\subsection{Landau Theory of Fermi Liquid}

Let us start by explaining the important points of the LFL theory
\cite{lanl1}. The LFL theory rests on the notion of quasiparticles which
represent elementary excitations of a Fermi liquid. Therefore
these are appropriate excitations to describe the low temperature
thermodynamic properties. In the case of an electron system, these
are characterized by the electron's quantum numbers and effective
mass $M^*$. The ground state energy of the system in question is a 
functional of the quasiparticle occupation numbers (or quasiparticle 
distribution function) $n({\bf p},T)$, just like the free energy 
$F[n({\bf p},T)]$, entropy $S[n({\bf p},T)]$, and other thermodynamic 
functions. From the condition that the free energy $F=E-TS$ should be  
minimal, we can find the distribution function \beq \frac{\delta(F-\mu 
N)}{\delta n({\bf p},T)}=\varepsilon({\bf p},T) 
-\mu(T)-T\ln\frac{1-n({\bf p},T)}{n({\bf p},T)}=0.\eeq
Here  $\mu$ is the chemical potential,
while \beq \varepsilon({\bf p},T)\ =\ \frac{\delta E[n({\bf p},T)]}{\delta
n({\bf p},T)}\ , \eeq is the quasiparticle energy. This energy is a
functional of $n({\bf p},T)$ just like the total energy $E[n({\bf 
p},T)]$. The entropy $S[n({\bf p},T)]$ is given by the familiar 
expression \cite{lanl1} \beq S[n({\bf p},T)]=-2\int\left[n({\bf p},T)\ln n({\bf 
p},T) +(1-n({\bf p},T))\ln(1-n({\bf p},T))\right]\frac{d{\bf 
p}}{(2\pi)^3}, \eeq which stems from purely combinatorial 
considerations. Equation (2.1) is usually presented as the Fermi-Dirac 
distribution \beq n({\bf p},T)\ =\ 
\left\{1+\exp\left[\frac{(\varepsilon({\bf p},T)-\mu)}
{T}\right]\right\}^{-1}. \eeq At $T\to 0$, one
gets from Eqs. (2.1) and (2.4) the standard
solution $n(p,T\to0)\to\theta(p_F-p)$, with $\theta(p_F-p)$
is the step function,
$\varepsilon(p\simeq p_F)-\mu=p_F(p-p_F)/M^*_L$, where
$M^*_L$ is the Landau effective mass
\cite{lanl1} \beq \frac1{M^*_L}\ =\
\frac1p\,\frac{d\varepsilon(p,T=0)}{dp}|_{p=p_F}\ .\eeq It is
implied that $M^*_L$ is positive and finite at the Fermi momentum
$p_F$. As a result, the $T$-dependent corrections to $M^*_L$, to
the quasiparticle energy $\varepsilon({\bf p})$, and to other
quantities, start with $T^2$-terms. The effective mass is given by the
well-known Landau equation
\beq \frac{1}{M^*_L}=
\frac{1}{M}+\sum_{\sigma_1}\int \frac{{\bf p}_F{\bf p_1}}{p_F^3}
F_{\sigma,\sigma_1}({\bf p_F},{\bf p}_1) \frac{\partial
n_{\sigma_1}({\bf p}_1,T)}{\partial {p}_1} \frac{d{\bf
p}_1}{(2\pi)^3}. \eeq Here $F_{\sigma,\sigma_1}({\bf p_F},{\bf
p}_1)$ is the Landau amplitude depending on the momenta ${\bf p}$,
spins $\sigma$, and $M$ is the bare mass
of an electron. For the sake of simplicity,
we omit the spin dependence of the effective mass since in the
case of a homogeneous liquid and weak magnetic fields $M^*_L$
does not noticeably depend on the spins.

Applying Eq. (2.6) at $T=0$ and taking into
account that $n({\bf p},T=0)$ becomes the step function $\theta(p_F-p)$,
we obtain the standard result
$$ \frac{M^*_L}{M}=\frac{1}{1-N_0F^1(p_F,p_F)/3}.$$
Here $N_0$ is the density of states of the free Fermi gas and
$F^1(p_F,p_F)$ is the $p$-wave component of the Landau
interaction amplitude. Since in the LFL theory 
$x=p_F^3/3\pi^2$, the Landau 
amplitude can be written as $F^1(p_F,p_F)=F^1(x)$. Assume that at
some critical point $x_{FC}$ the denominator
$(1-N_0F^1(p_F,p_F)/3)$ tends to zero, that is
$(1-N_0F^1(x)/3)\propto(x-x_{FC})+a(x-x_{FC})^2+...\to 0$.
As a result, one obtains
that $M^*_L(x)$ behaves as \cite{shag1,khod1}
\begin{equation}
\frac{M^*_L(x)}{M}\simeq A+\frac{B}{x-x_{FC}}\propto\frac{1}{r}.
\end{equation}
Here $A$ and $B$ are constants and
$r=(x-x_{FC})$ is the ``distance'' from the QCP of FCQPT
taking place at $x_{FC}$. The observed behavior is in good
agreement with recent experimental observations (see e.g.
\cite{skdk,cas1}), and calculations \cite{krot,sarm1,sarm2},
see Section 7 as well. In the
case of electronic systems,  Eq. (2.7) is valid at $x>x_{FC}$ when $r>0$
\cite{ksk,ksz}. Such a behavior of the effective mass can be
observed in the HF metals with a quite flat, narrow conduction
band, corresponding to the large effective mass $M^*_L(x\simeq
x_{FC})$, with strong electron correlations and the effective
Fermi temperature $T_k\sim p_F^2/M^*_L(x)$ of the order of a few
Kelvin or even lower (see e.g. \cite{ste}).

\subsection{Fermion condensation quantum phase transition}

As we have seen above at $T=0$ when $r=(x-x_{FC})\to 0$, the
effective mass diverges, $M^*_L(r)\to\infty$, and eventually beyond the
critical point $x_{FC}$ the distance $r$ becomes negative making
the effective mass negative, as it follows from Eq. (2.7). To escape the
possibility of being in unstable and in essence meaningless states with
negative values of the effective mass, the system is to undergo a
quantum phase transition at the critical point
$x=x_{FC}$. Because the
kinetic energy near the Fermi surface is proportional to the
inverse effective mass, this phase transition is triggered by the
frustrated kinetic energy and can be recognized as FCQPT
\cite{ms,shag3}. Therefore behind the critical point $x_{FC}$ of this
transition, the quasiparticle distribution represented by
the step function does not minimize the Landau
functional $E[n({\bf p})]$. As a result, at $x<x_{FC}$ the
quasiparticle distribution is determined by the standard equation
that is used to search for the minimum of the energy functional \cite{ks}
\begin{equation} \frac{\delta E[n({\bf p})]}{\delta n({\bf
p},T=0)}=\varepsilon({\bf p})=\mu; \,p_i\leq p\leq p_f.
\end{equation}
Equation (2.8) determines the quasiparticle distribution function
$n_0({\bf p})$, which minimizes the ground 
state energy $E$. Being determined by Eq. (2.8),  
$n_0({\bf p})$ does not coincide with the step function in the
region $(p_f-p_i)$, so that $0<n_0({\bf p})<1$, while outside the
region it coincides with the step function. It follows from Eq.
(2.8) that the single particle spectrum or the band is completely
flat over the region. Such a state was called the state with
FC because quasiparticles located in the
region $(p_f-p_i)$ of momentum space are pinned to the chemical
potential $\mu$ \cite{ksk,ks,vol}. We note that this behavior
was obtained for the band and quasiparticle distribution functions using 
some exactly solvable models \cite{irk,lid}.

\begin{figure}[!ht]
\begin{center}
\includegraphics[width=0.47\textwidth]{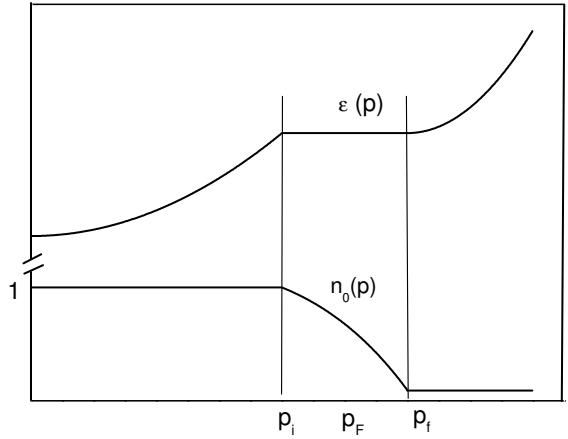}
\end{center}
\caption{The quasiparticle distribution function
$n_0(p)$ and energy
$\varepsilon(p)$. Since $n_0(p)$ is the solution of Eq. (2.8) it implies
$n_0(p<p_i)=1$, $n_0(p_i<p<p_f)<1$ and $n_0(p>p_f)=0$,
while $\varepsilon(p_i<p<p_f)=\mu$. The Fermi momentum $p_F$ obeys the 
condition $p_1<p_F<p_f$.} \label{Fig1} \end{figure}
The possible solution $n_0({\bf p})$ of Eq. (2.8) and the
corresponding single particle spectrum $\varepsilon({\bf p})$ are 
shown in Fig. 1.

As we shall see in Section 3
the relevant order parameter of the FC state, $\kappa({\bf
p})=\sqrt{n_0({\bf p})(1-n_0({\bf p}))}$, is the order parameter of
the superconducting state with the infinitely small value of the
superconducting gap. Therefore the entropy of this state is zero, 
$S(T=0)=0$ \cite{ksk}. Thus, this state is of pure 
quantum nature and cannot 
exist at any finite temperatures. This quantum state with FC is driven 
by the density $x$: at $x>x_{FC}$ the system is on the disordered side 
of FCQPT; at $x=x_{FC}$, Eq. (2.8) has the non-trivial solutions 
$n_0({\bf p})$ with $p_i=p_F=p_f$; at $x<x_{FC}$, the system is on the 
ordered side. We note that the solutions $n_0({\bf p})$ of Eq. (2.8)
can be viewed as new to the LFL theory.
Indeed, at $T=0$, the standard
solution $n({\bf p},T\to0)\to\theta(p_F-p)$ is not the
only one possible. The  "anomalous" solutions of Eq. (2.1) can exist
because the logarithm on the right hand side of
Eq. (2.1) is finite when $p$ belongs to the region $(p_f-p_i)$
and $0<n_0({\bf p})<1$.
Therefore in the region 
$T\ln[(1-n_0({\bf p},T))/n_0({\bf p},T)]_{|T\to0}\to 0$, 
and we again arrive at Eq. (2.8).

Let us assume that with the decrease of the density (or with the 
growth of the interaction strength) FC has just taken place. It means 
that $p_i\to p_f\to p_F$, and the deviation $\delta n({\bf p})=n_0({\bf 
p})-\theta(p_F-p)$ is small.  Expanding the functional $E[n({\bf p})]$ 
in Taylor's series with respect to $\delta n({\bf p})$ and retaining 
the leading terms, one obtains from Eq. (2.8) the following relation 
\beq \mu\ =\ \varepsilon({\bf p})\ =\ \varepsilon_0({\bf p})+\int 
F({\bf p},{\bf p}_1)\delta n({\bf p_1}) \frac{d{\bf p}_1}{(2\pi)^2}\ ; 
\quad p_i\leq p \leq p_f\ , \eeq where $F({\bf p},{\bf p}_1)=\delta^2 
E/\delta n({\bf p})\delta n({\bf p}_1)$ is the Landau amplitude. Both 
quantities, the interaction and the single-particle energy
$\varepsilon_0({\bf p})$ are calculated at $n({\bf p})=\theta(p_F-p)$.
Equation (2.9)
acquires nontrivial solutions at some density
$x=x_{FC}$. Thus, FCQPT takes
place if the Landau amplitudes depending on the density are
positive and sufficiently large, so that the potential energy is
bigger than the kinetic energy. Then the transformation of the
Fermi step function $n({\bf p})=\theta(p_F-p)$ into the smooth function
defined by Eq. (2.9) becomes possible \cite{ks,ksk}.
The system in question can be considered as
a strongly correlated Fermi liquid behind the point $x_{FC}$.
It is seen from
Eq. (2.9) that the FC quasiparticles form a collective state, since
their energies are defined by the macroscopical number of
quasiparticles within the momentum region $(p_f-p_i)$. The shape of
the excitation spectra related to FC is not affected by the Landau
interaction, which, generally speaking, depends on the system's
properties, including the collective states, 
the irregularity of the composition, impurities, etc. The
only thing determined by the interaction is the width of the FC
region $(p_f-p_i)$ provided the interaction is sufficiently strong
to produce the FC phase transition at all. Thus, we can conclude
that the spectra related to FC are of a universal form, being
dependent, as we shall see in Subsections 2.3  
and 3.1, on the temperature and the superconducting gap. 
The existence of such spectra can be viewed as the
characteristic feature of the "quantum protectorate" \cite{rlp,pa}.

\subsection{The ``shadow'' of the fermion condensate at finite temperatures}

According to Eq. (2.1), the single-particle energy
$\varepsilon({\bf p},T)$
within the interval $(p_f-p_i)$ at $T\ll
T_f$ is linear in $T$ \cite{kcs}. At the Fermi level,
one obtains by expanding $\ln(...)$ in
terms of $n({\bf p})$ \beq \varepsilon({\bf p},T)-\mu(T)\ =\
T\ln\frac{1-n({\bf p})}{n({\bf p})}\ \simeq\
T\frac{1-2n({\bf p})}{n({\bf p})}\bigg|_{p\simeq
p_F}\ . \eeq Here $T_f$ is the temperature, above which FC effects
become insignificant \cite{dkss} \beq \frac{T_f}{\varepsilon_F}\ \sim\
\frac{p_f^2-p_i^2}{2M\varepsilon_F}\ \sim\
\frac{\Omega_{FC}}{\Omega_F}\ . \eeq In this formula $\Omega_{FC}$
is the FC volume, $\varepsilon_F$ is the Fermi energy, and
$\Omega_F$ is the volume of the Fermi sphere. We note that at
$T\ll T_f$ the occupation numbers $n({\bf p})$ are approximately
independent of $T$, being given by Eq. (2.8). At finite
temperatures according to Eq. (2.10), the dispersionless plateau
$\varepsilon({\bf p})=\mu$ is slightly turned counter-clockwise about $\mu$.
As a result, the plateau is just a little tilted and rounded off at the
end points. According to Eqs. (2.5) and (2.10),
the effective mass $M^*_{FC}$ related
to FC is given by, \beq M^*_{FC}\ \simeq\ p_F\frac{p_f-p_i}{4T}\ . \eeq
To obtain Eq. (2.12) an approximation for the derivative $dn(p)/dp\simeq
-1/(p_f-p_i)$ was used. It is seen from Eq. (2.12)  that  at $0<T\ll T_f$,
the heavy electron liquid  behaves
as if it were placed at QCP, in fact it is placed at the quantum
critical line $x<x_{FC}$, that is the critical behavior is observed at
$T\to0$ for all $x\leq x_{FC}$.

Having in mind that $(p_f-p_i)\ll p_F$ and using Eqs. (2.11) and (2.12),
the following estimates for the effective mass $M^*_{FC}$ are
obtained: \beq \frac{M^*_{FC}}{M}\ \sim\ \frac{N(0)}{N_0(0)}\
\sim\ \frac{T_f}T\ . \eeq Eqs. (2.12) and (2.13) show the temperature
dependence of $M^*_{FC}$. In Eq. (2.13) $N_0(0)$ denotes
the density of states of noninteracting
electron gas, and $N(0)$ is the density of states at the Fermi
level. Multiplying both sides of Eq. (2.12) by $(p_f-p_i)$, we obtain
the energy scale $E_0$ separating the slow dispersing low energy
part related to the effective mass $M^*_{FC}$ from the faster
dispersing relatively high energy part defined by the effective
mass $M^*_{L}$ \cite{ms,shb,ars}, \beq E_0\ \simeq\ 4T\ . \eeq It is
seen from Eq. (2.14) that the scale $E_0$ does not depend on the
condensate volume. The single particle excitations are defined
according to Eq. (2.12) by the temperature and by $(p_f-p_i)$,
given by Eq. (2.8). Thus, we conclude that the one-electron spectrum
is of universal form and has the features 
of the "quantum protectorate".

It is pertinent to note that outside the FC region the single
particle spectrum is not strongly affected by the temperature, being
defined by $M^*_L$. Thus, we come to the conclusion that a system
with FC is characterized by two effective masses: $M^*_{FC}$ which
is related to the single particle spectrum at lower energy scale
and $M^*_L$ describing the spectrum at higher energy scale.  The
existence of two effective masses is manifested by a break (or kink)
in the quasiparticle dispersion, which can be approximated by two
straight lines intersecting at the energy $E_0$.  This break takes
place at temperatures $T_c\leq T\ll T_f$, which is in accord with 
experimental data \cite{blk}, and, as we will see, at $T\leq T_c$
which is also in accord with the experimental facts \cite{blk,krc}.
Here $T_c$ is the critical tempereture of the
superconducting phase transition.
The quasiparticle formalism is applicable to this problem since
the width $\gamma$ of single particle excitations is not large
compared to their energy, being proportional to the temperature,
$\gamma\sim T$ at $T>T_c$ \cite{dkss}. The lineshape can be
approximated by a simple Lorentzian \cite{ars}. This is consistent with
experimental data obtained from scans at a constant binding energy
\cite{vall} (see Sec. 4).

It is essential to have in mind, that the onset of the charge
density wave instability in a many-electron system, such as
an electron liquid, which takes place as soon as the effective
inter-electron constant reaches its critical value $r_s=r_{cdw}$,
is preceded by the unlimited growth of the effective mass, which 
is demonstrated Section 7. 
Here $r_s=r_0/a_B$ with $r_0$ being the average
distance between electrons, while $a_B$ is the Bohr radius. 
Therefore the FC occurs before the onset of the charge density
wave. Hence, at $T=0$, when $r_s$ reaches its critical value
$r_{FC}$ corresponding to $x_{FC}$,
$r_{FC}<r_{cdw}$, FCQPT already 
inevitably takes place \cite{ksz}.  It is 
pertinent to note that this growth of the effective mass with
decreasing electron density was observed experimentally in a
metallic 2D electron system in silicon at $r_s\simeq 7.5$
\cite{skdk}. Therefore we can take this value 
as an estimate, $r_{FC}\sim 7.5$. On the other
hand, there exist charge density waves or strong fluctuations of
charge ordering in underdoped high-$T_c$ superconductors
\cite{grun,tass}. Thus, the formation of FC in high-$T_c$ compounds can
be thought more as a general property of low density electron liquid 
embedded in these solids rather than an unusual 
and anomalous solution of Eq. (2.8) \cite{ksz}. Beyond the point of
FCQPT, the condensate volume is proportional to $(r_s-r_{FC})$ as
well as $T_f/\varepsilon_F\sim (r_s-r_{FC})/r_{FC}$ at least when
$(r_s-r_{FC})/r_{FC}\ll 1$. Therefore we obtain \beq
\frac{r_s-r_{FC}}{r_{FC}}\sim \frac{p_f-p_{i}}{p_{F}}
\sim \frac{x_{FC}-x}{x_{FC}}. \eeq FC
serves as a stimulator that creates new phase transitions, which
eliminates the degeneracy of the spectrum. For example FC can generate
spin density waves or antiferromagnetic phase transition, thus
leading to a whole variety of new properties of the system under
consideration. Then, the onset of the charge density wave is
preceded by FCQPT, and both of these phases can coexist at the
sufficiently low density when $r_s\geq r_{cdw}$.
The transition to superconductivity is strongly
assisted by FC because both of the phases are characterized by the
same order parameter. As a result, the superconductivity by removing
the spectrum degeneracy ``wins'' the competition with other
phase transitions up to the critical
temperature $T_c$, see Section 3. We
now turn to the consideration of the superconducting state and
quasiparticle dispersions  at $T\leq T_c$.

\section{The superconducting state with fermion condensate}
\setcounter{equation}{0}

In this Section we consider the superconducting state of 2D
heavy electron liquid since the high-$T_c$ superconductors 
are predominantly represented by 2D structures. On the other hand, our 
consideration can be easily adopted to the 3D case.

\subsection{Superconducting state at $T=0$}

At $T=0$, the ground state energy $E_{gs}[\kappa({\bf p}),n({\bf
p})]$ of a 2D electron liquid is a functional of both the order parameter
of the superconducting state $\kappa({\bf p})$ and
the quasiparticle occupation numbers $n({\bf p})$. 
This energy is 
determined by the well-known equation of
the Bardeen-Cooper-Schrieffer (BCS) weak-coupling theory of
superconductivity, see e.g. \cite{bcs,til} \beq E_{gs}\ =\ E[n({\bf
p})]+ \lambda_0\int V({\bf p}_1,{\bf p}_2) \kappa({\bf p}_1)
\kappa^*({\bf p}_2) \frac{d{\bf p}_1d{\bf p}_2}{(2\pi)^4}\ . \eeq
Here  $E[n({\bf p})]$ is the ground-state energy of a normal Fermi
liquid, and  \beq n({\bf p})=v^2({\bf p});\,\,\, \kappa({\bf p})=v({\bf
p})u({\bf p})=\sqrt{n({\bf p})(1-n({\bf p}))},\eeq
Where $u({\bf p})$ and $v({\bf p})$ are the
normalized coherence factors,  $v^2({\bf p})+u^2({\bf p})=1$.
It is assumed that the pairing
interaction $\lambda_0V({\bf p}_1,{\bf p}_2)$ is weak.
We define the superconducting gap
\beq \Delta({\bf p})=-\lambda_0\int
V({\bf p},{\bf p}_1)\kappa({\bf p}_1)
\frac{d{\bf p}_1}{4\pi^2}.\eeq
Minimizing
$E_{gs}$ with respect to $v({\bf p})$ we obtain the equation
connecting the single-particle energy $\varepsilon({\bf p})$ to
$\Delta({\bf p})$, \beq \varepsilon({\bf p})-\mu=\Delta({\bf
p}) \frac{1-2v^2({\bf p})} {2\kappa({\bf p})}, \eeq where the
single-particle energy $\varepsilon({\bf p})$ is determined by the
Landau equation (2.2). With insertion of the value
of $\kappa({\bf p})$ into Eq. (3.3), we obtain the known equation for
$\Delta({\bf p})$
\beq
\Delta({\bf p})
=-\frac{\lambda_0}{2}\int V({\bf p},{\bf p}_1)
\frac{\Delta({\bf p}_1)}{\sqrt{(\varepsilon({\bf p}_1)-\mu)^2
+\Delta^2({\bf p}_1)}} \frac{d{\bf p}_1}{4\pi^2}. \eeq
If $\lambda_0\to 0$, then the maximum value $\Delta_1$ of the
superconducting gap $\Delta({\bf p})$ tends to zero, and
Eq. (3.4) reduces to Eq. (2.8)
\beq \frac{\delta E[n({\bf p})]}{\delta n({\bf
p})}=\varepsilon({\bf
p})-\mu\ =\ 0,\,\, \mbox{provided that}\,\,\, 0<n({\bf p})<1;\,\,
\mbox{or}\,\,\kappa({\bf p})\neq 0;\: p_i\leq
p\leq p_f\ . \eeq
It is seen from Eq. (3.6) that at $x<x_{FC}$, the
function $n({\bf p})$ is determined by the standard equation
to search the minimum of functional $E[n({\bf p})]$ \cite{dkss,ks}.
Equation (3.6) determines the quasiparticle distribution function
$n_0({\bf p})$ which delivers the minimum  value to the ground
state energy $E_{gs}[\kappa({\bf p}),n({\bf p})]$ 
in the $\lambda_0\to0$ limit.
Now we can study the relationships between the state
defined by Eq. (3.6), or by Eq. (2.8), and the superconducting state. At
$T=0$, Eq. (3.6) defines the particular state of a Fermi-liquid with
FC, for which the modulus of the order parameter $|\kappa({\bf
p})|$ has finite values in the range of momenta $p_i\leq
p\leq p_f$, while $\Delta_1\to 0$.
Such a state can
be considered as superconducting, with an infinitely small value
of $\Delta_1$, so that the entropy of this state is equal to zero.
It is obvious that this state, being driven by the quantum phase
transition, disappears at $T>0$ \cite{ms,shb}. Any quantum phase
transition, which takes place at temperature $T=0$, is determined
by a control parameter other than the temperature, for instance, by
pressure,  magnetic field, or  the number 
density of mobile charge carriers $x$.
In the  case of FCQPT, as we have shown in Section 2,
the control parameter is the density $x$ of
the system, which determines the strength of the Landau amplitudes.
FCQPT occurs at the quantum
critical point $x=x_{FC}$. 

As any phase transition, FCQPT
is related to the order parameter, which induces a broken symmetry.
As it follows from Eq. (3.2),
the order parameter of the state with FC is $\kappa({\bf p})$.
Thus, the solutions $n_0({\bf p})$ of Eq. (3.6) represent
a new class of the solutions of both BCS equations and LFL 
equations.  In contrast to the conventional BCS solutions \cite{bcs}, 
the new ones are characterized by the infinitesimal value of the 
superconducting gap, $\Delta_1\to 0$, while the order parameter 
$\kappa({\bf p})$ is finite.  At the same time, 
in contrast to the standard solutions of 
the LFL theory, the new ones are characterized by the superconducting order 
parameter $\kappa({\bf p})$, that is at $T\to0$, the quasiparticle 
distribution function does not tend to the step function 
$\theta(p_f-p)$, being the solution of Eq. (3.6). Thus, we can conclude that 
the solutions of Eq. (3.6) can be viewed as common solutions of both 
BCS and LFL equations, while Eq. (3.6) can be derived 
starting from either BCS or LFL theory.  As we shall see, 
on the basis of the peculiarities of this new class of solutions, 
we can define the notion of the strongly correlated Fermi liquid
and explain the main properties of
the high-$T_c$ superconductivity and HF metals.
It is essential that the existence of state with FC can be revealed
experimentally since the order parameter
$\kappa({\bf p})$ is suppressed
by a magnetic field $B$.
Destroying the state with FC, the magnetic field $B$ converts the
strongly correlated Fermi liquid into the normal LFL.
In that case, the magnetic field plays the role of  control parameter.

When $p_i\to p_F\to p_f$, Eq. (3.6) determines the critical
density $x_{FC}$ at which FCQPT takes place.
When $x<x_{FC}$, the system becomes divided into two quasiparticle
subsystems: the first subsystem is in the $(p_f-p_i)$ range and is 
characterized  by the quasiparticles with the effective mass
$M^*_{FC}\propto1/\Delta_1$, as it follows from Eq. (3.4), 
while the second one is occupied by
quasiparticles with finite mass $M^*_L$ and momenta $p<p_i$.
When $\lambda_0\to0$, the
density of states near the Fermi level tends to infinity,
$N(0)\sim M^*_{FC}\sim 1/\Delta_1$. The quasiparticles with
$M^*_{FC}$ occupy the same energy level and form pairs with
binding energy of the order of $\Delta_1$ and with average
momentum $p_0$, $p_0/p_F\sim (p_f-p_i)/p_F\ll 1$. Therefore, this
state strongly resembles the Bose-Einstein condensate, in which 
quasiparticles occupy the same energy level. But the FC 
quasiparticles have to be
spread over the range $(p_f-p_i)$ in momentum space due to the
exclusion principle. In contrast to the Bose-Einstein
condensation, the fermion condensation temperature  is $T_c=0$.
And in contrast to the ordinary superconductivity, FC
is formed due to the Landau repulsive 
interaction $F({\bf p},{\bf p}_1)$ rather 
than by the relatively weak attractive quasiparticle-quasiparticle pairing
interaction $\lambda_0V({\bf p}_1,{\bf p}_2)$.

If $\lambda_0\neq 0$, $\Delta_1$ becomes finite, leading to a
finite value of the effective mass $M^*_{FC}$, which
can be obtained from Eq. (3.4) upon differentiating both parts of this 
equation with respect to the momentum $p$ and using Eq. (2.5) 
\cite{ms,shb,ars} \beq M^*_{FC}\ \simeq\
p_F\frac{p_f-p_i}{2\Delta_1}\ . \eeq As to the energy scale, it is
determined by the parameter $E_0$: \beq E_0\ =\ \varepsilon({\bf
p}_f)-\varepsilon({\bf p}_i)\ \simeq\ 
p_F\frac{(p_f-p_i)}{M^*_{FC}}\ \simeq\ 2\Delta_1\ . \eeq

\subsection{Superconducting state at finite temperatures}

Let us assume that the range of FC is small, that is $(p_f-p_i)/p_F\ll1$,
and $2\Delta_1\ll T_f$ so that the order parameter $\kappa({\bf
p})$ is governed mainly by FC \cite{ms,shb}. To solve Eq. (3.5)
analytically, we take the BCS
approximation for the interaction \cite{bcs}:  $\lambda_0V({\bf
p},{\bf p}_1)=-\lambda_0$ if $|\varepsilon({\bf p})-\mu|\leq
\omega_D$, i.e. the interaction is zero outside this region, with
$\omega_D$ being the characteristic energy, e.g. that of phonon.
As a result,
the gap becomes dependent only on the temperature, $\Delta({\bf
p})=\Delta_1(T)$, being independent of the momentum, and Eq. (3.5)
takes the form \beq 1\ =\
N_{FC}\lambda_0\int\limits_0^{E_0/2}\frac{d\xi}
{\sqrt{\xi^2+\Delta^2_1(0)}}
+N_{L}\lambda_0\int\limits_{E_0/2}^{\omega_D}\frac{d\xi}
{\sqrt{\xi^2+\Delta^2_1(0)}}\ . \eeq Here we set
$\xi=\varepsilon({\bf p})-\mu$ and introduce the density of states
$N_{FC}$ in the $(p_f-p_i)$ range, or $E_0$ range. It follows from Eq.
(3.7), that $N_{FC}=(p_f-p_F)p_F/2\pi\Delta_1(0)$. The density of states
$N_{L}$ in the range $(\omega_D-E_0/2)$ has the standard form
$N_{L}=M^*_{L}/2\pi$. If the energy scale $E_0\to 0$, Eq. (3.9)
reduces to the BCS equation. On the other hand, assuming that
$E_0\leq2\omega_D$ and omitting the second integral on the right
hand side of Eq. (3.9), we obtain \beq \Delta_1(0)\ =\
\frac{\lambda_0 p_F(p_f-p_F)}{2\pi}\ln\left(1+\sqrt2\right)\ =\
2\beta\varepsilon_F \frac{p_f-p_F}{p_F}\ln\left(1+\sqrt2\right) ,
\eeq where the Fermi energy $\varepsilon_F=p_F^2/2M^*_L$, and the
dimensionless coupling constant $\beta$ is given by the relation
$\beta=\lambda_0 M^*_L/2\pi$. Taking the usual values of 
$\beta$ for conventional superconductors as 
$\beta\simeq 0.3$, and assuming
$(p_f-p_F)/p_F\simeq 0.2$, we get
from Eq. (20) a large value of $\Delta_1(0)\sim 0.1\varepsilon_F$,
while for conventional superconductors  one has 
a much smaller gap, $\Delta_1(0)\sim
10^{-3}\varepsilon_F$. Taking into account the omitted above integral,
we obtain \beq \Delta_1(0)\ \simeq\ 2\beta\varepsilon_F
\frac{p_f-p_F}{p_F}\ln\left(1+\sqrt2\right)\left(1+\beta
\ln\frac{2\omega_D}{E_0}\right). \eeq It is seen from Eq. (3.11)
that the correction due to the second integral is small, provided
$E_0\simeq2\omega_D$. Below we shall show that $2T_c\simeq \Delta_1(0)$,
which leads to the conclusion that the isotope effect is absent
since $\Delta_1$ is independent of $\omega_D$. But this effect is
restored as $E_0\to 0$. 

Assuming $E_0\sim\omega_D$ but
$E_0>\omega_D$, we see that Eq. (3.11) has no standard solutions
$\Delta(p)=\Delta_1(T=0)$ because $\omega_D<\varepsilon(p\simeq
p_f)-\mu$ and the interaction vanishes at these momenta. The only
way to obtain solutions is to restore the condition
$E_0<\omega_D$. For instance, we can define such a momentum
$p_D<p_f$ that \beq \Delta_1(0)\ =\ 2\beta\varepsilon_F\
\frac{p_D-p_F}{p_F} \ln\left(1+\sqrt2\right)\ =\ \omega_D\ , \eeq
while the other part in the $(p_f-p_i)$ range can be occupied by a
gap $\Delta_2$ of the different sign, $\Delta_1/\Delta_2<0$. It
follows from Eq. (3.12) that the isotope effect is preserved, while
both gaps can have $s$-wave symmetry.

At $T\simeq T_c$, Eqs. (3.7) and (3.8) are replaced by Eqs. (2.12) and 
(2.14), which is valid also at $T_c\leq T\ll T_f$ \beq M^*_{FC}\simeq 
p_F\frac{p_f-p_i}{4T_c},\,\,\, E_0\simeq 4T_c;\,\,{\mathrm 
{if}}\,\,T_c\leq T\, {\mathrm {then}}\,, M^*_{FC}\simeq 
p_F\frac{p_f-p_i}{4T},\,\,\, E_0\simeq 4T\ . \eeq Equation (3.9) is 
replaced by its conventional finite temperature generalization 
\begin{eqnarray}
1 & = & N_{FC}\lambda_0\int_0^{E_0/2}
\frac{d\xi}
{\sqrt{\xi^2+\Delta^2_1(T)}}
\tanh\frac{\sqrt{\xi^2+\Delta^2_1(T)}}{2T}\ +
\nonumber\\
& + & N_{L}\lambda_0\int_{E_0/2}^{\omega_D}
\frac{d\xi}{\sqrt{\xi^2+\Delta^2_1(T)}}
\tanh\frac{\sqrt{\xi^2+\Delta^2_1(T)}}{2T}\ .
\end{eqnarray}
Putting $\Delta_1(T\to T_c)\to 0$, we obtain from Eq. (3.14) \beq
2T_c\ \simeq\ \Delta_1(0)\ , \eeq with $\Delta_1(T=0)$ being given
by Eq. (3.11). Upon comparing Eqs. (3.7), (3.8) and (3.13), (3.15), we 
see that $M^*_{FC}$ and $E_0$ are almost temperature independent at 
$T\leq T_c$.

\subsection{Bogoliubov quasiparticles}

It is seen
from Eq. (3.5) that the superconducting gap depends on the
single-particle spectrum $\varepsilon({\bf p})$. On the other
hand, it follows from Eq. (3.4) that $\varepsilon({\bf p})$
depends on $\Delta({\bf p})$, provided that at $\lambda_0\to 0$ Eq.
(3.6) has the solution determining the existence of FC. Let us
assume that $\lambda_0$ is small enough, so that the particle-particle
interaction $\lambda_0 V({\bf p},{\bf p}_1)$ can only lead to a
small perturbation of the order parameter $\kappa({\bf p})$
determined by Eq. (3.6). It
follows from Eq. (3.7) that the effective mass and the density of
states $N(0)\propto M^*_{FC}\propto 1/\Delta_1$ are finite.
As a result, we conclude that in contrast to the conventional theory of
superconductivity the single-particle spectrum $\varepsilon({\bf
p})$ strongly depends on the superconducting gap and we have to
solve Eqs. (2.2) and (3.5) in a self-consistent way. On the other
hand, let us assume that Eqs. (2.2) and (3.5) are solved, and
the effective mass $M^*_{FC}$ is determined. Now one
can fix the quasiparticle dispersion $\varepsilon({\bf p})$ by choosing
the effective mass $M^*$ of the system in question equal to the obtained
$M^*_{FC}$ and then solve Eq. (3.5) as it is done in the case of
the conventional BCS theory of superconductivity \cite{bcs}. As a
result, one observes that the superconducting state is
characterized by the Bogoliubov quasiparticles (BQ) \cite{bogol} with
the dispersion
$$E({\bf p})=\sqrt{(\varepsilon({\bf p})-\mu)^2+\Delta^2({\bf p})},$$
and the normalization condition for the coherence
factors $v({\bf p})$ and $u({\bf p})$ is
held.  We arrive to the conclusion that the observed features agree
with the behavior of BQ predicted from the BCS theory. This observation
suggests that the superconducting state with FC is BCS-like and
implies the basic validity of the BCS formalism in describing the
superconducting state \cite{asjetpl}. Although the maximum value
of the superconducting gap given by Eq. (3.11) and
other exotic properties are determined
by the presence of the fermion condensate.
It is exactly the case that was
observed experimentally in the high-$T_c$ cuprate
Bi$_2$Sr$_2$Ca$_2$Cu$_3$O$_{10+\delta}$ \cite{mat}.

We have returned back to the LFL theory since the high
energy degrees of freedom are eliminated and the quasiparticles
are introduced. The only difference between the Landau Fermi-liquid,
which serves as a basis when constructing the superconducting state,
and Fermi liquid after FCQPT is that we have to expand the number of
relevant low energy degrees of freedom by introducing a new type of
quasiparticles with the effective mass $M^*_{FC}$ given by Eq. (3.7)
and the energy scale $E_0$ given by Eq. (3.8). Therefore, the dispersion
$\varepsilon({\bf p})$ is characterized by two effective masses $M^*_L$
and $M^*_{FC}$ and by the scale $E_0$, which define the low temperature
properties including the line shape of quasiparticle excitations
\cite{ms,shb,ams}, while the dispersion of BQ has the standard form.
We note
that both the effective mass $M^*_{FC}$ and the scale $E_0$ are
temperature independent at $T<T_c$, where $T_c$ is the critical
temperature of the superconducting phase transition \cite{ams}.
At $T>T_c$, the effective mass $M^*_{FC}$ and the scale $E_0$ are
given by Eqs. (2.12) and (2.14) respectively.
Obviously, we cannot directly relate these new Landau Fermi-liquid
quasiparticle excitations with the quasiparticle excitations of an
ideal Fermi gas because the system in question has undergone FCQPT.
Nonetheless, the main basis of the Landau Fermi liquid theory survives
FCQPT: the notion of order parameter is preserved, and the
low energy excitations of the strongly correlated liquid with
FC are quasiparticles.

As it was shown above, properties of these new quasiparticles are
closely related to the properties of the superconducting state. We
may say that the quasiparticle system in the range $(p_f-p_i)$
becomes very ``soft'' and is to be considered as a strongly
correlated liquid. On the other hand, the system's properties and
dynamics are dominated by a strong collective effect having its
origin in its proximity to FCQPT and determined by the macroscopic
number of quasiparticles forming FC in the range $(p_f-p_i)$. Such a 
system cannot be perturbed by the scattering of individual 
quasiparticles and has features of a ``quantum protectorate" and 
demonstrates the universal behavior \cite{ms,shb,rlp,pa}.  A few 
remarks related to the quantum protectorate and the universal behavior 
\cite{rlp} are in order here.  As the Landau theory of Fermi liquid, 
the theory of the high-temperature superconductivity based on FCQPT 
deals with the quasiparticles which are elementary excitations of low 
energy.  As a result, this theory gives general qualitative 
description of both the superconducting and normal states.  
Of course, one can choose the 
phenomenological parameters and obtain the 
quantitative consideration of the superconductivity as it can be done 
in the framework of the Landau theory when describing a normal 
Fermi-liquid, say $^3$He.  Therefore, we conclude that any theory, 
which is capable to describe FC and compatible with the BCS theory,  
will give the qualitative picture of the superconducting state and 
the normal state that coincides with the picture based on FCQPT.  Both 
of the approaches can be agreed at a numerical level provided the 
corresponding parameters are adjusted. For example, since the formation 
of  the flat band is possible in the Hubbard model \cite{irk}, 
generally speaking  one can repeat the results of the theory based on FCQPT 
in the Hubbard model.  It is appropriate to mention here that the 
corresponding numerical description limited to the case of $T=0$ has 
been obtained within the Hubbard model \cite{rand,pwa}.

\subsection{Pseudogap} 

Now let us discuss some special features of the superconducting
state with FC \cite{sh,ms1}.
Consider two possible types of the superconducting gap
$\Delta({\bf p})$, namely that  
given by Eq. (3.5) and defined by the interaction
$\lambda_0V({\bf p},{\bf p}_1)$. If this interaction is dominated
by an attractive phonon-mediated attraction,
the solution of Eq. (3.5)
with the $s$-wave, or the $s+d$ mixed waves will have the lowest
energy. Provided the pairing interaction $\lambda_0V({\bf
p}_1,{\bf p}_2)$ is the combination of both the attractive
interaction and sufficiently strong repulsive interaction, the
$d$-wave superconductivity can take place (see e.g.
\cite{kug,abr}). But both the $s$-wave 
symmetry and the $d$-wave  
one lead to approximately the same value of the gap $\Delta_1$ in
Eq. (3.11) \cite{ams}. Therefore the non-universal pairing
symmetries in high-$T_c$ superconductivity are likely the result of
the pairing interaction, while
the $d$-wave pairing symmetry is not
essential. This point of view is supported by the data
\cite{skin,bis,skin1,skin2,chen}.

We can define the critical temperature $T^*$, at which
the superconductivity vanishes, as the temperature when
$\Delta_1(T^*)\equiv 0$. At $T\geq T^*$, Eq. (3.14) has only the
trivial solution $\Delta_1\equiv 0$. On the other hand, the critical
temperature $T_c$ can
be defined as a temperature, at which the superconductivity
disappears while the gap occupies only part of the Fermi surface.
Thus, as we shall see there are  two different temperatures
$T_c$ and $T^*$ in the case of the $d$-wave symmetry of the gap.
It was shown \cite{ars,sh} that in the presence
of FC there exist nontrivial solutions of Eq. (3.14) at $T_c\leq
T\leq T^*$ when the BCS like interaction is replaced by the
pairing interaction $\lambda_0V({\bf p}_1,{\bf p}_2)$, which includes
the strong repulsive interaction leading to
$d$-wave superconductivity. In that case,
the gap $\Delta({\bf p})$ as a function
of angle $\phi$, $\Delta({\bf p})=\Delta(p_F,\phi)$ possesses new
nodes at $T>T_{node}$, as it is illustrated by  Fig. 2 \cite{ars}.

\begin{figure}[!ht]
\begin{center}
\includegraphics[width=0.77\textwidth]{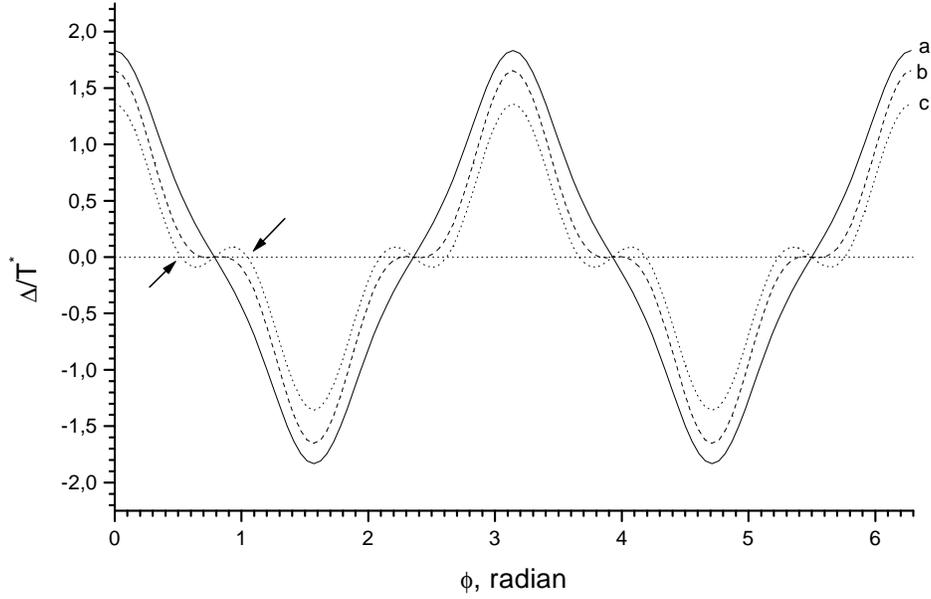}
\end{center}
\caption{The gap  $\Delta(p_F,\phi)$
as a function of $\phi$ calculated at three different temperatures
expressed in terms of $T_{node}\simeq T_c$, while $\Delta$ is
presented in terms of $T^*$. Curve (a), solid line, shows the gap
calculated at temperature $0.9 T_{node}$. Curve (b), dashed line,
presents the same at $T=T_{node}$, and curve (c), dotted line, shows the gap
calculated at temperature $1.2 T_{node}$.
Note the important difference in the
curves (b) and (c) compared with curve (a) due to the flattening of the 
curves (b) and (c) about the nodes. 
The two arrows indicate the area $\theta_c$ emerging at $T_{node}$.}
\label{Fig2} \end{figure}

We show in Fig. 2 the ratio $\Delta(p_F,\phi)/T^*$
calculated at three different
temperatures $0.9\,T_{node}$, $T_{node}$, and $1.2\,T_{node}$.
An important difference between curve (a) and (b) and (c)
is additional variations of the curves (b) and (c) marked by the two arrows.
As seen from Fig. 2, the flattening occurs
due to the appearance of the two new nodes. They appear at
$T=T_{node}$ and move apart with increase of the 
temperature confining the
area $\theta_c$ shown by the two arrows in Fig. 2. It is also seen
from Fig. 2 that the gap $\Delta$ is extremely small over the range
$\theta_c$.  It was recently shown in a number of papers (see, e.g.,
\cite{nm,th}) that there exists an interplay between the magnetism
and the superconductivity order parameters, leading to the damping of
the magnetism order parameter below $T_c$.  And vise versa, one can
anticipate the damping of the superconductivity order parameter by the
magnetism. Thus, we conclude that the gap in the range $\theta_c$
can be destroyed by strong antiferromagnetic correlations (or by spin
density waves) existing in optimally doped
and underdoped superconductors.
It is believed that impurities can easily destroy the gap
$\Delta({\bf p})$ in the considered area. As a result, the 
superconducting gap vanishes in the macroscopic area $\theta_c$ and 
causes the superconductivity to die out. We have to conclude  
that $T_c\simeq T_{node}$, with the exact value of $T_c$ defined by the 
competition between the antiferromagnetic correlations (or spin density 
waves) and the superconducting correlations over the range $\theta_c$. 
The behavior and the shape of the pseudogap is very similar to the ones 
of the superconducting gap as it is seen from Fig. 2.
The main difference seen from Fig. 2 is that the pseudogap vanishes along
segments $\theta_c$ of the Fermi surface, while the gap vanishes at 
the isolated $d$-wave nodes.  Our calculations show that the function 
$\theta_c(x)$ increases very fast at small values of $x$,
$\theta_c(x)\simeq\sqrt{x}$. Therefore
$T_c$ has to be close to $T_{node}$.
This result is in accord with numerical calculations of
the function $\theta_c([T-T_c]/T_c)$ plotted in Fig. 3.

\begin{figure}[!ht]
\begin{center}
\includegraphics[width=0.57\textwidth]{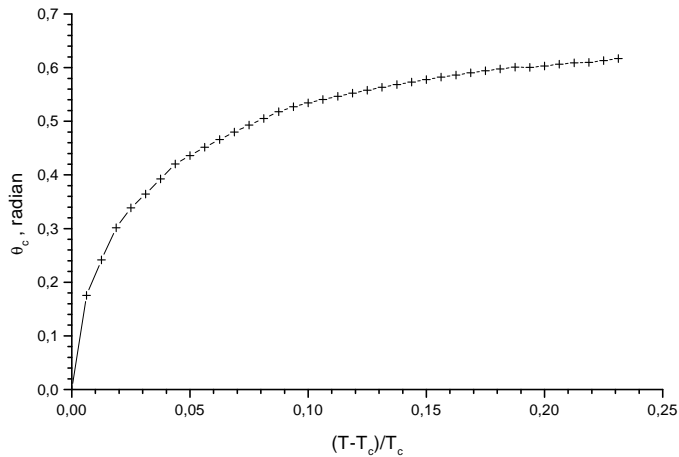}
\end{center}
\caption{ Calculated angle $\theta_c$, pulling apart
the two nodes, as a function of $(T-T_c)/T_c$.}
\label{Fig3} \end{figure}

Thus, we observe that the pseudogap state appears at temperatures
$T\geq T_c\simeq T_{node}$ and vanishes  at $T\geq T^*$ when Eq. (3.14)
has only the trivial solution $\Delta_1\equiv 0$. Quite naturally, one
has to recognize that $\Delta_1$ scales with $T^*$, while
Eq. (3.15) takes the form \beq 2T^*\ \simeq\ \Delta_1(0)\ . \eeq
It can then be concluded that the
temperature $T^*$ has the physical meaning of the BCS transition
temperature between the state with the order parameter $\kappa\neq 0$
and the normal state.

At $T<T_c$ quasiparticle excitations are characterized by sharp peaks.
When temperature becomes $T>T_c$ and  $\Delta(\theta)\equiv 0$ in the range
$\theta_c$, there appear normal quasiparticle excitations with the
width $\gamma$ along the segments $\theta_c$ of the Fermi surface.
There exists the pseudogap outside the segments $\theta_c$, and the Fermi 
level is occupied by the BCS-type excitations. Both types of 
excitations have the width of the same order of magnitude 
transferring their energy and momentum to the normal quasiparticle  
excitations. We now estimate $\gamma$.  For the entire Fermi level 
occupied by the normal state, the width is equal to $\gamma\approx 
N(0)^3 T^2/\varepsilon(T)^2$, with the density of states $N(0)\sim 
M^*(T)\sim  1/T$ (see Eq. (2.12)). The dielectric constant 
$\varepsilon(T)\sim N(0)$, and the width  $\gamma$  becomes $\gamma\sim 
T$ \cite{dkss}.  In our case, however, only a part of the Fermi surface 
within $\theta_c$ is occupied by the normal excitations.  Therefore, the 
number of states allowed for quasiparticles and for quasiholes is 
proportional to $\theta_c$, and the factor $T^2$ is replaced by 
$T^2\theta_c^2$.  Taking this into account, we obtain 
$\gamma\sim \theta_c^2 T\sim T(T-T_c)/T_c\sim (T-T_c)$. Here we have 
omitted the small contribution coming from the BCS-type excitations.
That is why the width $\gamma$ vanishes at $T=T_c$.  Thus, the
presented above analysis shows that in the pseudogap state at $T>T_c$, the
superconducting gap smoothly transforms into the pseudogap. The
excitations of that area of the Fermi surface 
that has the gap are of  the same 
width $\gamma\sim (T-T_c)$. The region occupied by the pseudogap
is shrinking with increasing temperature. It is worth noting that
the normal state resistivity $\rho(T)$  behaves as
$\rho(T)\propto T$ due to $\gamma\sim (T-T_c)$. Obviously at $T>T^*$
the behavior $\rho(T)\propto T$ remains valid up to
temperatures $T\sim T_f$,
and $T_f$ can be as high as the Fermi energy, provided that
FC occupies noticeable part of the Fermi volume, see Eq. (2.11).

The temperature $T_{node}$ is determined mainly by the repulsive 
interaction being part of the pairing interaction
$\lambda_0V({\bf p}_1,{\bf p}_2)$.  In its turn, the repulsive interaction
can depend on the properties of materials such as the composition, 
doping, etc.  Since the superconductivity is destroyed at $T_c$, the 
ratio $2\Delta_1/T_c$ can vary in a wide range and strongly depends 
upon the material's properties, as it follows from considerations given 
above  \cite{ars,sh,ms1}. The ratio $2\Delta_1/T_c$ can reach very high 
values. For instance, in the case of Bi$_2$Sr$_2$CaCu$_2$Q$_{6+\delta}$ 
where the superconductivity and the pseudogap are considered to be of 
the common origin, $2\Delta_1/T_c$ is about 28, while the ratio 
$2\Delta_1/T^*\simeq 4$, which is in agreement with the
experimental data for various cuprates \cite{kug}. Note that Eq.
(3.16) gives also good description of the maximum gap $\Delta_1$ in
the case of the d-wave superconductivity, because the different
regions with the maximum absolute value of $\Delta_1$ and the
maximal density of states can be considered as disconnected
\cite{abr}. Therefore the gap in this region is formed by the
attractive phonon interaction, which is approximately independent
of the momenta. We can also conclude, that in the case of
the $s$-wave pairing the pseudogap phenomenon is absent because 
there is no the sufficiently strong repulsive interaction.
Thus,  the transition from superconducting gap to pseudogap
can take place only in the case of the $d$-wave pairing,
so that the superconductivity is destroyed at
$T_c\simeq T_{node}$, with the superconducting gap being smoothly 
transformed into the pseudogap, which closes at some temperature 
$T^*>T_c$ \cite{ars,sh,ms1}. In the case of the $s$-wave pairing, we 
expect the absence of the pseudogap phenomenon in accordance with the 
observations (see e.g. \cite{chen} and references therein).

\subsection{Dependence of $T_c$ on the doping}

We now turn to consideration of the maximum value of the
superconducting gap $\Delta_1$ as a function of the number density $x$ of
the mobile charge carriers which is proportional to the doping.
Using  Eq. (2.15), we can rewrite Eq. (3.10) as follow
\beq\frac{\Delta_1}{\varepsilon_F}\sim\beta\frac{(x_{FC}-x)x}{x_{FC}}.\eeq
Here we take into account
that the Fermi level $\varepsilon_F\propto p_F^2$, the density
$x\sim p_F^2/(2M^*_L)$, and thus, $\varepsilon_F\propto x$. We can
reliably assume that $T_c\propto\Delta_1$ because the empirically
obtained simple bell-shaped curve of $T_c(x)$ in the high temperature
superconductors \cite{varma} should have only a smooth dependence upon $x$.
Then, $T_c(x)$ in accordance with the data has the same bell-shaped form
\cite{ams3} \beq T_c(x)\ \propto\ \beta (x_{FC}-x)x.
\eeq

As an example of the implementation of the previous analysis, let
us consider the main features of a room-temperature
superconductor. The superconductor has to be a quasi
two-dimensional structure like cuprates.
From Eq. (3.10) it follows, that $\Delta_1\sim \beta \varepsilon_F\propto
\beta/r_s^2$. Noting that FCQPT in 3D systems takes place at
$r_s\sim 20$ and in 2D systems at $r_s\sim 8$ \cite{ksz}, we can
expect that $\Delta_1$ of 3D systems comprises 10\% of the
corresponding maximum value of 2D superconducting gap, reaching a
value as high as 60 meV for underdoped crystals with $T_c=70$
\cite{mzo}. On the other hand, it is seen from Eq. (3.10), that
$\Delta_1$ can be even large, $\Delta_1\sim 75$ meV, and one can
expect $T_c\sim 300$ K in the case of the $s$-wave pairing as it
follows from the simple relation $2T_c\simeq \Delta_1$.  Indeed,
we can safely take $\varepsilon_F\sim 300$ meV, $\beta\sim 0.5$
and $(p_f-p_i)/p_F\sim0.5$. Thus, a possible room-temperature
superconductor has to be the $s$-wave superconductor in order to
get rid of the pseudogap phenomena, which can tremendously reduce the
transition temperature $T_c$. The density $x$ of the mobile charge
carriers must satisfy the condition $x\leq x_{FC}$ and be adjustable 
to reach the optimal doping level $x_{opt}\simeq x_{FC}/2$.

\subsection{The gap and specific heat near $T_c$}

Now we turn to the calculations of the gap and the specific heat
at the temperatures $T\to T_c$. It is worth noting that this
consideration is valid provided $T^*=T_c$. Otherwise the
considered below discontinuity in the specific heat is smoothed out 
over the temperature range $T^{*}\div T_c$. For the sake of simplicity, 
we calculate the main contribution to the gap and the specific heat 
coming from FC. The function $\Delta_1(T\to T_c)$ is found
from Eq. (3.14) by expanding the right hand side of the first
integral in powers of $\Delta_1$ and omitting the contribution
from the second integral on the right hand side of Eq. (3.14). This
procedure leads to the following equation \cite{ams} \beq
\Delta_1(T)\ \simeq\ 3.4T_c\sqrt{1-\frac{T}{T_c}}\ . \eeq Thus,
the gap in the spectrum of the single-particle excitations has
the usual behavior. To calculate the specific heat, the
conventional expression for the entropy $S$ \cite{bcs} can be used
\beq S\ =\ -2\int\left[f({\bf p})\ln f({\bf p}) +(1-f({\bf
p}))\ln(1-f({\bf p}))\right]\frac{d{\bf p}}{(2\pi)^2}\ , \eeq
where \beq f({\bf p})=\frac{1}{1+\exp[E({\bf p})/T]}\ ; \quad
E({\bf p}) =\sqrt{(\varepsilon({\bf p})-\mu)^2+\Delta_1^2(T)}\ .
\eeq The specific heat $C$ is determined by the equation
\begin{eqnarray}
C &=& T\frac{dS}{dT}\ \simeq\ 4\frac{N_{FC}}{T^2}\int\limits_0^{E_0}
f(E)(1-f(E))\left[E^2+T\Delta_1(T)
\frac{d\Delta_1(T)}{dT}\right]d\xi\ +
\nonumber\\
&+& 4\frac{N_{L}}{T^2}\int\limits_{E_0}^{\omega_D}
f(E)(1-f(E))\left[E^2+T\Delta_1(T)\frac{d\Delta_1(T)}{dT}\right]d\xi\ .
\end{eqnarray}
In deriving Eq. (3.22) we again used the variable $\xi$ and the
densities of states $N_{FC}$ and $N_{L}$, just as before in
connection with Eq. (3.9), and employed the notation
$E=\sqrt{\xi^2+\Delta_1^2(T)}$. Eq. (3.22) predicts the
discontinuity $\delta C=C_s-C_n$ in the specific heat $C$ at $T_c$ because
of the two last term in the square brackets on right hand side
of Eq. (3.22). Here, $C_s$ and $C_n$ are the
specific heat of the superconducting
state and the normal one respectively. Using Eq.
(3.19) to calculate the first term on
the right hand side of Eq. (3.22), we obtain \cite{ams}
\beq \delta\, C(T_c)
\simeq\ \frac3{2\pi^2}\ (p_f-p_i)\,p_F^2. \eeq This is
in contrast to the
conventional result where the discontinuity is a linear function of
$T_c$. $\delta C(T_c)$ is independent of the critical temperature $T_c$
because as seen from Eq. (3.13) the density of states varies
inversely with $T_c$. Note that in deriving Eq. (3.23) we took into
account the main contribution coming from  FC. This term
vanishes as soon as $E_0\to0$ and the second integral
on the right hand side of Eq. (3.22)
gives the conventional result.

A few remarks are in order here. As we shall 
demonstrate in Section 9, Eq. (9.4), 
the specific heat of systems with FC behaves as $C(T)\propto \sqrt{T/T_f}$.
The specific heat discontinuity given by Eq. (3.23) is
temperature independent. As a result, we obtain that
\beq \frac{\delta C(T_c)}{C_n(T_c)}
\sim\ \sqrt{\frac{T_f}{T_c}}\frac{(p_f-p_i)}{p_F}. \eeq
In contrast to the conventional case of normal superconductors, 
when $\delta C(T_c)/C_n(T_c)=1.43$ \cite{lanl1}, 
it is seen from Eq. (3.24) that the ratio $\delta C(T_c)/C_n(T_c)$
is not the constant and can
be very large provided that $T_c/T_f\ll 1$.

\section{The dispersion and lineshape of the single-particle spectra}
\setcounter{equation}{0}

The newly discovered additional
energy scale manifests itself as a break in the quasiparticle
dispersion near $(50-70)\,$meV, which results in a drastic change
of the quasiparticle velocity \cite{blk,krc,vall}. Such a behavior
is qualitatively different from what one could expect in a normal
Fermi liquid. Moreover, this behavior can hardly be understood in
the frames of either the Marginal Fermi Liquid theory or the quantum 
protectorate since there are no additional energy scales, or 
parameters, in these theories \cite{pa,var,varm1}. 
One could suggest that 
this observed strong self-energy effect, leading to the new energy 
scale, is due to the electron coupling with collective excitations. But 
in that case one has to give up the quantum protectorate idea, which 
would contradict observations \cite{rlp,pa}. 

As we have seen in Sections 2 and 3, Eqs. (2.12) and (3.7),
the  system with FC is characterized by
two effective masses: $M^*_{FC}$
that is related to the single particle spectrum at lower energy
scale, and $M^*_L$ describing the spectrum at higher energy scale.
These two effective masses manifest itself as a break in the
quasiparticle dispersion, which can be approximated by two straight
lines intersecting at the energy $E_0$, Eqs. (2.14) and (3.8).
This break takes place at temperatures $T\ll T_f$ when
the system both in its
superconducting state and normal one. This beahvior is in good agreement
with the experimental findings \cite{blk}.
It is pertinent to note that at $T<T_c$, the
effective mass $M^*_{FC}$ does not depend on the momenta $p_F$, $p_f$
and $p_i$ as it follows from Eqs. (3.7) and (3.10),
\beq M^*_{FC}\sim \frac{2\pi}{\lambda_0}.\eeq
Thus, it is seen from Eq. (4.1) that $M^*_{FC}$
does not depend on $x$. This result is in good agreement with 
experimental facts \cite{ino,zhou,padil}. 
The same is true for the dependence 
of the Fermi velocity $v_F=p_F/M^*_{FC}$ on $x$, because the Fermi 
momentum $p_F\sim\sqrt{n}$ only slightly depends on the electron density
$n=n_0(1-x)$ \cite{ino,zhou}.  Here $n_0$ is the single-particle
electron density at the half-filling.

Since $\lambda_0$ is the coupling constant defining the pairing
interaction, for example phonon-electron interaction,
it could be reasonable to expect that the break in the quasiparticle 
dispersion comes from the phonon-electron interaction. The phonon 
scenario could explain the persistence of the break at $T>T_c$, since 
phonons are $T$-independent. On the other hand, it was shown that the 
quasiparticle dispersion tends to recover 
to the conventional one-electron dispersion 
when the energy is well above the typical phonon energies \cite{vald}. 
The experimental observations do not show that the recovery to the 
one-electron dispersion takes place \cite{blk}.

The lineshape function $L(q,\omega)$ of the single-particle
spectrum is a function of two variables. Measurements
carried out at a fixed binding energy $\omega=\omega_0$,
with $\omega_0$ being the energy of a single-particle excitation, 
determine the lineshape $L(q,\omega=\omega_0)$ as a function of the 
momentum $q$.  We have shown above that $M^*_{FC}$ is finite and 
constant at $T\leq T_c$. Therefore, at excitation energies $\omega\leq 
E_0$, the system behaves like an ordinary superconducting Fermi liquid 
with the effective mass given by Eq. (3.7) \cite{ms,shb,ars}. At 
$T_c\leq T$ the low energy effective mass $M^*_{FC}$ is finite and is 
given by Eq. (2.12).  Once again, at the energies $\omega<E_0$, the 
system behaves as a Fermi liquid, the single-particle spectrum is well 
defined while the width of single-particle excitations is of the order 
of $T$ \cite{ms,shb,dkss}. This behavior was observed in experiments 
measuring the lineshape at a fixed energy \cite{vall,feng}.

The lineshape can also be determined as a function of $\omega$,
$L(q=q_0,\omega)$, at a fixed $q=q_0$.  At small $\omega$, the
lineshape resembles the considered above, and $L(q=q_0,\omega)$
has the characteristic maximum and width. At energies $\omega\geq
E_0$, the contribution coming form  quasiparticles with
the mass $M^*_{L}$ become important,
leading to the increase of $L(q=q_0,\omega)$. As a result, the
function $L(q=q_0,\omega)$ possesses
the known peak-dip-hump structure \cite{dess} directly defined by the
existence of the two effective masses $M^*_{FC}$ and $M^*_L$
\cite{ms,shb,ars}. We can conclude that in contrast to the Landau
quasiparticles, these quasiparticles have a more complicated lineshape.

To develop deeper quantitative and analytical insight into
the problem, we use the Kramers-Kr\"{o}nig transformation to
construct the imaginary part ${\mathrm{Im}}\Sigma({\bf
p},\varepsilon)$ of the single-particle 
self-energy $\Sigma({\bf p},\varepsilon)$ 
starting with the real one ${\mathrm{Re}}\Sigma({\bf
p},\varepsilon)$, which defines the effective mass \cite{mig} \beq
\left. \frac1{M^*}\ =\ \left(\frac{1}{M}+\frac{1}{p_F}
\frac{\partial{\mathrm{Re}}\Sigma}{\partial p}\right)\right/
\left(1-\frac{\partial{\mathrm{Re}}\Sigma}{\partial
\varepsilon}\right). \eeq Here $M$ is the bare mass, while the
relevant momenta $p$ and energies $\varepsilon$ obey the following
strong inequalities: $|p-p_F|/p_F\ll 1$, and
$\varepsilon/\varepsilon_F\ll 1$. We take
${\mathrm{Re}}\Sigma({\bf p},\varepsilon)$ in the simplest form
which accounts for the change of the effective mass at the energy
scale $E_0$: \beq \mbox{Re }\Sigma({\bf
p},\varepsilon)=-\varepsilon
\frac{M^*_{FC}}M\!+\!\left(\varepsilon-\frac{E_0}2\right)
\frac{M^*_{FC}\!-\!M^*_L}M\left[\theta\left(
\varepsilon\!-\!\frac{E_0}2\right)
+\theta\left(\mbox{-}\varepsilon\!-\!\frac{E_0}2\right)\right].
\eeq Here $\theta(\varepsilon)$ is the step function. Note that in
order to ensure a smooth transition from the single-particle
spectrum characterized by $M^*_{FC}$ to the spectrum defined by
$M^*_{L}$ the step function is to be substituted by some smooth
function. Upon inserting Eq. (4.3) into Eq. (4.2) we can check that
inside the interval $(-E_0/2,E_0/2)$ the effective mass 
can be estimated as $M^*\simeq
M^*_{FC}$, and outside the interval it is $M^*\simeq M^*_{L}$. By
applying the Kramers-Kr\"{o}nig transformation to
${\mathrm{Re}}\Sigma({\bf p},\varepsilon)$, we obtain the
imaginary part of the self-energy \cite{ams} \beq \mbox{Im
}\Sigma({\bf p},\varepsilon)\sim
\varepsilon^2\frac{M^*_{FC}}{\varepsilon_F M}+
\frac{M^*_{FC}-M^*_L}M\left(\!\varepsilon\ln\!\left|
\frac{\varepsilon\!+\!E_0/2}{\varepsilon\!-\!E_0/2}\right|\!+\!
\frac{E_0}2\ln\!\left|\frac{\varepsilon^2\!-\!E^2_0/4}{E^2_0/4}
\right|\right). \eeq We see from Eq. (4.4) that at
$\varepsilon/E_0\ll 1$ the imaginary part is proportional to
$\varepsilon^2$, at $2\varepsilon/E_0\simeq 1$
${\mathrm{Im}}\Sigma\sim \varepsilon$, and at $E_0/\varepsilon\ll 1$
the main contribution to the imaginary part is approximately
constant. This is the behavior that gives rise to the known
peak-dip-hump structure. It is seen from Eq. (4.4) that when
$E_0\to 0$ the second term on the right hand side tends to zero
and the single-particle excitations become  better defined,
resembling the situation in a normal Fermi-liquid, and the
peak-dip-hump structure eventually vanishes. On the other hand,
the so called renormaliztaion constant, or  
the quasiparticle renormalization factor, $a({\bf p})$ is given by \cite{mig}
\beq \frac1{a({\bf p})}\ =\ 1-\frac{\partial \mbox{ Re
}\Sigma({\bf p},\varepsilon)}{\partial\varepsilon}\ . \eeq
At $T\leq T_c$, as seen from Eqs. (4.3) and (4.5),
the quasiparticle amplitude at
the Fermi surface rises as the energy scale $E_0$ decreases.  It
follows from Eqs. (3.8) and (3.18) that $E_0\sim(x_{FC}-x)/x_{FC}$.
At $T>T_c$, it is seen from Eqs. (4.3) and (4.5) that the quasiparticle
amplitude rises as the effective mass $M^*_{FC}$ decreases. As seen
from Eqs. (2.12) and (2.15), $M^*_{FC}\sim (p_f-p_i)/p_F\sim
(x_{FC}-x)/x_{FC}$. As a result, we can conclude that the amplitude
rises as the level of doping increases, while the peak-dip-hump
structure vanishes and the single-particle excitations become better
defined in highly overdoped samples. At $x>x_{FC}$, the energy scale
$E_0=0$ and the quasiparticles are normal excitations of Landau Fermi
liquid. It is worth noting that such a behavior was observed
experimentally in highly overdoped Bi2212 where the gap size is about
10~meV \cite{val1}.  Such a small size of the gap verifies that the
region occupied by the FC is small since $E_0/2\simeq \Delta_1$. Then,
recent experimental data have shown that the Landau Fermi liquid does
exist in heavily overdoped non-superconducting
La$_{1.7}$Sr$_{0.3}$CuO$_4$ \cite{nakam,huss}.

\section {Field-induced LFL in heavy electron liquid with FC}
\setcounter{equation}{0}

In this Section we consider the behavior of the 
heavy-electron liquid with FC in
magnetic fields, assuming that the coupling constant is nonzero 
$\lambda_0\neq 0$ but  infinitely small. 
As we have seen in Section 3, 
at $T=0$ the superconducting order parameter $\kappa({\bf p})$ is
finite in the FC range, while the maximum value of the
superconducting gap $\Delta_1\propto \lambda_0$ is infinitely
small. Therefore, any small magnetic field $B \neq 0$ can be
considered as a critical field and will destroy the coherence of
$\kappa({\bf p})$ and thus FC itself. To define the type of FC
rearrangement, simple energy arguments are sufficient. On one
hand, while the field is zero in the sample until then the state with FC
is not destroyed. The energy gain $\Delta E_B$
due to removing the FC state
is $\Delta E_B\propto B^2$ and tends to zero with $B\to 0$. On the
other hand, occupying the finite range $(p_f-p_i)$ in the momentum
space, the function $n_0({\bf p})$ given by Eq. (2.8), or by Eq. (3.6),
leads to a finite gain in the ground state
energy \cite{ks}. Thus, a new ground state replacing FC should
have almost the same energy as the former one. Such a state is
given by the multiconnected Fermi spheres resembling an onion,
where the smooth quasiparticle
distribution function $n_0({\bf p})$ in the $(p_f-p_i)$ range is
replaced by a multiconnected distribution $\nu({\bf p})$
\cite{asp,pogsh}
\begin{equation}
\nu({\bf p})=\sum\limits_{k=1}^n\theta (p-p_{2k-1})\theta (p_{2k}-p).
\end{equation}
Here the parameters $p_i\leq p_1<p_2<\ldots <p_{2n}\leq p_f$ are
adjusted to obey the normalization condition:
$$
\int_{p_{2k-1}}^{p_{2k+3}}\nu({\bf p})
\frac{d{\bf p}}{(2\pi)^3}=\int_{p_{2k-1}}^{p_{2k+3}}
n_0({\bf p})\frac{d{\bf p}}{(2\pi)^3}.
$$
The corresponding multiconnected distribution is shown in Fig. 4.

\begin{figure}[!ht]
\begin{center}
\includegraphics[width=0.47\textwidth]{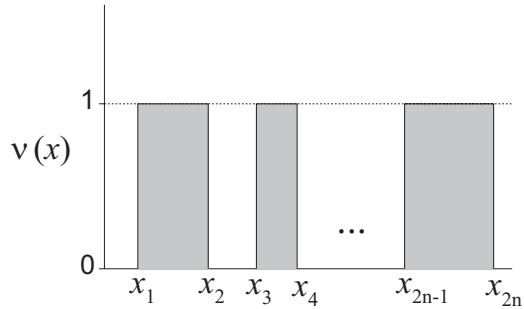}
\end{center}
\caption{ The function  $\nu({\bf p})$ for the multiconnected distribution
replacing the function $n_0({\bf p})$ in the range $(p_f-p_i)$ 
occupied by FC so that $p_i<p_F<p_f$.} \label{Fig4} \end{figure}

For definiteness, let us consider the most interesting case of a 3D
system, while the consideration of a 2D system also goes along the same
line. We note that the idea of multiconnected Fermi spheres, with
production of new, interior segments of the Fermi surface, has
been considered some time ago \cite{llvp,zb}. Let us assume that the
thickness $\delta p$ of each interior block is approximately the same
$ \delta p\simeq p_{2k+1}-p_{2k}$ and $\delta p$ is defined by $B$.
Using the Simpson's rule, we
obtain that the minimum loss in the ground state energy due to
formation of the blocks is about $(\delta p)^4$. This result can
be understood by considering that the continuous FC function
$n_0({\bf p})$ delivers the minimum value to the energy functional
$E[n({\bf p})]$, while the approximation of $\nu({\bf p})$ 
by the steps 
of size $\delta p$ produces the minimum error of the order of
$(\delta p)^4$. Taking into account that the gain due to the
magnetic field is proportional
to $B^2$ and equating both of the contribution we obtain
\beq \delta p\propto \sqrt{B}.\eeq
Thus, at $T\to 0$ when $B\to 0$,
the thickness $\delta p$ goes to zero as well, 
$\delta p\to0$, while the NFL behavior of the FC 
state is replaced by the  behavior of 
LFL with the Fermi momentum $p_f$.  
It follows from Eq. (3.6) that $p_f>p_F$, while  
the number density $x$ of itinerant electrons remains constant. 
As we shall see, this observation plays 
important role when considering the Hall coefficient $R_H(B)$
as a function of $B$ at low
temperatures in the HF metals with FC.

To calculate the effective mass $M^*(B)$ as function of
the applied magnetic field $B$,
we observe that at $T=0$ the application of finite
magnetic field $B$ splits the FC
state into the Landau levels and suppresses the superconducting
order parameter $\kappa({\bf p})$ thus destroying the FC state.
Therefore the LFL behavior is expected to be restored \cite{shag4,shag}.
The Landau levels at the Fermi surface can be approximated by a
single block whose thickness in momentum space  is $\delta p$.
Approximating the dispersion of quasiparticles within this block
by $\varepsilon(p)\sim (p-p_F+\delta p)(p-p_F)/M$, we obtain that
the effective mass $M^*(B)\sim M/(\delta p/p_F)$. The energy loss
$\Delta E_{FC}$ due to rearrangement of the FC state related to
this block  can be estimated using the Landau formula \cite{lanl1}
\begin{equation}
\Delta E_{ FC}=\int(\varepsilon({\bf p})-\mu)\delta n({\bf
p})\frac{d{\bf p}^3}{(2\pi)^3}.
\end{equation}
The region occupied by the variation $\delta n({\bf p})$ has the
thickness $\delta p$, while $(\varepsilon({\bf p}) -\mu)\sim
(p-p_F)p_F/M^*(B)$. As a result, we have $\Delta E_{ FC}\sim\delta
p^2/M^*(B)$. On the other hand, there is a gain $\Delta E_B\sim
(B^2\mu_{B})^2M^*(B)p_F$ due to the application of the magnetic
field and coming from the Zeeman splitting. Equating $\Delta E_B$
to $\Delta E_{FC}$ and taking into account that in this case
$M^*(B)\propto 1/\delta p$, we obtain the following relation
\begin{equation} \frac{\delta
p^2}{M^*(B)}\propto \frac{1}{(M^*(B))^3}\propto B^2M^*(B).
\end{equation} It follows from Eq.
(5.4) that the effective mass $M^*(B)$ diverges as
\begin{equation} M^*(B)\propto \frac{1}{\sqrt{B-B_{c0}}}.
\end{equation}
Here, $B_{c0}$ is the critical magnetic field which drives both a HF
metal to its magnetic field tuned QCP and the corresponding N\'eel
temperature toward $T=0$ \cite{shag4}. We note that in some cases
$B_{c0}=0$, for example, the HF metal CeRu$_2$Si$_2$ shows neither
evidence of the magnetic ordering, superconductivity down to the
lowest temperatures nor the LFL behavior  \cite{takah}.
Equation (5.5) shows that by applying the
magnetic field $B>B_{c0}$ the system can be driven back into
LFL with the effective mass $M^*(B)$ depending on the magnetic
field. This means that the following dependences are valid: 
for the coefficient
$A(B)\propto (M^*(B))^2$ \cite{ksch},
for the specific heat, $C/T=\gamma_0(B)\propto M^*(B)$, and for the
magnetic susceptibility $\chi_0(B)\propto M^*(B)$. 
The coefficient $A(B)$ 
determines the temperature dependent part of
the resistivity, $\rho(T)=\rho_0+\Delta\rho$, with $\rho_0$
being the residual resistivity and $\Delta\rho=A(B)T^2$.
Since the coefficient is
directly determined by the effective mass, we obtain from
Eq. (5.5) that
\beq A(B)\propto \frac{1}{B-B_{c0}}.\eeq
It is seen that
the well-known empirical Kadowaki-Woods (KW) ratio \cite{kadw},
$K=A/\gamma_0^2\simeq const$, is fulfilled. At this point, we stress
that the value of $K$ may be dependent on the degeneracy number of
quasiparticles as it was recently shown.
As a result, the grand-KW-relation produces good description of the
data for whole the range of degenerate HF systems \cite{tky}.
In the simplest case when the heavy electron
liquid is formed by quasiparticles with spin $1/2$ and the
degeneracy number is $2$, $K$ turns out to be close to the
empirical value \cite{ksch}, called as the KW relation
\cite{kadw}. Therefore, we come to the
conclusion that by applying magnetic fields the system is driven
back into the LFL state where the constancy of the
Kadowaki-Woods ratio is obeyed.

At finite temperatures, the system
remains in the LFL state, but
there exists a temperature $T^*(B)$, at which the NFL behavior
is restored. To calculate the function $T^*(B)$ , we observe that
the effective mass $M^*$ characterizing the single particle
spectrum cannot be changed at $T^*(B)$ since there are no any
phase transitions. In other words, at the
crossover point, we have to compare the effective mass
$M^*(T)$ defined by
$T^*(B)$, Eq. (2.12), and that $M^*(B)$
defined by the magnetic field $B$,
Eq. (5.5), $M^*(T)\sim M^*(B)$
\begin{equation} \frac{1}{M^*(T)}\propto T^*(B)\propto \frac{1}{M^*(B)}
\propto \sqrt{B-B_{c0}}.
\end{equation}
As a result, we obtain
\begin{equation}
T^*(B)\propto \sqrt{B-B_{c0}}.
\end{equation}
At temperatures $T\geq T^*(B)$, the system comes back to the
NFL behavior with $M^*$ defined by Eq. (2.12), and  
the LFL behavior disappear.  We can conclude 
that Eq. (5.8) determines the line 
in the $B-T$ phase diagram that separates the region of the $B$ 
dependent effective mass from the region of the $T$ dependent effective 
mass.  At the temperature $T^*(B)$, there occurs a crossover from the 
$T^2$ dependence of the resistivity to the $T$ dependence.  It follows 
from Eq. (5.8), that the heavy electron liquid at some temperature $T$ 
can be driven back into the Landau Fermi-liquid by applying a strong 
enough magnetic field $(B-B_{c0})\propto (T^*(B))^2$.  We can also 
conclude, that at finite temperature $T<T^*(B)$, the heavy electron 
liquid shows a more pronounced metallic behavior at the elevated
magnetic field $B$ since the effective mass decreases (see Eq. (5.5)). The
same behavior of the effective mass can be observed in the
Shubnikov --- de Haas oscillation measurements. 
We conclude that one obtains a unique possibility to
control the essence of the strongly correlated liquid by
magnetic fields which induce the change of the NFL
behavior to the LFL liquid behavior.

Let us briefly consider the case when
the system is very near FCQPT being on the ordered side and 
therefore $\delta p_{FC}=(p_f-p_i)/p_F\ll 1$. Since
$\delta p\propto M^*(B)$, it follows from Eqs. (5.2) and (5.5) that
\begin{equation}
\frac{\delta p}{p_F}\sim a_c\sqrt{\frac{B-B_{c0}}{B_{c0}}},
\end{equation}
where $a_c$ is a constant, which is expected to 
be of the order of a unit, 
$a_c\sim 1$. At bigger magnetic field $B$,
the value of $\delta p/p_F$ becomes comparable with $\delta p_{FC}$, and
the distribution function $\nu({\bf p})$ vanishes being replaced by the
conventional Zeeman  splitting.
As a result, we are dealing with the heavy electron liquid located on the
disordered side of FCQPT. As we shall see in Section 9, the behavior 
of this system is quite different from that of the system with FC.  It 
follows from Eq. (5.9) that relatively weak magnetic field $B_{cr}$ \beq 
(\delta p_{FC})^2\sim \frac{B_{cr}-B_{c0}}{B_{c0}},
\eeq
removes the system from the ordered side of the phase transition
provided that $\delta p_{FC}\ll 1$.

\section {Magnetic-field induced Landau Fermi liquid in high-$T_c$ metals}
\setcounter{equation}{0}

The LFL theory has revealed 
that the low-energy elementary excitations of a Fermi liquid look
like the spectrum of an ideal Fermi gas. These excitations are
described in terms of quasiparticles with an effective mass $M^*$,
charge $e$ and spin $1/2$. The quasiparticles define the major
part of the low-temperature properties of Fermi liquids.
As we have shown in Section 5, at
temperatures $T<T^*(B)$, the LFL behavior of the heavy electron liquid 
with FC is recovered by the application of magnetic field $B$ larger 
than the critical field $B_c$ suppressing the superconductivity. 
Thus, the heavy electron liquid with FC 
can be viewed as LFL induced by the magnetic field. In such a 
state, the Wiedemann-Franz (WF) law and the Korringa law are held and 
the elementary excitations are LFL quasiparticles. Our 
consideration is valid for relatively weak magnetic fields when the 
contributions coming from the magnetic field are proportional $B^2$ and 
the Zeeman splitting is much smaller then the Fermi momentum.

It was reported recently that in the normal state obtained by
applying a magnetic field greater than the upper critical filed
$B_c$, in a hole-doped cuprates at overdoped concentration
(Tl$_2$Ba$_2$CuO$_{6+\delta}$) \cite{cyr} and at optimal doping
concentration (Bi$_2$Sr$_2$CuO$_{6+\delta}$) \cite{cyr1}, there
are no any sizable violations of the WF law.
In the electron-doped copper oxide superconductor
Pr$_{0.91}$LaCe$_{0.09}$Cu0$_{4-y}$ ($T_c$=24 K) when
superconductivity is eliminated by a magnetic field,
it was found that the spin-lattice relaxation rate $1/T_1$
follows the $T_1 T=constant$ relation, known as the Korringa law 
\cite{korr}, down to temperature of $T=0.2$ K \cite{zheng}. At
higher temperatures and applied magnetic fields of 15.3 T
perpendicular to the CuO$_2$ plane, $1/T_1T$ as a function of $T$ is
a constant below $T=55$ K.
At $300$ K $>T>50$ K, $1/T_1T$  decreases with
growing $T$ \cite{zheng}. Recent measurements for strongly
overdoped non-superconducting La$_{1.7}$Sr$_{0.3}$CuO$_4$ have shown
that the resistivity $\rho$ exhibits $T^2$ behavior,
and the WF law holds perfectly \cite{nakam,huss}. 
Since the validity of the WF 
and the Korringa laws are the robust signature of LFL, these
experimental facts demonstrate that the observed elementary
excitations cannot be distinguished from the Landau quasiparticles.
This imposes strong constraints for models describing the hole-doped
and electron-doped high-temperature superconductors. For example, in
the cases of a Luttinger liquid \cite{kane}, spin-charge separation
(see e.g.  \cite{sen}), and in some solutions of $t-J$ model
\cite{hough} a violation of the WF law was predicted.

As any phase transition, FCQPT is related to the order parameter,
which induces a broken symmetry. It was shown in 
Section 3 that the relevant order
parameter is the superconducting order parameter $\kappa({\bf
p})$, which is 
suppressed by the critical magnetic field $B_c$, when $B_c^2\sim
\Delta_1^2$. If the coupling constant $\lambda_0\to 0$, the
critical magnetic field $B_c\to 0$ will destroy the state with FC
converting the strongly correlated Fermi liquid into  LFL. 
The magnetic field plays the role of the control parameter
determining the effective mass $M^*(B)$ as it follows from Eq. (5.5).

If $\lambda_0$ is finite, the critical
field is also finite, and Eq. (5.5) is valid at $B>B_c$.
In that case the system is driven back to LFL and
has the LFL behavior induced by the magnetic field.
Then, the low energy elementary excitations
are characterized by $M^*(B)$ and cannot be distinguished from
Landau quasiparticles. As a result, at $T\to 0$, the WF law is held
in accordance with experimental facts \cite{cyr,cyr1}.  On the hand,
in contrast to the LFL theory, the effective mass  $M^*(B)$
depends on the magnetic field.

Equation (5.5) shows that by applying a magnetic field $B>B_c$ the
system can be driven back into LFL with the effective mass
$M^*(B)$ which is finite and independent of the temperature.
This means that the low temperature properties depend on the
effective mass in accordance with the LFL theory. In particular,
the resistivity $\rho(T)$ as a function of the temperature behaves
as $\rho(T)=\rho_0+\Delta\rho(T)$ with $\Delta\rho(T)=AT^2$, and the
factor $A\propto (M^*(B))^2$. Taking into account that in the case of
the high-$T_c$ superconductors $B_{c0}$ is expected to be zero,
we obtain from Eq. (5.5)
that \beq \gamma_0\sqrt{B}=const.\eeq
Here $\gamma_0=C/T$ with $C$ is the specific heat. Taking into account
Eqs. (5.6) and Eq. (6.1), we obtain
\beq \gamma_0\sim A(B)\sqrt{B}.\eeq

At finite temperatures, the system
remains LFL, but there is the crossover from the LFL
behavior to the non-Fermi liquid behavior at temperature
$T^*(B)\propto \sqrt{B}$. At $T>T^*(B)$, the effective mass
starts to depend on the temperature $M^*\propto 1/T$, and the
resistivity possesses the non-Fermi liquid behavior with a
substantial linear term, $\Delta\rho(T)\propto T$ \cite{ms,shb,pogsh}.
Such a behavior of the resistivity was observed in the
cuprate superconductor Tl$_2$Ba$_2$CuO$_{6+\delta}$
($T_c<$ 15 K) \cite{mac}. At B=10 T, $\Delta\rho(T)$ is a
linear function of the temperature
between 120 mK and 1.2 K, whereas at $B=18$ T, the temperature
dependence of the resistivity is consistent with 
$\Delta\rho(T)=AT^2$ in the same temperature range \cite{mac}.

In LFL, the nuclear spin-lattice relaxation rate $1/T_1$
is determined by the quasiparticles near the Fermi level whose
population is proportional to $M^* T$, so that $1/T_1T\propto M^*$
is a constant \cite{zheng,korr}. When the
superconducting state is removed by the application of a magnetic
field, the underlying ground state can be seen as the field induced
LFL with effective mass depending on the magnetic field.  As a
result, the rate $1/T_1$ follows the $T_1T=constant$ relation, that
is the Korringa law is held.  Unlike the behavior of LFL, as it
follows from Eq. (5.5), $1/T_1T\propto M^*(B)$ decreases with
increasing the magnetic field at $T<T^*(B)$.  At $T>T^*(B)$,
we observe that $1/T_1T$ is a decreasing function of the temperature,
$1/T_1T\propto M^*\propto 1/T$.  These observations are in a
good agreement with the experimental facts \cite{zheng}.  Since
$T^*(B)$ is an increasing function of the
magnetic field, see Eq. (5.8), the
Korringa law retains its validity to higher temperatures at elevated
magnetic fields. We conclude, that at temperature $T_0\leq
T^*(B_0)$ and bigger magnetic fields $B>B_0$ the system shows a
more pronounced metallic behavior, since the effective mass decreases
with increasing $B$ (see Eq. (5.5)). Such a behavior of the effective
mass can be observed in the de-Haas van Alphen-Shubnikov studies,
$1/T_1T$  and the resistivity measurements. These
experiments can shed light on the physics of high-$T_c$ metals and
reveal relationships between high-$T_c$ metals and heavy-electron
metals \cite{aspla}.

The existence of FCQPT can also be revealed experimentally because
at densities $x>x_{FC}$, or beyond the FCQPT point, the system
should be LFL at sufficiently low temperatures \cite{shag}. Recent
experimental data have shown that this liquid exists in the heavily
overdoped non-superconducting compound La$_{1.7}$Sr$_{0.3}$CuO$_4$
\cite{nakam,huss}. It is remarkable that up to $T=55$ K the resistivity
exhibits the $T^2$ behavior with no additional linear term, and the
WF law is verified to within the experimental resolution
\cite{nakam,huss}. While at elevated temperatures, a strong deviations
from the LFL behavior are observed. We anticipate that in this case 
the system can be again driven back to the LFL behavior by the 
application of sufficiently strong magnetic fields.

Thus, the mentioned above striking measurements,
which were used in studies of the nature of
the high-$T_c$ superconductivity,
suggest that FCQPT and the
emergence of the novel quasiparticles with effective mass strongly 
depending on the magnetic field and temperature  and resembling the 
Landau quasiparticles are qualities intrinsic to the electronic system 
of the high-$T_c$ superconductors.

\section{ Appearance of FCQPT in different Fermi liquids}
\setcounter{equation}{0}

It is widely believed that unusual properties of 
correlated liquids observed in the high-temperature
superconductors, heavy-fermion metals, 2D $^3$He and etc., are
determined by quantum phase transitions.
Therefore, immediate experimental studies of relevant quantum phase
transitions and of their quantum critical points
are of crucial importance for understanding the physics of the
high-$T_c$ superconductivity and HF systems. In
case of the high-$T_c$ superconductors, these studies are
difficult to carry out, because all the corresponding area is occupied
by the superconductivity. On the other hand, recent experimental data
on different highly correlated Fermi liquids, when the system
in question is approaching FCQPT from the disordered side,
can help to illuminate both the nature of this point
and the control parameter, by which this point is driven. 
We shall call Fermi liquids approaching FCQPT from the disordered side
as highly correlated ones to distinguish them from strongly correlated 
liquids which have undergone FCQPT. Detailed explanations on this point 
are given in Section 8.  

Experimental facts on high-density 2D $^3$He 
\cite{mor,cas} show that the effective mass diverges when the density, 
at which 2D $^3$He liquid begins to solidify, is approached
\cite{cas}. Then, a sharp increase of the effective mass
in a metallic 2D electron system was observed, when the
density tends to the critical density of the metal-insulator
transition point, which occurs at sufficiently low densities
\cite{skdk}. Note, that
there is no ferromagnetic instability in both Fermi systems and the
relevant Landau amplitude $F^a_0>-1$ \cite{skdk,cas}, in accordance
with the almost localized fermion model \cite{pfw}.

Now let us consider the divergence of the effective mass in
2D and 3D highly correlated
Fermi liquids at $T=0$, when the density $x$ approaches FCQPT from
the side of normal Landau Fermi liquid, that is from the disordered
phase.  First, we calculate the divergence of $M^*$ as a function of
the difference $(x-x_{FC})$ in case of 2D Fermi liquid. For this purpose we
use the equation for $M^*$ obtained in \cite{ksz}, where the divergence
of the effective mass $M^*$ due to the onset of FC in different Fermi
liquids (such as 2D and 3D electron and $^3$He liquids)
was predicted. At $x\to x_{FC}$, the effective
mass $M^*$ can be approximated as
\beq\frac{1}{M^{*}}\simeq\frac{1}{M}+\frac{1}{4\pi^{2}}
\int\limits_{-1}^{1}\int\limits_0^{g_0}
\frac{v(q(y))}{\left[1-R(q(y),\omega=0,g)
\chi_0(q(y),\omega=0)\right]^{2}}\frac{ydydg}{\sqrt{1-y^{2}}}.
\eeq Here we adopt the notation $p_F\sqrt{2(1-y)}=q(y)$ with
$q(y)$ being the transferred momentum, $M$ is the bare mass,
$\omega$ is the frequency, $v(q)$ is the bare interaction, and the
integral is taken over the coupling constant $g$ from zero to its
real value $g_0$. In Eq.  (7.1), both $\chi_0(q,\omega)$ and
$R(q,\omega)$ are the linear response function of
a noninteracting Fermi liquid and the effective interaction
respectively. They define the linear response function of the system in
question \beq \chi(q,\omega,g)=\frac{\chi_0(q,\omega)}
{1-R(q,\omega,g)\chi_0(q,\omega)}. \eeq
In the vicinity of the charge
density wave instability, occurring at the density $x_{cdw}$, the
singular part of the function $\chi^{-1}$ on the disordered side
is of the well-known form (see  e.g.  \cite{varma})
\beq\chi^{-1}(q,\omega,g)\simeq
a(x_{cdw}-x)+b(q-q_c)^2+c(g_0-g),\eeq where
$a$, $b$ and $c$ are constants and $q_c\simeq 2p_F$ is the
wavenumber of the charge density wave order. Upon substituting Eq.
(7.3) into Eq. (7.1) and integrating, the equation for the
effective mass $M^*$ can be presented in 
the following form \cite{shag1,khod1} 
\beq\frac{1}{M^*}= \frac{1}{M}-\frac{C}{\sqrt{x-x_{cdw}}}\,,\eeq
with $C$ being some positive constant. The behavior of the effective mass
as a function of the electron number density $x$
in a silicon MOSFET is shown in Fig. 5. The fitting parameters are
$x_{cdw}=0.7\times10^{-11}$cm$^{-2}$,
$C=2.14\times10^{-6}$ cm$^{-1}$ and
$x_{FC}=0.9\times10^{-11}$cm$^{-2}$ \cite{khod1}.
It is seen from Fig. 5 that Eq. (7.4)
describes rather good the data.

\begin{figure}[!ht]
\begin{center}
\includegraphics[width=0.47\textwidth]{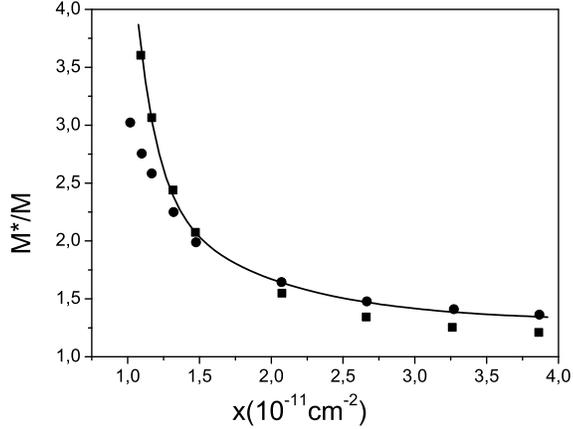}
\end{center}
\caption{The ratio $M^*/M$ in a silicon MOSFET versus the carrier 
number density $x$.  The filled squares denote the Shubnikov --– de 
Haas oscillations experimental data,  
the data obtained by the 
application of a parallel magnetic field are shown by the filled 
circles \cite{skdk,krav}.} \label{Fig5} \end{figure}

It is seen from Eq. (7.4)
that $M^*$ diverges at some point $x_{FC}$ referred to as the
critical point, at which FCQPT occurs, as a function of the
difference $(x-x_{FC})$ \cite{shag1,khod1}
\beq \frac{M^*}{M}\simeq A+ \frac{B}{x-x_{FC}},\eeq
where $A$ and $B$ are constants.
It follows from the derivation of Eqs. (7.4) and (7.5) that
their forms are
independent of the bare interaction $v(q)$, which 
effects however $A$, $B$ and $x_{FC}$ values. This result is in
agreement with Eq. (2.7) which exhibits the same type of divergence
independent of the  interaction. Therefore
both of these equations are also applicable to 2D $^3$He liquid
or to another Fermi liquid. It is also seen from Eqs. (7.4) and
(7.5) that FCQPT precedes the formation of charge-density waves. As
a consequence of this, the effective mass diverges at high
densities in case of 2D $^3$He, and at low densities in
case of 2D electron systems, in accordance with experimental facts
\cite{skdk,cas}. Note, that in both cases the difference
$(x-x_{FC})$ has to be positive, because $x$ approaches $x_{FC}$
when the system is on the disordered side of FCQPT
with the effective mass $M^*(x)>0$.
Thus, in considering the 2D $^3$He liquid
we have to replace $(x-x_{FC})$ by $(x_{FC}-x)$
on the right hand side of Eq. (7.5). In case of a 3D
system, at $x\to x_{FC}$, the effective mass is given by \cite{ksz}
\beq\frac{1}{M^{*}}\simeq\frac{1}{M}+\frac{p_F}{4\pi^{2}}
\int\limits_{- 1}^{1}\int\limits_0^{g_0}
\frac{v(q(y))ydydg}{\left[1-R(q(y),\omega=0,g)
\chi_0(q(y),\omega=0)\right]^{2}}. \eeq

\begin{figure}[!ht]
\begin{center}
\includegraphics[width=0.47\textwidth]{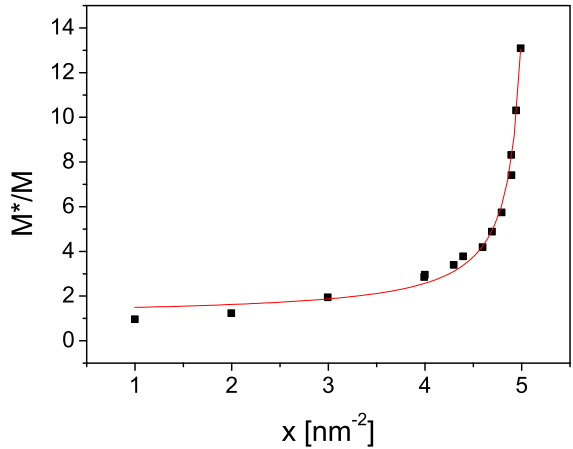}
\end{center}
\caption{The ratio $M^*/M$ in a 2D $^3$He versus the fluid number 
density $x$ inferred from the heat capacity measurements and the 
magnetization measurements \cite{cas}.  The solid line represents 
$M^*/M=A+\frac{B}{(x_{FC}-x)}$, with A=1.09, B=1.68 $\rm nm^{-2}$ and 
$x_{FC}=5.11$ $\rm nm^{-2}$.} \label{Fig6} \end{figure}

The comparison of Eq. (7.1)
and Eq. (7.6) shows that there is no fundamental difference between
these equations, and along the same way we again arrive at Eqs.
(7.4) and (7.5). The only difference between 2D electron systems and
3D ones is that in the latter FCQPT occurs at densities which are well 
below those corresponding to 2D systems. For bulk $^3$He, FCQPT cannot 
probably take place since it is absorbed by the first order 
solidification \cite{cas}.
The apparent divergence of the effective mass $M^*(x)$ obtained in 
measurements on 2D $^3$He \cite{cas} is shown in Fig. 6. This is in 
good agreement with
the divergence given by Eqs. (7.5) and (2.7).

\section{Heavy fermion metals with highly correlated electron liquid}
\setcounter{equation}{0}

In HF metals with strong electron correlations,
quantum phase transitions at zero temperature may strongly influence
the measurable quantities up to relatively high temperatures.  These
quantum phase transitions have recently attracted much attention
because the behavior of HF metals
is expected to follow universal patterns defined by the quantum
mechanical nature of the fluctuations taking place at quantum
critical points (see e.g. \cite{sac,voj}).
Only recently, there appeared experimental facts which
deliver experimental grounds to understand the nature of quantum
phase transition producing the universal behavior of HF metals.
It has been demonstrated that at low temperatures the
main properties of HF metals such as the
magnetoresistance, resistivity, specific heat, magnetization,
susceptibility, volume thermal expansion, etc, strongly depend on
temperature $T$  and applied magnetic field $B$. As a result,
these properties can be controlled by placing these metals at the
special point of the field-temperature $B-T$ phase diagram. In the
LFL theory, considered as the main
instrument when investigating quantum many electron physics, the
effective mass $M^*$ of quasiparticle excitations determining the
thermodynamic properties of electronic systems is practically
independent of temperature and applied magnetic fields.
Therefore, the observed anomalous behavior is uncommon and can 
hardly be understood within the framework of the conventional LFL theory
based on the notion of quasiparticles.
As a result, it is necessary to use theories that are based on
the Landau concept of the order parameter which is introduced  to
classify phases of the state of matter. 
These theories connect the anomalous 
behavior with critical fluctuations of the magnetic order
parameter. These fluctuations suppressing the quasiparticles are
attributed to CQPT taking place
when the system in question approaches its
QCP. As it was noted in Introduction,
the universal behavior is only observable if the
electron system of HF metal is very near QCP, for example, when
the correlation length is much larger than the microscopic length
scales. In the case of CQPT, the physics
is dominated by thermal
and quantum fluctuations of the critical state, which is
characterized by the absence of quasiparticles. It is believed
that the absence of quasiparticle-like excitations is the main
cause of the NFL behavior and other types of the critical behavior
in the quantum critical region. However, theories based on CQPT
fail to explain the experimental observations related to the
divergence of the effective mass $M^*$ at the magnetic field tuned
QCP, the specific behavior of the spin susceptibility and its
scaling properties, the thermal expansion behavior, etc, see e.g.
\cite{geg,geg1,pag,takah,bi,cust,pepin,bud,
pag1,kuch,movsh,senth1,senth3,zhu,ishida}.

The LFL theory rests on the notion of quasiparticles which
represent elementary excitations of a Fermi liquid. Therefore
these are appropriate excitations to describe the low temperature
thermodynamic properties. The 
inability of the LFL theory to explain the experimental
observations which point to the dependence of $M^*$ upon the
temperature $T$ and applied magnetic field $B$ may lead to the
conclusion that the quasiparticles do not survive near QCP, and
one might be further led to the conclusion that the heavy electron
does not retain its integrity as a quasiparticle excitation (see
e.g. \cite{cust,senth1,senth3,col1,col2}).

The mentioned above inability to explain the behavior of HF
metals at QCP within the framework of theories  based on CQPT may
also lead to the conclusion that the other important Landau
concept of the order parameter fails as well, see e.g.
\cite{senth1,senth3,col1,col2}. Thus, we are left without the most
fundamental principles of many body quantum physics while a great
deal of interesting NFL phenomena related to the anomalous
behavior and the experimental facts collected in measurements on
the HF metals remain out of reasonable theoretical explanations.

On the other hand, it is the very nature of HF metals that suggests
that their unusual properties
are defined by a quantum phase transition
related to the unlimited growth of the effective
mass at its QCP. Moreover, a divergence to
infinity of the effective electron mass
was observed  at a magnetic field-induced QCP \cite{geg,pag,cust}.
In Section 2, we have demonstrated  that such a
quantum phase transition is to be FCQPT, an essential
feature of which is the divergence of the effective mass $M^*$
at its QCP, see Eq. (2.7).

\subsection{Highly correlated heavy electron liquid}

When a Fermi system approaches
FCQPT from the disordered phase
it remains the Landau Fermi liquid with the effective mass $M^*$
strongly depending on the distance $r=(x-x_{FC})$,
temperature $T$ and a magnetic field $B$.  This
state of the system, with $M^*$ essentially depending on $T$, $r$
and $B$, resembles the strongly correlated liquid described in Section 2.
However, in contrast to the
strongly correlated liquid, there is no energy scale $E_0$ given 
by Eq. (2.14) and the system under consideration is the Landau Fermi 
liquid at sufficiently low temperatures with the effective mass 
$M^*\propto1/r$ (see Eqs. (2.7) and (7.5)). 
Therefore this liquid can be called the 
{\it highly correlated liquid} which is obviously to have uncommon properties 
\cite{shag2,shag1}.  Again, we use the heavy electron liquid model to 
study the universal behavior of the HF metals at low temperatures.  As it was 
mentioned in Section 2, it 
is possible, since we consider processes related to the power-low 
divergence of the effective mass. This divergence is determined by 
small momenta transferred as compared to momenta of the order of the 
reciprocal lattice cell, and the contribution 
coming due to the lattice structure can be 
ignored. Thus, we may usefully ignore the complications related to the 
lattice and get rid of the specific peculiarities of a HF metal 
regarding the medium as homogeneous heavy electron isotropic liquid.

The effective mass $M^*$ of quasiparticle excitations controlling
the density of states determines the thermodynamic properties of
electronic systems.
To study the behavior of the effective mass $M^*(T,B)$ as a function of
temperature and magnetic field, we use the Landau
equation determining $M^*(T,B)$. In the case of a homogeneous
liquid, at finite temperatures and low magnetic fields, this equations
reads 
\cite{lanl1} \beq \frac{1}{M^*(T,B)}=
\frac{1}{M}+\sum_{\sigma_1}\int \frac{{\bf p}_F{\bf p_1}}{p_F^3}
F_{\sigma,\sigma_1}({\bf p_F},{\bf p}_1) \frac{\partial
n_{\sigma_1}({\bf p}_1,T,B)}{\partial {p}_1} \frac{d{\bf
p}_1}{(2\pi)^3}. \eeq Here $F_{\sigma,\sigma_1}({\bf p_F},{\bf
p}_1)$ is the Landau amplitude depending on the momenta $p$ and
spins $\sigma$, $p_F$ is the Fermi momentum, $M$ is the bare mass
of an electron and $n_{\sigma}({\bf p},T)$ is the quasiparticle
distribution function. Since HF metals are predominantly three
dimensional (3D) structures we treat the homogeneous heavy
electron liquid as a $3$D liquid also. For the sake of simplicity,
we omit the spin dependence of the effective mass since in the
case of a homogeneous liquid and weak magnetic fields, $M^*(T,B)$
does not noticeably depend on the spins. The quasiparticle
distribution function is of the form
\begin{equation} n_{\sigma}({\bf p},T)=\left\{ 1+\exp
\left[ \frac{(\varepsilon({\bf p},T)-\mu_{\sigma})}T \right]
\right\} ^{-1},
\end{equation}
where $\varepsilon({\bf p},T)$ is determined by Eq. (2.2).
In our case, the single-particle spectrum does not
noticeably depend on the spin, while the chemical potential may
have a dependence due to the Zeeman splitting. We will show
explicitly the spin dependence of a physical value when this
dependence is of importance for understanding.

Replacing $n_{\sigma}({\bf p},T,B)$ by $n_{\sigma}({\bf
p},T,B)\equiv \delta n_{\sigma}({\bf p},T,B)+n_{\sigma}({\bf
p},T=0,B=0)$ where $\delta n_{\sigma}({\bf p},T,B)=
n_{\sigma}({\bf p},T,B)-n_{\sigma}({\bf p},T=0,B=0)$, Eq. (1)
takes the form \beq \frac{M}{M^*(T,B)}=
\frac{M}{M^*(x)}+\frac{M}{p_F^2}\sum_{\sigma_1}\int \frac{{\bf
p}_F{\bf p_1}}{p_F} F_{\sigma,\sigma_1}({\bf p_F},{\bf p}_1)
\frac{\partial \delta n_{\sigma_1}({\bf p}_1,T,B)}{\partial {p}_1}
\frac{d{\bf p}_1}{(2\pi)^3}. \eeq
We assume that the heavy electron liquid is near FCQPT,
therefore the distance $r$ is small so that $M/M^*(x)\ll1$,
as it is seen from Eq. (2.7). 
In the case of normal metals with the effective mass of 
the order of a few bare electron masses, $M/M^*(x)\sim 1$, 
and up to temperatures $T\sim 100$ K, 
the second term on the right hand side of Eq. (8.3) is 
of the order of $T^2/\mu^2$ and is much smaller than the first 
term. Thus, the system in question demonstrates the LFL behavior 
with the effective mass being practically independent of 
temperature, that is the corrections are proportional to $T^2$. 
One can check that the same is true when magnetic field up to 
$B\sim 30$ T is applied. Near the critical point $x_{FC}$, when 
$M/M^*(x\to x_{FC})\to 0$, the behavior of the effective mass
changes drastically, because the first term
on the right hand side of Eq. (8.3) vanishes and the second
term determines the effective mass itself rather than small
corrections to $M^*(x)$ related to $T$ and $B$. In that case, Eq.
(8.3) becomes a homogeneous equation and determines the
effective mass as a function of $B$ and $T$. As we will see, Eq.
(8.3) describes both the NFL behavior and the LFL one with the
presence of quasiparticles. In contrast to the conventional Landau
quasiparticles these are characterized by the effective mass that
strongly depends on both the magnetic field and the temperature.

Let us turn to a qualitative analysis of solutions of Eq. (8.3)
when $x\simeq x_{FC}$. We start with the case when $T=0$ and $B$
is finite. The application of magnetic field leads to the Zeeman
splitting of the Fermi surface and the difference $\delta p$
between the Fermi surfaces with ``spin up'' and ``spin down''
becomes  $\delta p=p_F^{\uparrow}-p_F^{\downarrow}\sim
\mu_{B}BM^*(B)/p_F$ with $\mu_{B}$ being the Bohr magneton. Upon
taking this into account, we observe that the second term in Eq.
(8.3) is proportional to $(\delta p)^2 \propto
(\mu_{B}BM^*(B)/p_F)^2$, and Eq. (8.3)
takes the form \cite{shag4,shag,ckhz}
\beq
\frac{M}{M^*(B)}=
\frac{M}{M^*(x)}+c\frac{(\mu_{B}BM^*(B))^2}{p_F^4}, \eeq where $c$
is a constant. Note that the effective mass $M^*(B)$ depends on
$x$ as well and this dependence disappears at $x=x_{FC}$. At the
point $x=x_{FC}$, the term $M/M^*(x)$ vanishes, Eq. (8.4) becomes
homogeneous and can be solved analytically \cite{shag1,shag,ckhz} \beq
M^*(B)\propto \frac{1}{(B-B_{c0})^{2/3}}. \eeq Here 
$B_{c0}$ is the critical magnetic field which drives both a HF
metal to its magnetic field tuned QCP and the corresponding N\'eel
temperature toward $T=0$ \cite{shag4}. We recall that in some cases
$B_{c0}=0$, for example, the HF metal CeRu$_2$Si$_2$ shows neither
evidence of the magnetic ordering, superconductivity down to the
lowest temperatures nor the LFL behavior  \cite{takah}.

Equation (8.5) shows the universal power low behavior of the effective 
mass which does not depend on the inter-particle interaction. We 
illustrate this behavior by calculations using a model functional 
\cite{ksk,ksn} \beq E[n(p)]=\int \frac{{\bf p}^2}{2M}\frac{d{\bf 
p}}{(2\pi)^3}+\frac{1}{2}\int V({\bf p}_1-{\bf p}_2)n({\bf p}_1)n({\bf 
p}_2) \frac{d{\bf p}_1d{\bf p}_2}{(2\pi)^6},\eeq with the 
inter-particle interaction \beq V({\bf p})=g_0\frac{\exp(-\beta_0|{\bf 
p}|)}{|{\bf p}|}. \eeq We normalized the effective mass by $M$, 
$M^*=M^*(B)/M$, temperature $T_0$  and magnetic field $H$ by the Fermi 
energy $\varepsilon^0_F$, $T=T_0/\varepsilon^0_F$, $H=(\mu_B 
B)/\varepsilon^0_F$, and use the dimensionless coupling constant 
$g=(g_0 M)/(2/pi^2)$ and $\beta=\beta_0 p_F$.  FCQPT takes place when 
the parameters reach their critical values, $\beta=b_c$ and $g=g_c$. On 
the other hand, FCQPT takes place when effective mass $M^*\to\infty$. 
This condition allows to relate $b_c$ and $g_c$ \cite{ksk,ksn} $$ 
\frac{g_c}{b_c^3}(1+b_c)\exp(-b_c) [b_c\cosh(b_c)-\sinh(b_c)]=1. $$ It 
follows from this equation that the critical point of FCQPT  can be 
reached by changing $g_0$ if $\beta_0$ and $p_F$ are fixed, or changing 
$p_F$ if $\beta_0$ and $g_0$ are fixed, etc. For simplicity, we shall 
change $g$ to reach FCQPT and investigate the properties of the system 
behind the critical point.

In Fig. 7 we present our calculations (triangles and squares) of the
magnetic-field dependence of the effective mass at the critical point 
of FCQPT.  At $\beta=b_c=3$, FCQPT takes place when $g=g_c=6.7167$.  It 
is seen from Fig. 7 that the calculated power-low divergence of the 
effective mass is in accordance with Eq. (8.5). 

\begin{figure}[!ht]
\begin{center}
\includegraphics[width=0.47\textwidth]{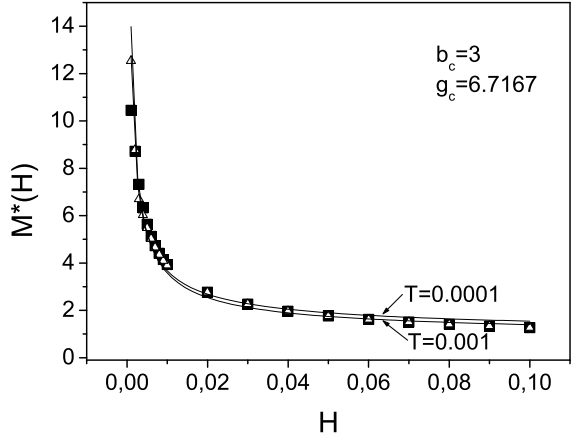}
\end{center}
\caption{Calculated magnetic-field dependent effective mass
$M^*$ at fixed temperatures. Arrows indicate the temperatures. The 
vertical axis represents the normalized effective mass $M^*$. The 
horizontal axis is the normalized magnetic field $H$. Details of 
normalization are explained in the text.  Solid lines represents 
$M^*(H)\propto H^{-2/3}$.} \label{Fig7} \end{figure}

At densities $x>x_{FC}$, $M^*(x)$ is finite and we are dealing with the
conventional Landau quasiparticles provided that the magnetic
field is weak, so that $M^*(x)/M^*(B)\ll 1$ with $M^*(B)$  given
by Eq. (8.5). In that case, the second term on the right hand side
of Eq. (8.4) is proportional to $(BM^*(x))^2$ and represents small
corrections. In the opposite case, when $M^*(x)/M^*(B)\gg 1$, the
heavy electron liquid behaves as at QCP. Since in
the LFL regime the main thermodynamic properties of the system is
determined by the effective mass, it follows from Eq. (8.5) that we
obtain a unique possibility to control the magnetoresistance,
resistivity, specific heat, magnetization, volume thermal
expansion, etc. At this point, we note that the large effective
mass leads to the high density of states provoking a large number
of states and phase transitions to emerge and compete with one
another. Here we assume that these can be suppressed by the
application of a magnetic field and concentrate on the
thermodynamical properties.

To consider the qualitative behavior of $M^*(T)$ at elevated
temperatures, we simplify Eq. (8.3) by omitting the variable $B$ and
simulating the influence of the applied magnetic field by the
finite effective mass entering the denominator of the first term
on the right hand side of Eq. (8.3). This effective mass becomes a
function of the distance $r$, $M^*(r)$, which  is determined by
both $B$ and $(x-x_{FC})$. If the magnetic field vanishes the
distance is $r=(x-x_{FC})$. We integrate the second term over the
angle variable, then over $p_1$ by parts and substitute the
variable $p_1$ by $z$, $z=(\varepsilon(p_1)-\mu)/T$. In the case
of the flat and narrow band, we use the approximation
$(\varepsilon(p_1)-\mu)\simeq p_F(p_1-p_F)/M^*(T)$ 
and finally obtain \beq
\frac{M}{M^*(T)}=\frac{M}{M^*(r)} +\alpha\int^{\infty}_{0}
F(p_F,p_F(1+\alpha z)) \frac{1}{1+e^z}dz
-\alpha\int^{1/\alpha}_{0} F(p_F,p_F(1-\alpha z))
\frac{1}{1+e^z}dz. \eeq Here the notations are 
used: $F\sim M d(F^1p^2)/dp$,
the factor $\alpha=TM^*(T)/p_F^2= TM^*(T)/(T_kM^*(r))$, 
$T_k=p_F^2/M^*(r)$ The
Fermi momentum is defined as $\varepsilon(p_F)=\mu$. We first
assume that $\alpha\ll 1$. Then omitting terms of the order of
$\exp(-1/\alpha)$, we expand the upper limit of the second
integral on the right hand side of Eq. (8.8) to $\infty$ and observe
that the sum of the second and third terms represents an even 
function of $\alpha$. These are the typical expressions with 
Fermi-Dirac functions as integrands. 
They can be calculated using  standard 
procedures (see e.g. \cite{lanl2}). Since we need only an
estimation of the integrals, we represent Eq. (8.8) as \beq
\frac{M}{M^*(T)}\simeq
\frac{M}{M^*(r)}+a_1\left(\frac{TM^*(T)}{T_kM^*(r)}\right)^2
+a_2\left(\frac{TM^*(T)}{T_kM^*(r)}\right)^4+...\eeq Here $a_1$
and $a_2$ are constants of the order of units. Equation (8.9) can be
considered as a typical equation of the LFL theory with the only
exception being the effective mass $M^*(r)$ which strongly depends
on the distance $r=x-x_{FC}\geq 0$ and diverges at $r\to 0$.
Nonetheless, it follows from Eq. (8.9) that when $T\to 0$ the
corrections to $M^*(r)$ start with the $T^2$ terms provided that
\beq \frac{M}{M^*(r)}\gg
\left(\frac{TM^*(T)}{T_kM^*(r)}\right)^2\simeq
\frac{T^2}{T_k^2}, \eeq and the system exhibits the LFL behavior.
It is seen from Eq. (10) that when $r\to 0$,  $M^*(r)\to\infty$,
and the LFL behavior disappears. The free term on the right hand side
of Eq. (8.8) vanishes, $M/M^*(r)\to 0$, and Eq. (8.8) in itself
becoming homogeneous
determines the value and universal behavior of the effective mass.

At some temperature $T_1\ll T_k$, the value of the sum on the
right hand side of Eq. (8.9) is determined by the second term. Then
Eq. (8.10) is not valid, and upon keeping only the second term in
Eq. (8.9) this can be used to determine $M^*(T)$ in a transition
region \cite{ckhz,shag5} \beq M^*(T)\propto \frac{1}{T^{2/3}}.\eeq
The variation as $T^{-2/3}$ exponent  with the temperature growth 
deserves a comment. Equation (8.11) is valid if the second term in
Eq. (8.9) is much larger than the first one, that is \beq
\frac{T^2}{T_k^2} \gg \frac{M}{M^*(r)},\eeq and this term is
grater than the third one, \beq \frac{T}{T_k} \ll
\frac{M^*(r)}{M^*(T)}\simeq 1.\eeq Obviously, both Eqs. (8.12) and
(8.13) can be simultaneously satisfied if $M/M^*(r)\ll 1$ and $T$ is
finite. The range of temperatures, over which Eq. (8.11) is valid, 
shrinks to zero as soon as $r\to 0$ because $T_k\to 0$. Thus, if
the system is very near QCP, $x\to x_{FC}$, it is possible to
observe the behavior of the effective mass given by Eq. (8.11) in a
wide range of temperatures provided that the effective mass
$M^*(r)$ is diminished by the application of high magnetic fields,
that is, the distance $r$ becomes larger under 
the action of $B$. When $r$ is 
finite the $T^{-2/3}$ behavior can be observed at relatively high
temperatures. To estimate the transition temperature $T_1(B)$, we
observe that the effective mass is a continuous function of the
temperature, thus $M^*(B)\sim M^*(T_1)$. Taking into account Eqs.
(8.7) and (8.11), we obtain $T_1(B)\propto B$.

\begin{figure}[!ht]
\begin{center}
\includegraphics[width=0.95\textwidth]{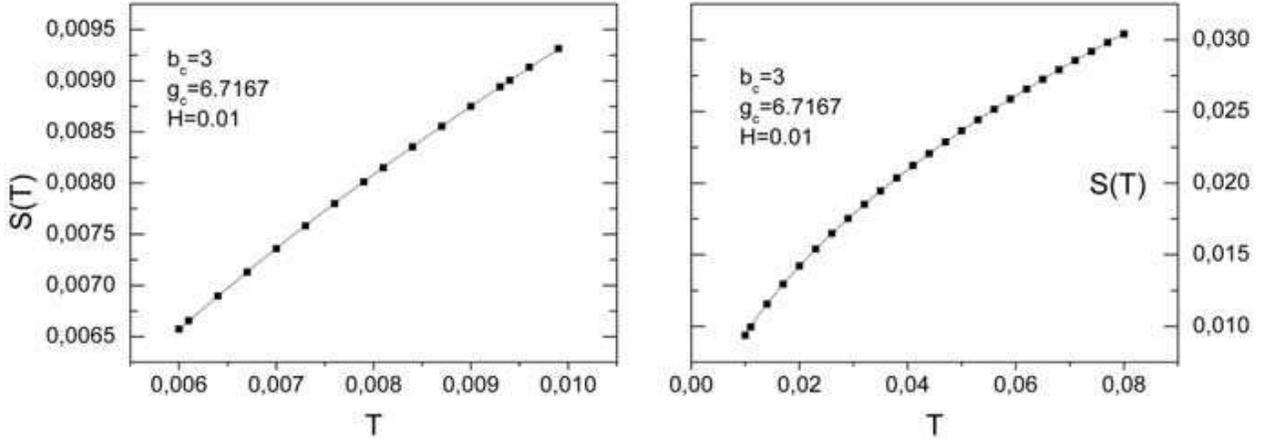}
\end{center}
\caption{Calculated entropy $S(T)$ as a function
of temperature at fixed magnetic field $H=0.01$.
Solid line represents $S(T)$ at the 
transition region $T=T_1(B)$ (left panel). 
At low temperatures $T<0.007$, the system exhibits the LFL 
behavior, $S(T)\propto T$.  
The other solid (right panel)
line represents $S(T)\propto T^{1/2}$.}
\label{Fig8} \end{figure}

Then, at elevated temperatures, the system enters into a different
regime. The coefficient $\alpha$ becomes $\alpha\sim 1$, the upper
limit of the second integral in Eq. (8.8) cannot be expanded to
$\infty$, and odd terms come into play. As a result, Eq. (8.9) is no
longer valid, but the sum of both the first integral and the
second one on the right hand side of Eq. (8.8) is proportional to
$M^*(T)T$. Upon omitting the first term $M/M^*(r)$ and
approximating the sum of the integrals by $M^*(T)T$, we solve Eq.
(8.8) and obtain \beq M^*(T)\propto \frac{1}{\sqrt{T}}.\eeq

We illustrate the above consideration by numerical calculations 
(shown by filled squares in Fig. 8) based
on the model functional (8.6). In Fig. 8, we show the evolution of the low
temperature entropy  from  the transition region $T\sim T_1(B)$ to
$S(T)\propto \sqrt{T}$ at higher temperatures 
upon applying magnetic field $H=0.01$. The calculated behavior  
of $S(T)/T\propto M^*(T)$ is in accord with Eq. (8.14).  
Details of the normalization $T$ and $H$ are given at Fig. 7. 

Thus, we can conclude that at higher 
temperatures when $x\simeq x_{FC}$ 
the system exhibits three types of regimes: the LFL behavior at
$\alpha\ll 1$, when Eq. (8.10) is valid; the $M^*(T)\propto
T^{-2/3}$ behavior and $S(T)\propto M^*(T)T\propto T^{1/3}$,
when Eqs. (8.12) and (8.13) are valid; and the
$1/\sqrt{T}$ behavior of the effective mass at $\alpha\sim 1$,
while the entropy $S(T)\propto M^*(T)T\propto \sqrt{T}$, and the specific
heat $C(T)=T(\partial S(T)/\partial T)\propto \sqrt{T}$.

In the absence of magnetic field, calculated evolutions
of $M^*(T)$, $S(T)$ and $C(T)$ based on
the model functional (8.6) are shown by filled squares in Fig. 9,
Fig. 10, and Fig. 11 respectively.

\begin{figure}[!ht]
\begin{center}
\includegraphics[width=0.47\textwidth]{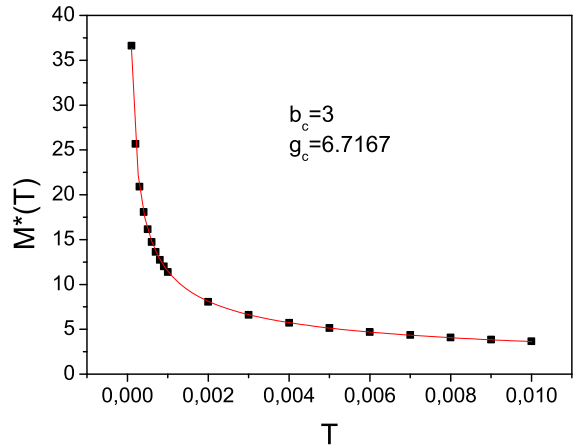}
\end{center}
\caption{Calculated effective mass $M^*(T)$ as a function
of temperature. Solid line represents $M^*(T)\propto 1/\sqrt{T}$.}
\label{Fig9} \end{figure}

\begin{figure}[!ht]
\begin{center}
\includegraphics[width=0.47\textwidth]{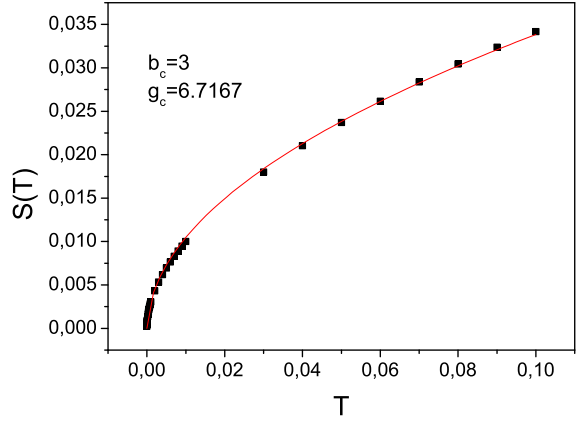}
\end{center}
\caption{Calculated entropy $S(T)$ as a function
of temperature. Solid line represents $S(T)\propto \sqrt{T}$.}
\label{Fig10} \end{figure}

\begin{figure}[!ht]
\begin{center}
\includegraphics[width=0.47\textwidth]{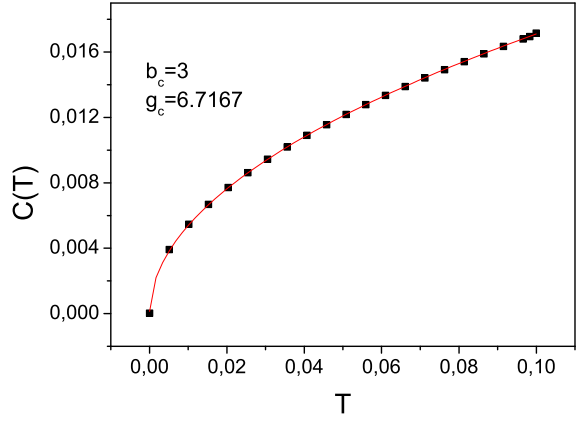}
\end{center}
\caption{Calculated specific heat $C(T)$ as a function
of temperature. Solid line represents $C(T)\propto \sqrt{T}$.}
\label{Fig11} \end{figure}

Let us estimate the quasiparticles width
$\gamma(T)$. Within the framework of the LFL theory it is
given by \cite{lanl1} \beq \gamma\sim |\Gamma|^2(M^*)^3T^2,\eeq
where $\Gamma$ is the particle-hole amplitude. In the case of a
strongly correlated system with its large density of states
related to the huge value of the effective mass, the amplitude
$\Gamma$ cannot be approximated by the bare particle interaction
but can be estimated within the ladder approximation which gives
$|\Gamma|\sim 1/(p_FM^*(T))$ \cite{dkss}. 
As a result, we obtain that 
in the LFL regime $\gamma(T)\propto T^2$, in the $T^{-2/3}$ regime
$\gamma(T)\propto T^{4/3}$, and in the $1/\sqrt{T}$ regime
$\gamma(T)\propto T^{3/2}$. We observe that in all the cases the
width is small as compared to the quasiparticle characteristic energy
which is of the order of $T$, 
so the notion of a quasiparticle is 
meaningful. We can conclude that when the heavy
electron liquid is near the QCP of FCQPT, being on the disordered side, 
its low energy excitations are quasiparticle excitations with the
effective mass $M^*(B,T)$. At this point we note that at $x\to
x_{FC}$, the quasiparticle renormalization factor $a({\bf p})$ remains
finite and approximately constant, and the divergence of the
effective mass given by Eq. (2.7) is not related to vanishing $a({\bf p})$ 
\cite{khod2}. Thus the notion of the quasiparticles is
preserved and these are the relevant excitations when considering
the thermodynamical properties of the heavy electron liquid.

\subsection{Resistivity of heavy fermion metals}

Since the resistivity, $\rho(T)=\rho_0+\Delta\rho(T)$, is
directly determined by the effective mass, because the coefficient
$A(B,T)\propto (M^*(B,T))^2$ \cite{ksch}, the above mentioned 
temperature dependences can 
be observed in measurements of the resistivity of HF metals.

At $T\ll T_1$, the system in question demonstrates the LFL regime,
the divergence of the effective mass at $x\to x_{FC}$ is
described by Eq. (8.5) and the coefficient $A(B)$ diverges as \beq
A(B)\propto \frac{1}{(B-B_{c0})^{4/3}}.\eeq Thus, the resistivity, 
$\rho(B)\propto A(B)T^2$, as a 
function of magnetic field diverges as given by Eq. (8.16).  
As a function of temperature, the resistivity  behaves as 
$\Delta\rho_1=c_1T^2/(B-B_{c0})^{4/3}\propto T^2$. The second  
is NFL regime which is determined by Eq. (8.11) and characterized by 
$\Delta\rho_2=c_2T^2/(T^{2/3})^2\propto T^{2/3}$. At 
$T>T_1(B)$, the third NFL regime is given by Eq. (8.14) and represented by
$\Delta\rho_3=c_3T^2/(\sqrt{T})^2\propto T$. Here $c_1,c_2,c_3$
are constants. It is remarkable that all temperature 
dependences corresponding to these regimes were
observed in measurements on the HF metals
$\rm CeCoIn_5$ and $\rm YbAgGe$
\cite{bud,pag1,pag,movsh}. If we consider the ratio
$\Delta\rho_2/\Delta\rho_1\propto ((B-B_{c0})/T)^{4/3}$, we come 
to a very interesting conclusion that the ratio is a function of
only the variable $(B-B_{c0})/T$, thus representing the scaling
behavior. This result is in excellent agreement with experimental
facts \cite{pag}.

\subsection{Magnetic susceptibility}

The magnetic susceptibility
is proportional to the effective mass, $\chi\propto M^*$,
with $M^*$ given by Eq. (8.5).
Therefore, at $T\ll T_1$,
\beq \chi(B)\propto M^*(B)\propto (B-B_{c0})^{-2/3},\eeq
while the static magnetization $M_B(B)$ is given by
\beq M_B(B)\propto BM^*(B)\propto (B-B_{c0})^{1/3}.\eeq
At $T\gg T_1$, as it follows from Eq. (8.14), Eq. (8.17) has to be
rewritten as \beq \chi(T)\propto M^*(T)\propto
\frac{1}{\sqrt{T}}.\eeq The observed behavior of $\chi(B)$ and $M_B(B)$
and the behavior of $\chi(T)$ are in accord with results 
of measurements on CeRu$_2$Si$_2$ with the critical field
$B_{c0}\to 0$ \cite{takah}.

Consider the state of the system when $r\to 0$. Its properties
are determined by magnetic fields $B$ and temperature $T$
because there are no other parameters to describe such a state. 
At the transition temperatures $T\simeq T_1(B)$, the
effective mass depends on both $T$ and $B$, while at $T\ll T_1(B)$,
the system is LFL with the effective mass being given by Eq. (8.5),
and at $T\geq T_1(B)$, the mass is defined by Eq. (8.11).  Instead of
solving Eq. (8.3), it is possible to construct a simple interpolation
formula to describe the behavior of the effective mass over all 
regions, \beq M^*(B,T)=\frac{1}{c_1(B-B_{c0})^{2/3}+
c_2f(y)T^{2/3}}.\eeq Here, $f(y)$ is a universal monotonic function
of $y=(T/(B-B_{c0}))^{2/3}$ such that $f(y\sim 1)=1$, and
$f(y\ll 1)=0$.  It is seen from Eq. (8.20) that the behavior of the
effective mass can be represented by a universal function of
only one variable $y$ if the temperature  is measured in the units of
the transition temperature $T_1(B)$, and the effective 
mass is measured in the units of $M^*(B)$ given by Eq. (8.5).
This representation describes the scaling behavior of the effective mass.
As seen from Eqs. (8.18) and (8.20), the scaling behavior of the
magnetization can be represented
in the same way,
provided the magnetization is normalized
by the saturated value at each field given
by Eq. (8.5)
\beq \frac{M_B(B,T)}{M_B(B)}\propto\frac{1}{1+ c_3f(y)y},\eeq
where $c_3$ is a constant.
It is seen from Eq. (8.21), that
magnetization is a monotonic function of $y$.
Upon using the definition of susceptibility, $\chi=\partial
M_B/\partial B$, we come to the conclusion that the susceptibility
also exhibits the scaling behavior and can be presented as a
universal function of only one variable $y$, provided it is
normalized by the saturated value at each field given by Eq. (8.17)
\beq \frac{\chi(B,T)}{\chi(B)}\propto\frac{1}{1+ c_3f(y)y}
+2c_3y\frac{f(y)+ydf(y)/dy}{(1+ c_3f(y)y)^2}.\eeq
It is of importance
to note that the susceptibility is not a monotonic function of $y$
because the derivative is the sum of two contributions
with different behavior.  The second 
contribution on the right hand side of Eq. (8.22) gives the
susceptibility a maximum \cite{shag4,shag5,ckhz}. As shown in Fig. 12,
the above behaviors of the
magnetization  and susceptibility are in accord with numerical 
calculations of the susceptibility and magnetization \cite{ckhz} and 
with the facts observed in measurements on CeRu$_2$Si$_2$ 
\cite{takah}.

\begin{figure}[!ht]
\begin{center}
\includegraphics[width=0.69\textwidth]{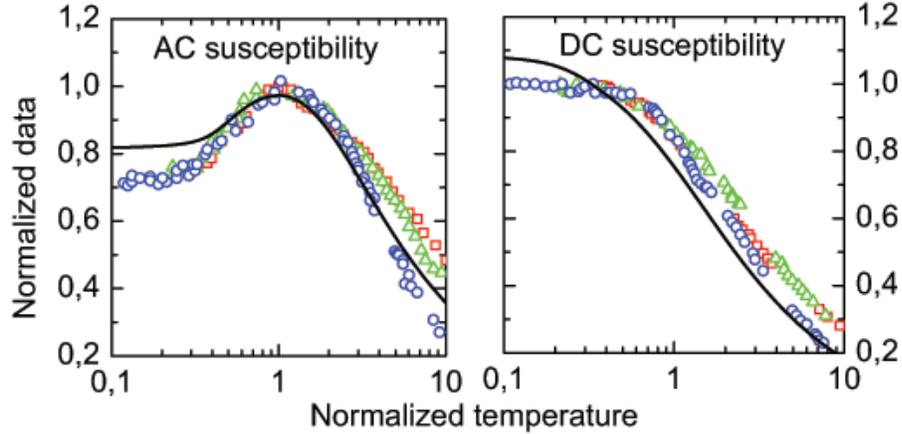}
\end{center}
\caption{Normalized magnetic susceptibility 
$\chi(B,T)/\chi(B,T_P)$ (left panel) and normalized
magnetization ${M_B}(B,T)/{M_B}(B,T_P)$ (right panel) for
CeRu$_2$Si$_2$ in magnetic fields 0.20 mT (squares),
0.39 mT (triangles), and 0.94 mT (circles), plotted
against normalized temperature $T/T_P$ \cite{takah},
where $T_P$ is the temperature at peak susceptibility.
The solid curves trace the calculated universal behavior \cite{ckhz}.}
\label{Fig12}
\end{figure}

It is seen from Fig. 12 that for finite $B$,  the curve describing
$\chi(B,T)$ acquires a maximum at some temperature $T_P$.
This behavior is in good agreement with experimental facts \cite{takah}
and associated with  the
suppression of the divergent NFL terms $\sim T^{-2/3}$ in $\chi(B,T)$
and recovery of the LFL behavior  at static magnetic fields in which the
Zeeman energy splitting  $\mu_BB$ exceeds $T$ \cite{ckhz}. 
Note that magnetization 
$M_B(B,T)$ does not exhibits a maximum. In Fig. 12, the temperature
is normalized by $T_P$, the susceptibility is normalized by its
peak height $\chi(B,T_P)$ and the magnetization by the saturated value
at each field.

\subsection{Magnetoresistance}

Using the results just presented, we consider
the behavior of magnetoresistance (MR)
\beq \rho_{mr}(B,T)=\frac{\rho(B,T)-\rho(0,T)}{\rho(0,T)},\eeq
as a function of magnetic
field $B$ and $T$. Here the resistivity,  
$\rho(B,T)=\rho_0+\Delta\rho(B,T)+\Delta\rho_{mr}(B)$, is 
measured at the magnetic field
$B$ and temperature $T$. We assume that the contribution
$\Delta\rho_{mr}(B)$
coming from the magnetic field $B$ can be treated within
the low field approximation and given by the
well-known Kohler's rule,
\beq\frac{\Delta\rho_{mr}(B)}{\rho(0,T)}
\simeq \lambda_{\bot} 
\left(\frac{B\rho(0,\Theta_D)}{B_0\rho(0,T)}\right)^2 ,\eeq with 
$\Theta_D$ is the Debye temperature, $B_0$ is the characteristic field 
and $\lambda_{\bot}$ is a constant. Note, that the low field 
approximation implies that $\Delta\rho_{mr}(B)\ll 
\rho(0,T)\equiv\rho(T)$.  We also assume that that temperature is not 
too low so that $\rho_0\leq \Delta\rho(B=0,T)$, while $B\geq B_{c0}$.  
Substituting Eq. (8.24)
into Eq. (8.23), we find that
\beq \rho_{mr}(B,T)\sim
\frac{c(M^*(B,T))^2T^2+\Delta\rho_{mr}(B)-c(M^*(0,T))^2T^2}
{\rho(0,T)}.\eeq
Here $M^*(B,T)$ denotes the effective mass which now depends
on both the magnetic field and the temperature, and $c$ is a constant
determining the temperature dependent part of the resistivity,
$c(M^*(B,T))^2 T^2=\Delta\rho(B,T)$.

Consider MR given by Eq. (8.25) as a function
of $B$ at some temperature $T=T_0$.
At low magnetic fields when $T_0>T_1(B)\propto B$, the main contribution
to MR comes from $\Delta\rho_{mr}(B)$ since the effective mass depends 
mainly on temperature. Therefore, the ratio 
$|M^*(B,T)-M^*(0,T)|/M^*(0,T)\ll 1$, the main contribution is given 
by $\Delta\rho_{mr}(B)$, and MR is an increasing function of $B$.  When 
$B$ becomes so large that $T_1(B)\sim T_0$, the difference 
$(M^*(B,T)-M^*(0,T))$ becomes negative, and MR as the function
of $B$ reaches its maximum value at $T_1(B)\sim T_0$. We recall that
$T_1(B)$ determines the crossover from $T^2$ dependence
of the resistivity to the $T$ dependence.
At elevated $B$ when $T_1(B)>T_0$,
the ratio $(M^*(B,T)-M^*(0,T))/M^*(0,T)\sim -1$ and MR becomes 
negative being a decreasing function of $B$.

Consider now MR as a function of $T$ at some $B_0$.
At $T\leq T_1(B_0)$, we have LFL.
At low temperatures $T\ll T_1(B_0)$, it follows
from Eqs. (8.5) and (8.14) that $M^*(B_0)/M^*(T)\ll 1$, and
MR is determined by the resistivity $\rho(0,T)$.
Note, that $B_0$ has to be comparatively high to ensure the inequality,
$M^*(B_0)/M^*(T)\ll 1$. As a result,
$\rho_{mr}(B_0,T\to0) \sim -1$,
because $\Delta\rho_{mr}(B)/\rho(0,T)\ll 1$.
Differentiating the function
$\rho_{mr}(B_0,T)$
with respect to $B_0$ we can check that its slope becomes steeper
as $B_0$ is decreased, being proportional $\propto (B_0-B_{c0})^{-7/3}$.
At $T\simeq T_1(B_0)$, MR possesses a node because at this
point the effective mass $M^*(B_0)\simeq M^*(T)$,
and $\rho(B_0,T)\simeq \rho(0,T)$.
We can conclude that the crossover from the $T^2$
resistivity to the $T$ resistivity, which occurs at $T\sim T_1(B_0)$,
manifests itself in
the transition from negative MR to positive MR.
At $T\geq T_1(B_0)$,
the main contribution
to MR comes from $\Delta\rho_{mr}(B_0)$, and MR reaches its maximum value.
Upon using Eqs. (8.14) and (8.25) and taking into account that at this point
$T\propto (B_0-B_{c0})$, we obtain that the maximum value
$\rho^m_{mr}(B_0)$ of MR is
$\rho^m_{mr}(B_0)\propto 1/(B_0-B_{c0})$.
Thus, the maximum value is a decreasing function of $B_0$.
At $T_1(B_0)\ll T$, MR is a decreasing function of the temperature.
At these temperatures, MR becomes small comparatively to its maximum value
$\rho^m_{mr}(B_0)$ because $|M^*(B,T)-M^*(0,T)|/M^*(0,T)\ll 1$ and
$\Delta\rho_{mr}(B_0)/\rho(T)\ll 1$.

The recent paper \cite{pag} reports the measurements of
the CeCoIn$_5$ resistivity in
magnetic fields.
The both transitions from negative MR
to positive MR with increasing $T$ and from positive MR to 
negative one with increasing $B$ were  
observed \cite{pag}. Thus, the above observed behavior of MR
is in good agreement with the experimental facts. We believe that an 
additional analysis of the data \cite{pag} can reveal that the 
crossover from $T^2$ dependence of the resistivity to the $T$ 
dependence occurs at $T\propto (B-B_{c0})\div (B-B_{c0})^{4/3}$, this 
analysis could reveal the above described supplementary peculiarities 
of MR as well \cite{shag2}.

\section{HF metals with strongly correlated electronic liquid}
\setcounter{equation}{0}

As we have seen in Section 2, at $T=0$ when $r=(x-x_{FC})\to 0$, the
effective mass $M^*(r)\to\infty$ and eventually beyond the
critical point $x=x_{FC}$ the distance $r$ becomes negative making
the effective mass negative as follows from Eq. (2.7).  
As it was shown in Section 2, the system is to undergo FCQPT. 
Therefore behind the critical point $x_{FC}$ of this
transition, the quasiparticle distribution function represented by
the step function does not deliver the minimum to the Landau
functional $E[n({\bf p})]$. As a result, at $x<x_{FC}$ the
quasiparticle distribution is determined by Eq. 
(2.8) to search the minimum of a functional, which
determines the quasiparticle distribution function
$n_0({\bf p})$ delivering the lowest possible value to the ground
state energy $E$. 
It was shown in Section 3 that the relevant order parameter $\kappa({\bf
p})=\sqrt{n_0({\bf p})(1-n_0({\bf p}))}$ is 
at the same time the order parameter of 
the superconducting state
when the maximum value $\Delta_1$ of the
superconducting gap is infinitely small. Thus, this state cannot exist at
any finite temperatures and driven by the parameter $x$: at
$x>x_{FC}$ the system is on the disordered side of FCQPT,
while at $x<x_{FC}$, the system is on the
ordered side. In Section 2 we have shown that this ordered state has a
strong impact on the properties of the system at finite temperatures
$T\ll T_f$: the effective mass
$M^*(T)$ diverges as $1/T$, see Eq. (2.12), and the electronic system 
with FC is characterized by the energy scale $E_0$ given by Eq. (2.14).  
As a result, we can consider FCQPT as
the phase transition, which separates the {\it highly correlated} heavy 
electron liquid from the {\it strongly correlated} one. 
It was shown in Section 8, that the highly correlated liquid at $T\to0$ 
and $x>x_{FC}$ behaves as LFL, therefore, FCQPT separates the regions 
of LFL and strongly correlated liquids. Obviously, the 
strongly correlated electron liquid demonstrates the NFL behavior  down 
to zero temperature.

At $0<T\ll T_f$, the function $n_0({\bf p})$
determines the entropy $S_{NFL}(T)$ of the heavy electron
liquid in its NFL state. Inserting into Eq. (2.3) the function 
$n_0({\bf p})$, one can check that behind the point of FCQPT there is a 
temperature independent contribution 
$S_0(r)\sim(p_f-p_i)/p_F\sim|r|/x_{FC}$, where $r=x-x_{FC}$.  Another 
specific contribution is related to the spectrum $\varepsilon ({\bf 
p})$ which insures the connection between the dispersionless region 
$(p_f-p_i)$ occupied by FC and the normal quasiparticles located at 
$p<p_i$ and at $p>p_f$, and therefore it is of the form $\varepsilon 
({\bf p})\propto (p-p_f)^2\sim (p_i-p)^2$.  Such a form of the spectrum 
can be verified in exactly solvable models for systems with FC and 
leads to the contribution of this spectrum to the specific heat 
$C\sim\sqrt{T/T_f}$ \cite{ks}.
Thus at $0<T\ll T_f$,
the entropy can be approximated as
\beq
S_{NFL}(T) \simeq S_0(r)+a\sqrt{\frac{T}{T_f}}+b\frac{T}{T_f}, \eeq
with $a$ and $b$ are constants. The
third term on the right hand side of Eq. (9.1) comes from the
contribution of the temperature independent part of the spectrum
$\varepsilon({\bf p})$ and gives a relatively small contribution to the
entropy.

The calculated evolution of $S(T)$, $C(T)$ and $M^*(T)$ based on
the model functional (8.6) are shown by filled symbols in Fig. 13,
Fig. 14, and Fig. 15 respectively. The calculations were carried out
for $g=7,8,12$ and $\beta=b_c=3$, while the critical value $g=g_c=6.7167$.

\begin{figure}[!ht]
\begin{center}
\includegraphics[width=0.47\textwidth]{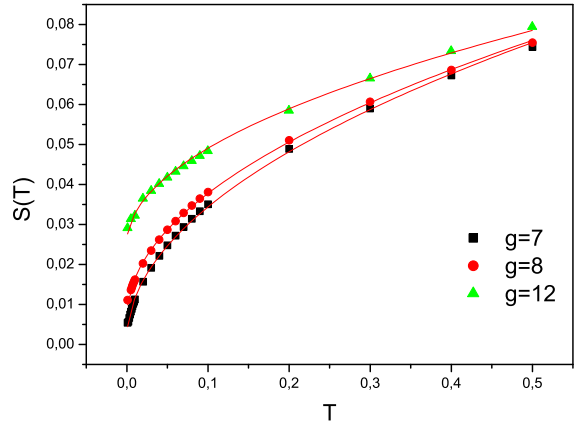}
\end{center}
\caption{Calculated entropy $S(T)$ as a function
of temperature. Solid lines represent $S(T)$ given
by Eq. (9.1).}
\label{Fig13} \end{figure}

It is seen from Fig. 13 that $S_0(r)$ increases when the system being 
on the ordered side moves away from FCQPT. 

\begin{figure}[!ht]
\begin{center}
\includegraphics[width=0.47\textwidth]{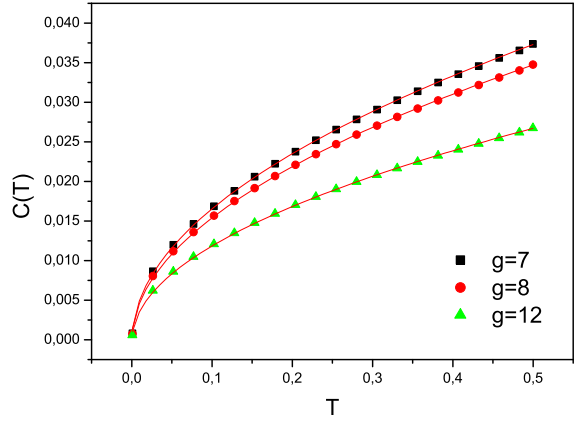}
\end{center}
\caption{Calculated specific heat $C(T)$ as a function
of temperature. Solid lines represent $C(T)\propto \sqrt{T}$.}
\label{Fig14} \end{figure}

\begin{figure}[!ht]
\begin{center}
\includegraphics[width=0.47\textwidth]{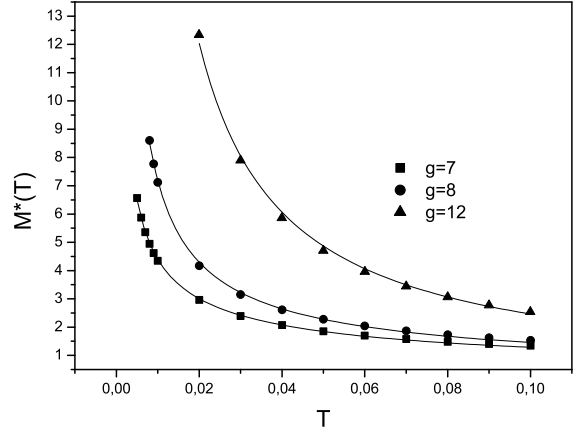}
\end{center}
\caption{Calculated effective mass $M^*(T)$ as a function
of temperature. Solid lines represent $M^*(T)\propto 
a_1/T+a_2\sqrt{T}+a_3$.} \label{Fig15} \end{figure}

Obviously, the term $S_0(r)$ 
does not contribute to the specific heat. As a result the specific heat 
demonstrates the anomalous behavior $C(T)\propto \sqrt{T}$
as it is seen from Fig. 14.
As to the effective mass $M^*(T)\propto S(T)/T$, it demonstrates
the divergence $M^*(T)\propto 1/T$ in accordance with
Eq. (2.5) and contains all terms 
defining the behavior of the entropy, as shown in Fig. 15.

\subsection{The Gr\"uneisen ratio and Hall coefficient in heavy 
fermion metals; $T-B$ phase diagram}

The temperature independent term $S_0(r)$ determines the specific NFL
behavior of the system.
For example, the existence of the temperature independent term
$S_0(r)$  of the entropy can be illuminated by
calculating the thermal expansion coefficient $\alpha (T)$ \cite{alp,zver},
which is given by \cite{lanl1}
\begin{equation}
\alpha (T)=\frac 13\left( \frac{\partial (\log V)}{\partial T}\right) _P
=-\frac x{3K}\left( \frac{\partial (S/x)}{\partial x}\right) _T.
\end{equation}
Here, $P$ is the pressure and $V$ is the volume. The compressibility $K$ is
not expected to be singular at FCQPT and in systems with FC, because FC is
attached to the Fermi level, and it moves along as $\mu (x)$ changes, while
the compressibility $K=d\mu /d(Vx)$ is approximately constant \cite{noz}.
Inserting Eq. (9.1) into Eq. (9.2), we find that
\beq
\alpha _{FC}(T)\simeq a_0\sim \frac{M_{FC}^{*}T}{p_F^2K}. \eeq
Here, $a_0$ is a number independent of temperature.
When deriving Eq. (9.3) we keep only the main contribution coming from
$S_0(r)$.  On the other hand, the specific heat
\beq C(T)=T\frac{\partial S(T)}{\partial T}\simeq
\frac{a}{2}\sqrt{\frac{T}{T_f}}.\eeq
As a result, the Gr\"uneisen ratio $\Gamma(T)$
diverges as
\beq\Gamma(T)=\frac{\alpha(T)}{C(T)}\simeq
2\frac{a_0}{a}\sqrt{\frac{T_f}{T}}.\eeq

At this point, we consider how the behavior of the effective mass
given by Eqs. (2.12) and (8.14) correspond to experimental
observations. It was recently observed that the thermal expansion
coefficient $\alpha(T)/T$ measured on CeNi$_2$Ge$_2$ shows a
$1/\sqrt{T}$ divergence over two orders of magnitude in the
temperature range from 6 K down to at least 50 mK, while
measurements on YbRh$_2$(Si$_{0.95}$Ge$_{0.05}$)$_2$ demonstrate
that $\alpha/T \propto 1/T$ \cite{geg1}, contrary to  the LFL
theory which yields $\alpha(T)/T\propto M^*\simeq const$. Since
the effective mass depends on $T$, we obtain that the $1/\sqrt{T}$
behavior, Eq. (8.14), is in excellent agreement with the result for
the former system \cite{alp}, and the $1/T$ behavior, Eq. (2.12),
predicted in \cite{zver} corresponds to the latter HF metal.

We see that  at $0<T\ll T_f$, the heavy electron liquid with FC behaves
as if it were placed at QCP. In fact it is placed at the quantum
critical line $x<x_{FC}$, that is the critical behavior is observed at
$T\to0$ for all $x\leq x_{FC}$. At $T\to0$, the heavy electron liquid
undergoes a first-order quantum phase transition because the entropy is
not a continuous function: at finite temperatures the entropy is given
by Eq. (9.1), while $S(T=0)=0$. Therefore, the entropy undergoes a sudden
jump $\delta S=S_0(r)$ in the zero  temperature limit. We make up a
conclusion that due to the first order phase transition, the critical
fluctuations are suppressed at the quantum critical line and the
corresponding divergences, for example the divergence of ${\rm
\Gamma}(T)$,  are determined by the quasiparticles rather than by the
critical fluctuations as one could expect in the case of  CQPT, see
e.g. \cite{voj}. Note that according to the well known inequality,
$\delta Q\leq T\delta S$, the heat $\delta Q$ of the transition from
the ordered phase to the disordered one is equal to zero, because
$\delta Q\leq S_0(r)T\to 0$ at $T\to 0$.

To study the $B-T$ phase diagram of the heavy electron liquid with FC,
we consider the case when the NFL behavior arises by the suppression
of the antiferromagnetic (AF) phase upon
applying a magnetic field $B$, for example, as it
takes place in the HF metals $\rm YbRh_2Si_2$ and
YbRh$_2$(Si$_{0.95}$Ge$_{0.05}$)$_2$ \cite{geg,geg1}. The AF phase is
represented by the heavy electron LFL, with the entropy vanishing as
$T\to 0$. For magnetic fields exceeding the critical value $B_{c0}$ at
which the N\'eel temperature $T_N(B\to B_{c0})\to 0$ the weakly ordered
AF phase transforms into weakly polarized heavy electron LFL. As it
was discussed in Section 5, at $T=0$
the application of the magnetic field $B$ splits the FC state occupying
the region $(p_f-p_i)$ into the Landau levels and suppresses the
superconducting order parameter $\kappa({\bf p})$ destroying the FC
state.  Such a state is given by the multiconnected Fermi sphere, where
the smooth quasiparticle distribution function $n_0({\bf p})$ in the
$(p_f-p_i)$ range is replaced by a multiconnected distribution 
$\nu({\bf p})$, see Fig. 4.  Therefore the LFL behavior is restored 
being represented by the weakly polarized heavy electron LFL and 
characterized by quasiparticles with the effective mass $M^*(B)$ given 
by Eq. (5.5). At elevated temperatures $T>T^*(B-B_{c0})\propto 
\sqrt{B-B_{c0}}$, the NFL state is restored and the entropy of the 
heavy electron liquid is given by Eq. (9.1). This behavior is displayed 
in the $T-B$ phase diagram shown in Fig. 16.

\begin{figure}[!ht]
\begin{center}
\includegraphics[width=0.47\textwidth]{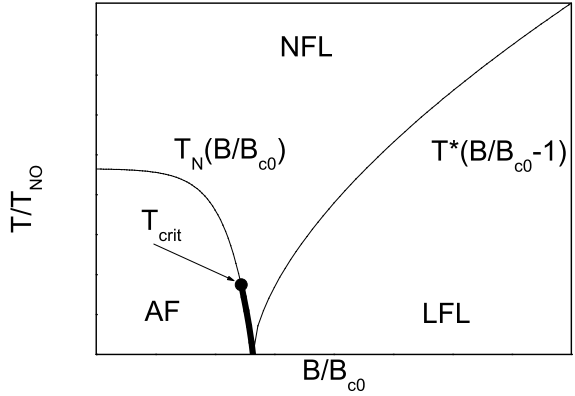}
\end{center}
\caption{$T-B$ phase diagram of the heavy electron liquid. The
$T_N(B/B_{c0})$ curve represents the field dependence of the N\'eel
temperature. Line separating the antiferromagnetic (AF) and the
non-Fermi liquid (NFL) state is a guide to the eye. The black dot at
$T=T_{crit}$ marked by the arrow is the critical temperature, at which
the second order AF phase transition becomes the first one. At
$T<T_{crit}$, the thick solid line represents the field dependence of
the N\'eel temperature when the AF phase transition is of the first
order.  The NFL state is characterized by the entropy $S_{NFL}$ given
by Eq. (9.1). The line separating the NFL state and the weakly polarized
heavy electron Landau Fermi Liquid (LFL) is $T^*(B/B_{c0}-1)\propto
\sqrt{B/B_{c0}-1}$.} \label{fig16} \end{figure}

In accordance with experimental facts we assume that at relatively
high temperatures $T/T_{NO}\sim 1$ the AF phase transition is of the
second order \cite{geg}, where $T_{NO}$ is the N\'eel temperature in
the absence of the magnetic field. In that case, the entropy and the
other thermodynamic functions are continuous functions at the
transition temperature $T_N(B)$. This means that the entropy of the AF
phase $S_{AF}(T)$ coincides with the entropy of the NFL state given by
Eq. (9.1), \beq S_{AF}(T\to T_N(B))=S_{NFL}(T\to T_N(B)).\eeq Since the
AF phase demonstrates the LFL behavior, that is $S_{AF}(T\to 0)\to0$,
Eq. (10) cannot be satisfied at sufficiently low temperatures $T\leq
T_{crit}$ due to the temperature-independent term $S_0(r)$, see Eq.
(9.1). Thus, the second order AF phase transition becomes the first order
one at $T=T_{crit}$ as it is shown in Fig. 16. At $T=0$, the critical
field $B_{c0}$, at which the AF phase becomes the heavy LFL, is
determined by the condition that the ground state energy of the AF
phase coincides with the ground state energy $E[n_0({\bf p})]$ of the
heavy LFL, that is the ground state of the AF phase becomes degenerated
at $B=B_{c0}$. Therefore, the N\'eel temperature $T_N(B\to B_{c0})\to
0$, and the behavior of the effective mass $M^*(B\geq B_{c0})$ is given
by Eq. (5.5), that is $M^*(B)$  diverges when $B\to B_{c0}$.  

We note that the corresponding quantum and 
thermal critical fluctuations vanish
at $T<T_{crit}$ because we are dealing with the first order AF  phase
transition. We can also reliably conclude that the critical behavior
observed at $T\to0$ and $B\to B_{c0}$ is determined by the
corresponding quasiparticles rather than by the critical fluctuations
accompanying second order phase transitions. When $r\to 0$ the heavy
electron liquid approaches FCQPT from the ordered phase. Obviously,
$T_{crit}\to0$ at the point $r=0$, and we are led to the conclusion
that the N\'eel temperature vanishes at the point when the AF second
order phase transition becomes the first order one. As a result, one
can expect that the contributions coming from the corresponding
critical fluctuations can only lead to the logarithmic corrections  to
the Landau theory of the phase transitions \cite{lanl2}, and the power
low critical behavior is again defined by the corresponding
quasiparticles. Thus, we conclude that the Landau paradigm based on
the notions of the quasiparticles and order parameter is
applicable when considering the heavy electron liquid.

Now we are in position to consider the recently observed jump in the
Hall coefficient at $B\to B_{c0}$ in the zero temperature limit
\cite{pash}. At $T=0$, the application of the critical magnetic field
$B_{c0}$ suppressing the AF phase (with the Fermi momentum
$p_{AF}\simeq p_F$) restores the LFL with the Fermi momentum $p_f>p_F$.
At $B<B_{c0}$, the ground state energy of the AF phase is lower then
that of the heavy LFL, while at $B>B_{c0}$, we are dealing with the
opposite case, and  the heavy LFL wins the competition.  At $B=B_{c0}$,
both AF and LFL have the same ground state energy being degenerated.
Thus, at $T=0$ and $B=B_{c0}$, the infinitesimal change in the magnetic
field $B$ leads to the finite jump in the Fermi momentum
because the distribution function becomes multiconnected, see Fig. 4,
while the overall Fermi volume remains constant.
That is, the
number of itinerant electrons does not change. In response
the Hall coefficient $R_H(B)\propto 1/x \propto 1/p_f^3$
undergoes the corresponding
sudden jump. Here we have assumed that the low temperature $R_H(B)$ can
be considered as a measure of the Fermi volume and, therefore, as a
measure of the Fermi momentum \cite{pash}. As a result, we obtain \beq
\frac{R_H(B=B_{c0}-\delta)}{R_H(B=B_{c0}+\delta)}\simeq
1+3\frac{p_f-p_F}{p_F}\simeq 1+d\frac{S_0(r)}{x_{FC}}.
\eeq Here $\delta$ is an infinitesimal magnetic field, $S_0(r)/x_{FC}$ is
the entropy per one heavy electron, and $d$ is a constant, $d\sim 5$.
It follows from Eq. (9.1) that the abrupt change in the Hall coefficient
tends to zero when $r\to0$ and vanishes when the system in question is
on the disordered side of FCQPT \cite{spa}.

Now consider the magnetic susceptibility
which is proportional to the effective mass given by Eq. (5.5).
Therefore, at $T\ll T^*(B)$,
the magnetic susceptibility given by Eq. (5.8) is of the form 
\beq \chi(B)\propto M^*(B)\propto \frac{1}{\sqrt{B-B_{c0}}},\eeq
while the static magnetization $M(B)$ is given by \cite{shag4}
\beq M(B)\propto  \sqrt{B-B_{c0}}.\eeq
\begin{figure}[!ht]
\begin{center}
\includegraphics[width=0.47\textwidth]{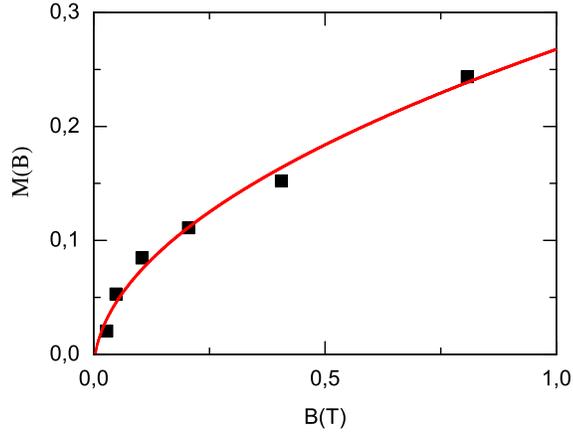} 
\end{center} 
\caption{Magnetization $M(B)$ shown by filled 
squares as a function of 
magnetic field $B$ \cite{cust}. The curve 
represents the field dependence of 
$M(B)=a_M\sqrt{B}$ given by Eq. (9.9) 
with $a_M$ is a costnat.}
\label{fig17} \end{figure}%
As seen from
Fig. 17, the field dependence of $M(B)$ given by Eq. (9.9)
is in good agreement with the data obtained in measurements on
YbRh$_2($Si$_{0.95}$Ge$_{0.05}$)$_2$ \cite{cust}.
We can also conclude that
Eqs. (8.21) and (8.22) determining
the scaling behavior of the effective mass, static magnetization and
the susceptibility are also valid in the case of
strongly correlated liquid, but the variable $y$ is now given by
$y=T/\sqrt{B-B_{c0}}$, while the function $f(y)$ can be dependent
on $(p_f-p_i)/p_F$.  This dependence comes from Eq. (2.12).
As a result, we can obtain that at $T<T^*(B)$, the factor
$d\rho/dT\propto A(B)T$ behaves as $A(B)T\propto T/(B-B_{c0})$, and
at $T>T^*(B)$, it behaves as $A(B)T\propto 1/T$. These observations
are in good agreement with the data obtained in measurements on
YbRh$_2($Si$_{0.95}$Ge$_{0.05}$)$_2$ \cite{cust}.

We note that, as in the case of the highly correlated liquid,
the susceptibility $\chi(B,T)$  of the strongly correlated liquid
is not a monotonic function of $y$ and possesses a maximum as a function 
of the temperature 
because the derivative $dM(B)/dB$ is the sum of two contributions.
As it was shown in Section 5,
the well-known empirical Kadowaki-Woods ratio \cite{kadw},
$K=A/\gamma_0^2\simeq const$, is also obeyed in the case
of the strongly correlated liquid. These results are in good agreement with
facts  \cite{geg,geg2,cust}.

\begin{figure}[!ht]
\begin{center}
\includegraphics[width=0.47\textwidth]{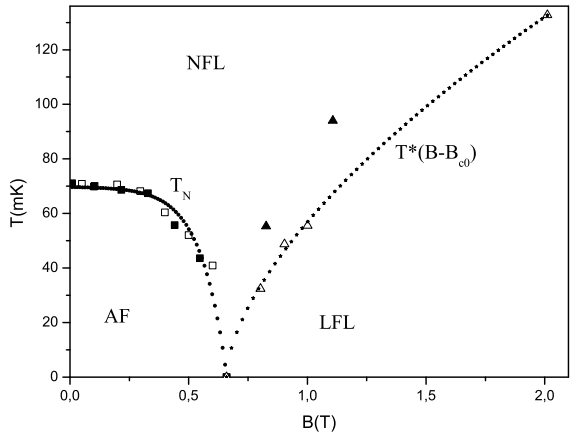}
\end{center}
\caption{$T-B$ phase diagram for ${\rm YbRh_2Si_2}$ \cite{geg,cust}.
The $T_N$ curve
represents the field dependence of the N\'eel temperature. Line
separating the antiferromagnetic (AF) and the non-Fermi liquid (NFL)
state is a guide to the eye.  The NFL state is characterized by the
entropy $S_{NFL}$ given by Eq. (9.1). Line separating the NFL state and
LFL is $T^*(B-B_{c0})=c\sqrt{B-B_{c0}}$, with
$c$ being an adjustable factor.} \label{fig18} \end{figure}

As an application of the above consideration we study the $T-B$ phase
diagram for the HF metal ${\rm YbRh_2Si_{2}}$ \cite{geg} shown in Fig.
18. The LFL behavior is characterized by the effective mass $M^*(B)$
which diverges as $1/\sqrt{B-B_{c0}}$ \cite{geg}. We can conclude that
Eq. (5.5) gives good description of this experimental fact, and $M^*(B)$
diverges at the point $B\to B_{c0}$ with $T_N(B=B_{c0})=0$. It is seen
from Fig. 18, that the line separating the LFL state and NFL can be
approximated by the function $c\sqrt{B-B_{c0}}$ with $c$ being a
parameter.  Taking into account that the behavior of YbRh$_2$Si$_{2}$
strongly resembles the behavior of YbRh$_2$(Si$_{0.95}$Ge$_{0.05}$)$_2$
\cite{geg1,cust,pepin,geg2}, we can conclude that in the NFL state the
thermal expansion coefficient $\alpha(T)$ does not depend on $T$ and
the Gr\"uneisen ratio as a function of temperature $T$ diverges
\cite{geg1}. We are led to the conclusion that the entropy of the NFL
state is given by Eq. (9.1). Taking into account that at relatively high
temperatures the AF phase transition is of the second order \cite{geg},
we predict that at lower temperatures this becomes the first order
phase transition.  Then, the described behavior of the Hall coefficient
$R_H(B)$ is in good agreement with experimental facts \cite{pash}.

Thus, we can conclude that the $T-B$ phase diagram of the heavy
electron liquid with FC is in good agreement with the experimental
$T-B$ phase diagram obtained in measurements on the HF metals $\rm
YbRh_2Si_2$ and YbRh$_2$(Si$_{0.95}$Ge$_{0.05}$)$_2$.

\subsection{Heavy fermion metals very near FCQPT on the ordered side}

Let us consider the case when
$\delta p_{FC}=(p_f-p_i)/p_F\ll 1$ and
the electronic system of HF metal is very near FCQPT being on the 
ordered side.  As we have seen in Section 5, Eq. (5.10), the 
application of magnetic field $(B-B_{c0})/B_{c0}\geq B_c$ removes the 
system from the ordered side placing it on the disordered one. As a 
result at $T\leq T_1(B)$, the effective mass $M^*(B)$ is given by Eqs.  
(8.5) and (8.11), while the corresponding
resistivity is described in Section 8.2. In the absence
of magnetic field or at $T_f\gg T>T_1(B)$, the system
demonstrates the NFL behavior, the effective mass $M^*(T)$
is given by Eq. (2.12), and
the entropy is given by Eq. (9.1). While the magnetic susceptibility
$\chi(T)\propto M^*(T)\propto 1/T$, the thermal expansion
coefficient $\alpha(T)$ is $T$-independent being determined
by Eq. (9.3) and making the Gr\"uneisen ratio divergent, 
as it follows from Eq. (9.5).  
Note, that the specific heat behaves as $C(T)\propto \sqrt{T}$ in both 
cases: when the electronic system is on the ordered side or 
on the disordered side of FCQPT,  
see Fig. 11 and Fig. 15. It follows from Eq. (2.12)  
that $\gamma(T)\propto T$.  Therefore, the temperature dependent part 
$\Delta \rho(T)$ of the resistivity behaves as $\Delta \rho(T)\propto 
\gamma(T)\propto T$.  Thus, the system demonstrates the NFL regime, 
$\Delta\rho(T)\propto T$, when it is either in the highly correlated 
or in the strongly correlated regimes.

At some temperature $T_c$, the system can undergo the superconducting
phase transition. In contrast to the
conventional superconductors where the discontinuity $\delta C(T_c)$ in 
the specific heat at $T_c$ is a linear function of
$T_c$, $\delta C(T_c)$ is independent of the critical temperature $T_c$. 
As seen from Eqs. (3.23) and (3.24), both the discontinuity $\delta C(T_c)$  
and the ratio $\delta C(T_c)/C_n(T_c)$ can be very large as compared 
to the conventional case \cite{ams,khzvya}.

Recent experiments show that the electronic system of the HF metal 
$\rm CeCoIn_5$ can be considered as the system located near FCQPT and 
containing FC. Indeed, 
under the application of magnetic field it 
behaves as the highly correlated liquid (see Section 8) with the 
effective mass $A(B)\propto (B-B_{c0})^{-4/3}$ \cite{pag}, 
as it is  seen from Eq. (8.16). 
While in the NFL regime, $\alpha(T) \propto const$ and the 
Gr\"uneisen ratio diverges \cite{oes}, see Eqs. (9.3) and (9.5), 
respectively.  Estimations of $\delta p_{FC}$ based on the evaluation 
of the magnetic susceptibility show that $\delta p_{FC}\simeq 0.044$ 
\cite{khzvya}. The obtained value of $\delta p_{FC}$ allows to explain 
the relatively big value of the discontinuity 
$\delta C(T_c)$ \cite{khzvya}  
observed at $T_c=2.3$ K in measurements on $\rm CeCoIn_5$ \cite{bau}.  
Thus, we can conclude that $B_{cr}\simeq 0.003$, as it follows from Eq. (5.10),
and the HF metal $\rm CeCoIn_5$ can be considered as placed near FCQPT.

\section{Dissymmetrical
tunnelling in HF metals and high-$T_c$ superconductors}
\setcounter{equation}{0}

Experiments on the HF metals explore
mainly their thermodynamic properties.
It would be desirable to probe the other properties of the
heavy electron liquid such as the probabilities of quasiparticle
occupations, which are not directly linked to the density of states
or to the behavior of the effective mass $M^*$ \cite{tun}.
Scanning tunnelling microscopy (STM)
being sensitive to both the density of states and the
probabilities of quasiparticle occupations is an ideal technique
for studying such effects at quantum level.

The tunnelling current $I$ through the point contact between two
ordinary metals is proportional to the driving voltage $V$ and to
the squared modulus of the quantum mechanical transition amplitude
$t$ multiplied by the difference
$N_1(0)N_2(0)(n_1(p,T)-n_2(p,T))$ \cite{zag}. Here
$n(p,T)$ is the quasiparticle distribution function and $N(0)$ is
the density of states of the corresponding metal. On the other
hand, the wave function calculated in the WKB approximation and
defining $t$ is proportional to $(N_1(0)N_2(0))^{-1/2}$. As a
result, the density of states is dropped out  and the tunnelling
current is independent of $N_1(0)N_2(0)$. Upon taking into
account that at $T\to 0$ the distribution $n(p,T\to0)\to n_F(p)$,
where $n_F(p)$ is the step function $\theta(p-p_F)$, 
one can check that within the LFL theory 
the differential tunnelling conductivity $\sigma_d(V)=dI/dV$ is a
symmetric function of the voltage $V$. 

In fact, the symmetry of
$\sigma_d(V)$ holds provided that so called particle-hole symmetry
is preserved as it is within the LFL theory. Therefore, the existence of
the $\sigma_d(V)$ symmetry is quite obvious and common in the case
of metal-to-metal contacts when these metals are ordinary metals and 
in their normal or superconducting states.

Now we turn to a consideration of the tunnelling current at low
temperatures which in the case of ordinary metals is given by
\cite{zag} \beq
I(V)=2|t|^2\int\left[n_F(z-\mu)-n_F(z-\mu+V)\right]dz.\eeq We use
an atomic system of units: $e=m=\hbar =1$, where $e$ and $m$ are
electron charge and mass, respectively, and
the energy $z$  belongs to the interval $E_0$ given by Eq. (2.14)
\beq\mu-2T\leq z \leq\mu+2T.\eeq
Since temperatures are low, 
we approximate the distribution function of ordinary metal by the
step function $n_F$. It follows from Eq. (10.1) that quasiparticles
with the energy $z$, $\mu-V\leq z\leq \mu$, contribute to the
current, while $\sigma_d(V)\simeq 2|t|^2$ is a symmetrical
function of $V$. In the case of the heavy electron liquid with FC,
the tunnelling current is of the form  \cite{tun} \beq
I(V)=2\int\left[n_0(z-\mu)-n_F(z-\mu+V)\right]dz.\eeq 
Here we have
replaced the distribution function of ordinary metal by $n_0$
that is the solution of Eq. (2.8). 
We have also normalized the transition amplitude $|t|^2$ such that
$|t|^2=1$. Assume that $V$ satisfies the condition, $|V|\leq 2T$,
while the current flows from the HF metal to the ordinary one.
Quasiparticles of the energy z, $\mu-V\leq z$, contribute to
$I(V)$, and the differential conductivity 
is givem by the relation $\sigma_d(V)\simeq 2n_0(z\simeq \mu-V)$. 
If the sign of the voltage is changed, the 
direction of the current is also changed. In that case,
quasiparticles of the energy $z$, $\mu+V\geq z$, contribute to
$I(V)$, and the differential conductivity $\sigma_d(-V)\simeq
2(1-n_0(z\simeq \mu+V))$. The dissymmetrical part $\Delta
\sigma_d(V)=(\sigma_d(-V)-\sigma_d(V))$ of the differential
conductivity is of the form \beq \Delta \sigma_d(V)\simeq
2[1-(n_0(z-\mu\simeq V)-n_0(z-\mu\simeq -V))].\eeq It is worth
noting that according to Eq. (10.4) we have  $\Delta \sigma_d(V)=0$ 
if the considered HF metal is replaced by an ordinary metal.
Indeed, the effective mass is finite at $T\to0$, then $n_0(T\to
0)\to n_F$ being given by Eq. (2.4), and $1-n(z-\mu\simeq
V)=n(z-\mu\simeq -V)$. One might say that the dissymmetrical part
vanishes due to the particle-hole symmetry. On the other hand,
there are no reasons to expect that $(1-n_0(z-\mu\simeq
V)-n_0(z-\mu\simeq -V))=0$. Thus, we conclude 
that the differential conductivity becomes a dissymmetrical
function of the voltage. 

To estimate $\Delta \sigma_d(V)$, we
observe that it is zero when $V=0$, because $n_0(p=p_F)=1/2$ as
it should be and it follows from Eq. (2.1) as well. It is seen from
Eq. (10.4) that $\Delta \sigma_d(V)$ is an even function of both
$(z-\mu)$ and  $V$.
Therefore we can assume that at low values of the voltage $V$ the
dissymmetrical part behaves as $\Delta \sigma_d(V)\propto V^2$.
Then, the natural scale to measure the voltage is $2T$, as it is
seen from Eq. (10.2). In fact, the dissymmetrical part is to be
proportional to $(p_f-p_i)/p_F$. As a result,  we obtain \beq
\Delta \sigma_d(V)\simeq
c\left(\frac{V}{2T}\right)^2\frac{p_f-p_i}{p_F}
\simeq c\left(\frac{V}{2T}\right)^2\frac{S_0(r)}{x_{FC}}.\eeq
Here, $S_0(r)$ is the temperature independent part of the
entropy (see Eq. (9.1)), $c$ is a
constant, which is expected to be of the order of a unit. This
constant can be evaluated by using analytical solvable models. For
example, calculations of $c$ within a simple model, when the
Landau functional $E[n(p)]$ is of the form \cite{ksk} \beq
E[n(p)]=\int \frac{p^2}{2M}\frac{d{\bf p}}{(2\pi)^3}+V_1\int
n(p)n(p)\frac{d{\bf p}}{(2\pi)^3},\eeq give $c\simeq 1/2$. It 
follows from Eq. (10.5), that when $V\simeq 2T$ and FC occupies a
noticeable part of the Fermi volume, 
$(p_f-p_i)/p_F\simeq 1$, the
dissymmetrical part becomes comparable with differential
tunnelling conductivity, $\Delta \sigma_d(V)\sim  V_d(V)$.

The dissymmetrical behavior of the tunnelling conductivity can be
observed in measurements on both high-$T_c$ metals in their normal 
state and the heavy fermion metals, for example, such as 
YbRh$_2$(Si$_{0.95}$Ge$_{0.05}$)$_2$ or YbRh$_2$Si$_2$ which are 
expected to have undergone FCQPT.  In the case of HF metals, upon the 
application of magnetic field $B$ the effective mass is to diverge
as given by Eq. (5.5).
Here $B_{c0}$ is the critical magnetic field which drives the HF
metal to its magnetic field tuned quantum critical point. The
value of the critical exponent $\alpha=-1/2$ is in good agreement
with experimental observations collected on these metals
\cite{geg,cust}. The measurements of $\Delta \sigma_d(V)$ have to
be carried out applying magnetic field $B_{c0}$ at temperatures
$T^*(B)<T\leq T_f$. In the case of these metals, $T_f$ is of the order of
few Kelvin. We note that at sufficiently low
temperatures, the application of magnetic field $B>B_{c0}$ leads
to the restoration of the LFL behavior with $M^*(B)$ given
by Eq. (5.5). As a result, the dissymmetrical
behavior of the tunnelling conductivity vanishes \cite{tun}.

The dissymmetrical differential conductivity $\Delta \sigma_d(V)$
can also be observed when both the high-$T_c$ metal and
the HF metal in question go from
normal to superconducting. The reason is that $n_0(p)$ is again
responsible for the dissymmetrical part of $\sigma_d(V)$. As we have seen
in Section 3, this 
$n_0(p)$ is not appreciably disturbed by the pairing interaction
which is relatively weak as compared to the Landau interaction
forming the distribution function $n_0(p)$. In
the case of superconductivity, we have to take into account that
the ratio,  \beq \frac{N_s(E)}{N(0)}=
\frac{|E|}{\sqrt{E^2-\Delta^2}},\eeq comes into the play because 
the density of states $N_s(E)$ of the superconducting metal is zero 
in the gap, that is when $|E|\leq |\Delta|$. Here
$E$ is the quasiparticle energy, while the normal state
quasiparticle energy is $\varepsilon-\mu=\sqrt{E^2-\Delta^2}$. Now
we can adjust Eq. (10.4) for the case of superconducting HF metal,  
multiplying the right hand side of Eq. (10.4) by $N_s/N(0)$ and
replacing the quasiparticle energy $z-\mu$ by
$\sqrt{E^2-\Delta^2}$ with $E$ being represented by the voltage
$V$. As a result, Eq. (10.5) can be presented in the following form
\beq \Delta \sigma_d(V)\simeq \left|\frac{V}{\Delta}\right|
\frac{\left(\sqrt{V^2-\Delta^2}\right)^2}
{|\Delta|\sqrt{V^2-\Delta^2}}
\frac{p_f-p_i}{p_F}\simeq
\sqrt{1-\left[\frac{\Delta} {V}\right]^2}
\left[\frac{V}{\Delta}\right]^2
\frac{S_0(r)}{x_{FC}}.\eeq
It is seen from Eqs. (10.5) and (10.8)
that, as in the case described by Eq. (9.7) when the abrupt change in
the Hall coefficient is defined by $S_0(r)$, the dissymmetrical part 
of the differential tunnelling conductivity is also proportional to the 
$S_0(r)$ vanishing at $r\to0$.
Note that the scale $2T$ entering Eq.
(10.5) is replaced by the scale $\Delta$ in Eq. (10.8). In the same
way, as Eq. (10.5) is valid up to $V\sim 2T$, Eq. (10.8) is valid up
to $V\sim 2|\Delta|$. It is seen from Eq. (10.8) that the
dissymmetrical part of the differential tunnelling conductivity
becomes  as large as the differential tunnelling conductivity at
$V\sim 2|\Delta|$ provided that FC occupies a large part of the
Fermi volume, $(p_f-p_i)/p_F\simeq 1$. In the case of a $d$-wave
gap, the right hand side of Eq. (10.8) has to be integrated over the
gap distribution. As a result, $\Delta \sigma_d(V)$ is expected to
be finite even at $V=\Delta_1$, where $\Delta_1$ is the maximum
value of the $d$-wave gap \cite{tun}.

\begin{figure}[!ht]
\begin{center}
\includegraphics[width=0.47\textwidth]{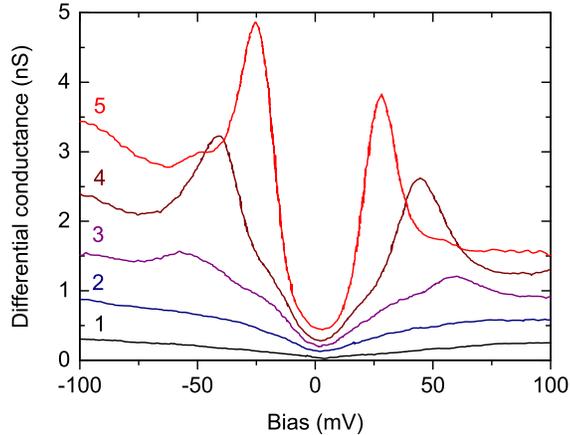}
\end{center}
\caption{Spatial variation of the tunnelling differential conductance 
spectra measured on $\rm Bi_2Sr_2CaCu_2O_{8+x}$.
Curves 1 and 2 are taken at positions
where the integrated LDOS is very small. The low
differential conductance and the absence of a superconducting gap are 
indicative of insulating behavior. Curve 3 is for a large gap of 65 meV, 
with low coherence peaks.  The integrated value of the LDOS at the 
position for curve 3 is small but larger than those in curves 1 and 2. 
Curve 4 is for a gap of 40 meV, which is close to the mean value of 
the gap distribution.  Curve 5, taken at the position with the 
highest integrated LDOS, is for the smallest gap of 25 meV with two 
very sharp coherence peaks \cite{pan}.} \label{Fig19} \end{figure}

The presence of an electronic inhomogeneity in
$\rm Bi_2Sr_2CaCu_2O_{8+x}$ was recently discovered 
in observations using scanning tunnelling microscopy
and spectroscopy \cite{pan}.
This inhomogeneity is manifested as spatial variations in the local
density of states (LDOS) spectrum, in the low-energy spectral
weight, and in the magnitude of the superconducting energy gap.
The inhomogeneity observed in the integrated
LDOS is not induced by impurities, but rather is intrinsic in nature.
The observations allowed to relate the
magnitude of the integrated LDOS to the local oxygen doping
concentration \cite{pan}. Spatial variation of the tunnelling differential
conductance spectrum are shown in Fig. 19. The curves 1 and 2 can
be considered as corresponding to the normal state of
high-$T_c$ superconductor. The other curves can be viewed as
corresponding to high-$T_c$ superconductors with different oxygen doping
concentrations. It is seen from Fig. 19 that the
tunnelling differential conductivity is strongly dissymmetrical in both
the normal state and the superconducting state of the
$\rm Bi_2Sr_2CaCu_2O_{8+x}$ compound.
Therefore, we can conclude that the dissymmetrical tunnelling described
by Eqs. (10.5) and (10.8) is in good qualitative agreement with the
experimental  facts displayed in Fig. 19.

The dissymmetrical tunnelling conductivity given by Eqs. (10.5) and (10.8)
can be observed in measurements on the heavy fermion metal
$\rm CeCoIn_5$ in its superconducting state and its NFL normal state.
As it was discussed in Subsection 9.2, this metal is
expected to have undergone FCQPT.

\section{Summary and conclusion} 

Through out this paper 
we have discussed the manifestations of the fermion condensation,
which can be compared  to the Bose-Einstein condensation. A number
of experimental evidences have been presented that are supportive
to the idea of the existence of FC in different 
natural and artificial substances. We have
demonstrated also that numerous experimental facts collected in 
a whole variety of 
materials,  belonging  to the high-$T_c$ superconductors, heavy
fermion metals and strongly correlated 2D structures, can be
explained within the framework of the theory based on FCQPT.

We have shown that FCQPT separates the regions of LFL and strongly
correlated liquids. Beyond the FCQPT point the
quasiparticle system is divided into two subsystems, one
containing normal quasiparticles, while the other being occupied by
FC localized at the Fermi level. In the
superconducting state the quasiparticle
dispersion $\varepsilon({\bf p})$ in systems with FC can be
approximated by two straight lines, characterized by effective masses
$M^*_{FC}$ and $M^*_L$, and intersecting near the binding energy $E_0$
which is of the order of the superconducting gap.  The same
quasiparticle picture and the energy scale $E_0$ persist in the normal
state. 

We have demonstrated that fermion systems with FC have features
of  "quantum protectorate"  and shown that the theory of high-$T_c$ 
superconductivity, based on FCQPT 
and on the conventional theory of 
superconductivity, permits the description of high values of $T_c$ and 
of the maximum value of the gap $\Delta_1$, which may be as big as
$\Delta_1\sim 0.1\varepsilon_F$ or even larger.  We have also traced
the transition from conventional superconductors to high-$T_c$ ones.
We have shown by a simple, although self-consistent analysis that
both the pseudogap state and the
general features of the shape of the critical temperature $T_c(x)$ as a
function of the number density $x$ of the mobile carriers in the high-$T_c$
compounds can be understood within the framework of the theory.

We have also shown that striking experimental results on the
magnetic-field induced LFL in high-$T_c$ metals,
which unveil the nature of
the high-$T_c$ superconductivity,
suggest that FCQPT and the
emergence of the novel quasiparticles with effective mass strongly 
depending on the magnetic field and temperature  and resembling the 
Landau quasiparticles are qualities intrinsic to the electronic system 
of the high-$T_c$ superconductors.

We have provided explanations of the experimental
data on the divergence of the effective mass in 2D electron liquid
and in 2D $^3$He, as well as shown that above the critical point of FCQPT
the system exhibits the LFL behavior.
The behavior of the heavy electron liquid
approaching FCQPT form the disordered phase can be viewed as the highly
correlated one because the effective mass is very large and strongly
depends on the density, temperature and magnetic fields.

At different temperatures, the behavior in magnetic
fields of the highly correlated electron
liquid approaching FCQPT from the disordered phase
has been considered. We have shown that at sufficiently high temperatures 
$T_1(B)<T$, 
the effective mass starts to depend on $T$,
$M^*\propto T^{-1/2}$. This $T^{-1/2}$ dependence of the effective
mass at elevated  temperatures leads to the non-Fermi liquid behavior
of the resistivity, $\rho(T)\propto T$. 
The application of  magnetic field $B$ restores
the common $T^2$ behavior of the resistivity.
If the magnetic field $(B-B_{c0})$ decreases to zero, 
the effective mass $M^*$ diverges as 
$M^*\propto(B-B_{c0})^{-2/3}$.
At finite magnetic fields, the regime NFL is restored at some temperature
$T_1(B)\propto(B-B_{c0})$ with $M^*(T)\propto T^{-2/3}$.
We have demonstrated that this $B-T$ phase diagram has a strong impact on 
MR of the highly correlated electron liquid. At fixed $B$, 
MR as a function of the temperature exhibits a transition from the
negative values of MR at $T\to 0$ to
the positive values at $T\propto (B-B_{c0})$.
While at low temperatures and elevated magnetic fields, MR goes from positive 
to negative. This behavior was observed in the heavy fermion metals.

We have demonstrated that the strongly correlated electron
liquid with FC, which exhibits strong deviations from the LFL 
behavior down to lowest temperatures,
can be driven into LFL by 
applying a magnetic field $B$. If the
magnetic field $(B-B_{c0})$ decreases to zero, the effective mass
$M^*$ diverges as $M^*\propto 1/\sqrt{B-B_{c0}}$ and
the N\'eel temperature $T_N$
of the AF phase transition tends to zero, $T_N(B\to B_{c0})\to 0$.
The NFL regime is restored
at some temperature $T^*(B)\propto\sqrt{B-B_{c0}}$.
In that case and at $T\to0$ and $B=B_{c0}$, 
the Gr\"uneisen ratio as a  function 
of temperature $T$ diverges. While the entropy $S(T)$ possesses 
the specific low temperature behavior,  
$S(T)\propto  S_0+a\sqrt{T}+bT$ with $S_0$, $a$ and $b$ 
are temperature independent constants.
We have shown that the obtained $T-B$ phase diagram
is in good agreement with the experimental
$T-B$ phase diagram obtained in measurements on the HF metals $\rm
YbRh_2Si_2$ and YbRh$_2$(Si$_{0.95}$Ge$_{0.05}$)$_2$.  We have also
demonstrated that the abrupt jump in the Hall coefficient
$R_H(B\to B_{c0},T\to0)$ is
determined by the presence of FC.  We have observed that
the second order AF phase transition
changes to the first order one below $T_{crit}<T_N(B=0)$
making the corresponding quantum and
thermal critical fluctuations vanish at the jump. Therefore, the abrupt
jump and the divergence of the effective mass taking place at $T_N\to0$
are defined by the behavior of quasiparticles rather than by the
corresponding thermal and quantum critical fluctuations.

We have predicted that the differential tunnelling
conductivity between a metallic point and an ordinary metal, which is
commonly symmetric as a function of the voltage, becomes noticeably
dissymmetrical when the ordinary metal is replaced by a HF metal,  
the electronic system of which has undergone FCQPT. This
dissymmetry can be observed when the HF metal is both normal and
superconducting. We have also discussed possible experiments to
study the dissymmetry in measurements on the HF metals. 
In the case of the high-$T_c$ superconductors,
our consideration is in good agreement with available data.

In conclusion, we have shown that
in contrast to covnetional quantum phase transitions,
whose physics is dominated by thermal
and quantum fluctuations and characterized by the absence of
quasiparticles, the physics of Fermi systems and the heavy electron liquid 
near FCQPT or 
undergone FCQPT is determined by quasiparticles
resembling the Landau quasiparticles.
We have shown that the Landau paradigm based on the notion
of quasiparticles and order parameters is still
applicable when considering the low temperature properties of the
heavy electron liquid, whose understanding has been problematic
largely because of the absence of theoretical guidance. In
contrast with the conventional Landau quasiparticles, the
effective mass of the considered quasiparticles
strongly depends on the temperature, applied magnetic
fields, the number density $x$, pressure, etc. 
These quasiparticles and the order parameter are well
defined and capable of describing both the LFL and the NFL
behaviors of both the high-$T_c$ superconductors and
the HF metals and their universal thermodynamic
properties down to the lowest temperatures.
This system of quasiparticles determines the
recovery of the LFL behavior under applied
magnetic fields and preserves the Kadowaki-Woods ratio.
Thus, we obtain a unique possibility to
control the essence of HF systems and accordingly  
of the HF metals by magnetic fields in a wide range of temperatures. 

Finally, our general consideration suggests that FCQPT and the
emergence of novel quasiparticles at QCP and behind QCP and resembling
the Landau quasiparticles are qualities intrinsic to strongly
correlated substances, while  FCQPT can be viewed as the 
universal cause of the non-Fermi liquid
behavior observed in different metals and liquids.

\section{Acknowledgments}

We are grateful to P. Coleman, V.A. Khodel and M. Norman for valuable 
discussions.

This work was supported in part by the Russian Foundation
for Basic Research, project No. 01-02-17189
and in part by INTAS, project No. 03-51-6170.
The visit of VRS to Clark Atlanta University has been supported by
NSF through a grant to CTSPS. VRS is grateful to the Racah
Institute of Physics
for the hospitality during his stay at the Hebrew University, Jerusalem.
MYaA is grateful to the Binational Science Foundation, grant 2002064  
and Israeli Science Foundation, grant 174/03. AZM is
supported by US DOE, Division of Chemical Sciences, Office of
Basic Energy Sciences, Office of Energy Research. PKG is supported by  
the grants of Russian Academy of Sciences.


\begin{thebibliography}{199}

\bibitem{ste} G.R. Stewart, Rev. Mod. Phys. {\bf 73}, 797 (2001).

\bibitem{varma} C.M. Varma, Z. Nussionov, and W. van Saarlos, Phys. Rep.
{\bf 361}, 267 (2002).

\bibitem{sac} S. Sachdev, {\it Quantum Phase transitions}
(Cambridge, Cambridge University Press, 1999).

\bibitem{voj} M. Vojta, Rep. Prog. Phys. {\bf 66}, 2069 (2003).

\bibitem{uzum} D.V. Shopova and D.I. Uzunov, Phys. Rep. {\bf 379}, 1 (2003).

\bibitem{geg1}  R. K\"uchler, N. Oeschler, P. Gegenwart,
T. Cichorek, K. Neumaier, O. Tegus,
C. Geibel, J. A. Mydosh, F. Steglich, L. Zhu, and Q. Si,
Phys. Rev. Lett. {\bf 91}, 066405 (2003).

\bibitem{alp} M.Ya. Amusia, A.Z. Msezane, and V.R. Shaginyan,
Phys. Lett. A {\bf 320}, 459 (2004).

\bibitem{shag4}  V.R. Shaginyan, JETP Lett. {\bf 79}, 286 (2004).

\bibitem{kadw} K. Kadowaki and S.B. Woods, Solid State Commun.
{\bf 58}, 507 (1986).

\bibitem{geg} P. Gegenwart, J. Custers, C. Geibel, K. Neumaier, T. Tayama,
K. Tenya, O. Trovarelli, and F. Steglich, Phys. Rev. Lett. {\bf 89},
056402 (2002).

\bibitem{cyr} C. Proust, E. Boaknin, R. W. Hill,
L. Taillefer, and A. P. Mackenzie,
Phys. Rev. Lett. {\bf 89}, 147003 (2002).

\bibitem{mill} A.J. Millis, A.J. Schofield, G.G. Lonzarich, and
S.A. Grigera, Phys. Rev. Lett. {\bf 88}, 217204 (2002).

\bibitem{bi} A. Bianchi, R. Movshovich, I. Vekhter,
P. G. Pagliuso, and J. L. Sarrao,
Phys. Rev. Lett. {\bf 91}, 257001 (2003);
F. Ronning, C. Capan, A. Bianchi, R. Movshovich, A. Lacerda, M. F. 
Hundley, J. D.  Thompson, P. G. Pagliuso, and J. L. Sarrao, Phys. Rev. 
B {\bf 71}, 104528 (2005).

\bibitem{pag1} J. Paglione, M.A. Tanatar, D.G. Hawthorn, E. Boaknin,
F. Ronning, R.W. Hill, M. Sutherland, L. Taillefer,
C. Petrovic, and P.C. Canfield, cond-mat/0405157.

\bibitem{bel1} S.T. Belyaev, Sov. Phys. JETP, {\bf 7}, 289 (1958).

\bibitem{bel2} S.T. Belyaev, Sov. Phys. JETP, {\bf 7}, 299 (1958).

\bibitem{ks} V.A. Khodel and V.R. Shaginyan,
JETP Lett. {\bf 51}, 553 (1990).;

\bibitem{ksn} V.A. Khodel and V.R. Shaginyan, Nucl. Phys. A
{\bf 555}, 33 (1993).

\bibitem{ksk} V.A. Khodel, V.R. Shaginyan, and V.V. Khodel,
Phys. Rep. {\bf 249}, 1 (1994).

\bibitem{dkss} J. Dukelsky, V.A. Khodel, P. Schuck, and V.R. Shaginyan,
Z. Phys. {\bf102}, 245 (1997); V.A. Khodel and V.R.  Shaginyan,
Condensed Matter Theories, {\bf12}, 222 (1997).

\bibitem{vsl} V.R. Shaginyan, Phys. Lett. A {\bf 249}, 237 (1998).

\bibitem{vol} G. E. Volovik,
JETP Lett. {\bf 53}, 222 (1991).

\bibitem{dzyal} I.E. Dzyaloshinskii, J. Phys. I (France) {\bf 6}, 119 (1996).

\bibitem{lid} D. Lidsky, J. Shiraishi, Y. Hatsugai, and M. Kohmoto,
Phys. Rev. B {\bf 57}, 1340 (1998).

\bibitem{irk} V.Yu. Irkhin, A.A. Katanin, and M.I. Katsnelson, Phys.
Rev.  Lett.  {\bf 89}, 076401 (2002).

\bibitem{shag2} V.R. Shaginyan, JETP Lett. {\bf 77}, 178 (2003).

\bibitem{lanl1}  E.M. Lifshitz and L.P. Pitaevskii,
{\it Statistical Physics}
(Part 2, Butterworth-Heinemann, Oxford, 1999).

\bibitem{pom} L.Ja. Pomeranchuk, JETP {\bf 35}, 524 (1958).

\bibitem{ms} M.Ya. Amusia and V.R. Shaginyan,
JETP Lett. {\bf 73}, 232 (2001);

\bibitem{shb} M.Ya. Amusia and V.R. Shaginyan,
Phys. Rev. B {\bf 63}, 224507
(2001); V.R. Shaginyan, Physica B {\bf 312-313C}, 413 (2002).

\bibitem{shag1} V.R. Shaginyan,
JETP Lett. {\bf 77}, 99 (2003).

\bibitem{khod1} V.M.Yakovenko and V.A. Khodel,
JETP Lett. {\bf 78}, 398 (2003); cond-mat/0308380.

\bibitem{cas1} A. Casey, H. Patel, J. Nye'ki,
B. P. Cowan, and J. Saunders,
Phys. Rev. Lett. {\bf 90}, 115301 (2003).

\bibitem{skdk} A.A. Shashkin, S.V. Kravchenko, V.T. Dolgopolov,
and T.M. Klapwijk, Phys. Rev. B {\bf 66}, 073303 (2002);
A.A. Shashkin, M. Rahimi, S. Anissimova, S.V. Kravchenko
V.T. Dolgopolov, and T. M. Klapwijk,
Phys. Rev. Lett. {\bf 91}, 046403 (2003).

\bibitem{krot} J. Boronat, J. Casulleras,
V. Grau, E. Krotscheck, and J. Springer,
Phys. Rev. Lett. {\bf 91}, 085302 (2003).

\bibitem{sarm1} Y. Zhang, V. M. Yakovenko, and S. Das Sarma,
Phys. Rev. B {\bf 71}, 115105 (2005).

\bibitem{sarm2} Y. Zhang and S. Das Sarma,
Phys. Rev. B {\bf 70}, 035104 (2004).

\bibitem{ksz} V.A. Khodel, V.R. Shaginyan, and M.V. Zverev,
JETP Lett. {\bf 65}, 253 (1997).

\bibitem{shag3}  V.R. Shaginyan, J.G. Han, and J. Lee,
Phys. Lett. A {\bf 329}, 108 (2004).

\bibitem{rlp} R.B. Laughlin and D. Pines, Proc. Natl. Acad. Sci. USA
{\bf 97}, 28 (2000).

\bibitem{pa} P.W. Anderson, cond-mat/0007185; cond-mat/0007287.

\bibitem{kcs} V.A. Khodel, J.W. Clark, and V.R. Shaginyan,
Solid Stat. Comm. {\bf 96}, 353 (1995).

\bibitem{ars} S.A. Artamonov and V.R. Shaginyan,
JETP {\bf 92}, 287 (2001).

\bibitem{blk} P. V. Bogdanov, A. Lanzara, S. A. Kellar, X. J. Zhou, E. D. Lu,
W. J. Zheng, G. Gu, J.-I. Shimoyama, K. Kishio, H. Ikeda,
R. Yoshizaki, Z. Hussain, and Z. X. Shen,
Phys. Rev. Lett. {\bf 85}, 2581 (2000).

\bibitem{krc} A. Kaminski, M. Randeria, J. C. Campuzano,
M. R. Norman, H. Fretwell,
J. Mesot, T. Sato, T. Takahashi, and K. Kadowaki,
Phys. Rev. Lett. {\bf 86}, 1070 (2001).

\bibitem{vall} T. Valla, A. V. Fedorov, P. D. Johnson, B. O. Wells,
S. L. Hulbert, Q. Li, G. D. Gu, and N. Koshizuka,
Science {\bf 285}, 2110 (1999);
\\ T. Valla, A. V. Fedorov, P. D. Johnson, Q. Li, G. D. Gu,
and N. Koshizuka, Phys. Rev. Lett. {\bf 85}, 828 (2000).

\bibitem{grun} G. Gr\"uner, {\it Density Waves in Solids} 
(Addison-Wesley, Reading, MA, 1994). 

\bibitem{tass} L. Tassini, F. Venturini, Q.-M. Zhang, R. Hackl, 
N. Kikugawa, and T. Fujita, 
Phys. Rev. Lett. {\bf 95}, 117002 (2005). 

\bibitem{bcs} J. Bardeen, L.N. Cooper, and J.R. Schrieffer,
Phys. Rev. {\bf108}, 1175 (1957).

\bibitem{til} D.R. Tilley and J. Tilley, {\it Superfluidity and
Superconductivity}, (Bristol, Hilger, 1985).

\bibitem{bogol} N.N. Bogoliubov, Nuovo Cimento {\bf 7}, 794 (1958).

\bibitem{asjetpl} M.Ya. Amusia and V.R. Shaginyan, JETP Lett. {\bf 77},
671 (2003).

\bibitem{mat} H. Matsui, T. Sato, T. Takahashi, S.-C. Wang, H.-B. 
Yang, H. Ding, T. Fujii, T. Watanabe, and A. Matsuda, Phys. Rev. Lett. 
{\bf 90}, 217002 (2003).

\bibitem{ams} M.Ya. Amusia, S.A. Artamonov, and V.R. Shaginyan,
JETP Lett. {\bf 74}, 435 (2001).

\bibitem{rand} A. Paramekanti, M. Randeria, and N. Trivedi,
Phys. Rev. Lett. {\bf 87}, 217002 (2001);
A. Paramekanti, M. Randeria, and N. Trivedi, cond-mat/0305611.

\bibitem{pwa} P. W. Anderson, P. A. Lee, M. Randeria,
T. M. Rice, N. Trivedi, and F. C. Zhang,
J Phys. Condens. Matter {\bf 16}, R755 (2004).

\bibitem{sh} V.R. Shaginyan, JETP Lett. {\bf 68}, 527 (1998).

\bibitem{ms1} M.Ya. Amusia and V.R. Shaginyan,
Phys. Lett. A {\bf 298}, 193 (2002).

\bibitem{kug} M. Kugler, M. Fischer, Ch. Renner, S. Ono, and Y. Ando,
Phys. Rev. Lett. {\bf 86}, 4911 (2001).

\bibitem{abr} A.A. Abrikosov, Phys. Rev. B {\bf 52},
R15738 (1995); A.A. Abrikosov, cond-mat/9912394.

\bibitem{skin} N.-C. Yeh, C.-T. Chen, G. Hammer, J. Mannhart, A. Schmeh,
C. W. Schneider, R. R. Schulz, S. Tajima, K. Yoshida, D. Garrigus, and 
M. Strasik, Phys. Rev. Lett. {\bf 87}, 087003 (2001).

\bibitem{bis} A. Biswas, P. Fournier, M. M. Qazilbash,
V. N. Smolyaninova, H. Balci, and R. L. Greene,
Phys. Rev. Lett. {\bf 88}, 207004 (2002).

\bibitem{skin1} J.A. Skinta, M.-S. Kim, T.R. Lemberger, T. Greibe, and 
M. Naito, Phys. Rev. Lett. {\bf 88}, 207005 (2002).

\bibitem{skin2} J.A. Skinta, T.R. Lemberger, T. Greibe, and M. Naito,
Phys. Rev. Lett. {\bf 88}, 207003 (2002).

\bibitem{chen} C.-T. Chen, P. Seneor, N.-C. Yeh, R.P. Vasquez, L.D. Bell,
C.U. Jung, J.Y. Kim, M.-S. Park, H.-J. Kim, and S.-I. Lee,
Phys. Rev. Lett. {\bf 88}, 227002 (2002).

\bibitem{nm} N. Metoki, Y. Haga, Y. Koike, and Y. Onuki,
Phys.  Rev.  Lett.  {\bf 80}, 5417 (1998).

\bibitem{th} T. Honma, Y. Haga, and E. Yamamoto, J. Phys.
Soc. Jpn.  {\bf 68}, 338 (1999).

\bibitem{ams3} M.Ya. Amusia and V.R. Shaginyan,
JETP Lett. {\bf 76}, 651 (2002).

\bibitem{mzo}  N. Miyakawa, J. F. Zasadzinski, L. Ozyuzer, P. Guptasarma,
D. G. Hinks, C. Kendziora, and K. E. Gray,
Phys. Rev. Lett. {\bf 83}, 1018 (1999).

\bibitem{var} C.M. Varma, P.B. Littlewood, S. Schmitt-Rink, E.
Abrahams, and A.E. Ruckenstein, Phys.  Rev.  Lett.  {\bf 63}, 1996
(1989). 

\bibitem{varm1} C.M.  Varma,  P.B. Littlewood, S. Schmitt-Rink, E.
Abrahams, and A.E. Ruckenstein, Phys.  Rev.  Lett. {\bf 64}, 497
(1990).

\bibitem{ino} A. Ino, C. Kim, M. Nakamura, T. Yoshida,
T. Mizokawa, A. Fujimori, Z.-X. Shen,
T. Kakeshita, H. Eisaki, and S. Uchida,
Phys. Rev. B {\bf 65}, 094504 (2002).

\bibitem{zhou} X.J. Zhou, T. Yoshida, A. Lanzara,
P.V. Bogdanov, S.A. Kellar, K.M. Shen,
W. L. Yang, F. Ronning, T. Sasagawa, T. Kakeshita,
T. Noda, H. Eisaki, S. Uchida, C. T. Lin, F. Zhou,
J.W. Xiong, W.X. Ti, Z.X. Zhao, A. Fujimori, Z. Hussain,
and Z.-X. Shen, Nature {\bf 423}, 398 (2003).

\bibitem{padil} W. J. Padilla, Y. S. Lee, M. Dumm, G. Blumberg, S. Ono, K. Segawa,
S. Komiya, Y. Ando, D. N. Basov, Phys. Rev. B {\bf 72}, 060511 (2005).

\bibitem{vald} T. Valla, A.V. Fedorov, P.D. Johnson, and S.L. Hulbert,
Phys. Rev. Lett. {\bf 83}, 2085 (1999).

\bibitem{feng} D.L. Feng, A. Damascelli, K.M. Shen, N. Motoyama,
D.H. Lu, H. Eisaki, K. Shimizu, J.-I. Shimoyama,
K. Kishio, N. Kaneko, M. Greven, G. D. Gu,
X. J. Zhou, C. Kim, F. Ronning, N.P. Armitage, and Z.-X Shen,
Phys. Rev. Lett. {\bf 88}, 107001 (2002).

\bibitem{dess} D.S. Dessau, B O. Wells, Z.–X. Shen, W. E. Spicer,
A.J. Arko, R.S. List,
D. B. Mitzi, and A. Kapitulnik,  Phys. Rev. Lett. {\bf 66}, 2160 (1991).

\bibitem{mig} A.B. Migdal, {\it Theory of Finite Fermi Systems and
Applications to Atomic Nuclei} (Benjamin, Reading, MA, 1977).

\bibitem{val1} Z.M. Yusof, B.O. Wells, T. Valla, A.V. Fedorov, P.D. 
Johnson, Q. Li, C. Kendziora, S. Jian, and D. G. Hinks, Phys. Rev. 
Lett. {\bf 88}, 167006 (2002).

\bibitem{nakam}  S. Nakamae, K. Behnia, N. Mangkorntong, M. Nohara,
H. Takagi, S. Yates, and N.E. Hussey, Phys. Rev. B {\bf 68}, 100502 (2003).

\bibitem{huss} N.E. Hussey, M. Abdel-Jawad, A. Carrington, A.P. Mackenzie,
and L. Balicas, Nature {\bf 425}, 814 (2003).

\bibitem{asp} S. A. Artamonov, V.R. Shaginyan, and Yu.G. Pogorelov,
JETP Lett. {\bf 68}, 942 (1998).

\bibitem{pogsh} Yu.G. Pogorelov and V.R. Shaginyan,
JETP Lett. {\bf 76}, 532 (2002).

\bibitem{llvp} M. de Llano and J. P. Vary,
Phys. Rev. C {\bf 19}, 1083
(1979); M. de Llano, A. Plastino, and J.G. Zabolitsky, Phys. Rev.
C {\bf 20}, 2418 (1979).

\bibitem{zb} M.V. Zverev and M. Baldo,
J. Phys. Condens. Matter {\bf 11},  2059 (1999).

\bibitem{shag} V.R. Shaginyan, M.Z. Msezane, and M.Ya. Amusia,
Phys. Lett. A {\bf 338}, 393 (2005).

\bibitem{takah} D. Takahashi, S. Abe, H. Mizuno, D. A. Tayurskii,
K. Matsumoto, H. Suzuki, and Y. Onuki, Phys. Rev. B {\bf 67}, 180407 (2003).

\bibitem{ksch} V.A. Khodel and P. Schuck, Z. Phys. B {\bf 104}, 505
(1997).

\bibitem{tky}  N. Tsujii, H. Kontani, and
K.  Yoshimura, Phys. Rev. Lett. {\bf 94}, 057201 (2005).

\bibitem{cyr1} R. Bel, K. Behnia, C. Proust, P. van der Linden, D.K. Maude,
and S.I. Vedeneev, Phys. Rev. Lett. {\bf 92}, 17703 (2004).

\bibitem{korr} J. Korringa, Physica (Utrecht)
{\bf 16}, 601 (1950).

\bibitem{zheng} G.-q. Zheng, T. Sato, Y. Kitaoka, M. Fujita, and K. Yamada, 
Phys. Rev. Lett. {\bf 90}, 197005 (2003).

\bibitem{kane} C.L. Kane and M.P.A. Fisher, Phys. Rev. Lett.
{\bf 76}, 3192 (1996).

\bibitem{sen} T. Senthil and M.P.A. Fisher, Phys. Rev. B
{\bf 62}, 7850 (2000).

\bibitem{hough} A. Houghton, S. Lee, and J.P. Marston, Phys. Rev B
{\bf 65}, 22503 (2002).

\bibitem{mac} A.P. Mackenzie, S.R. Julian, D.C. Sinclair, and C. T. Lin,
Phys. Rev. B {\bf 53}, 5848 (1996).

\bibitem{aspla} M.Ya. Amusia and V.R. Shaginyan, Phys. Lett.
A {\bf 315}, 288 (2003).

\bibitem{mor}  K.–D. Morhard, C. Ba\"uerle, J. Bossy,
Yu. Bunkov, S.N. Fisher, and H. Godfrin,
Phys. Rev. B {\bf 53}, 2658 (1996).

\bibitem{cas} A. Casey, H. Patel, J. Ny\'eki, B. P. Cowan, and J. Saunders,
Phys. Rev. Lett. {\bf 90}, 115301 (2003).

\bibitem{pfw} M. Pfitzner and P. W\"olfe, Phys. Rev. B {\bf 33}, 2003
(1986).

\bibitem{krav} S.V. Kravchenko and M.P. Sarachik, Rep. Prog. Phys. 
{\bf 67}, 1 (2004).

\bibitem{cust} J. Custers, P. Gegenwart, H. Wilhelm,
K. Neumaier, Y. Tokiwa, O. Trovarelli,
C. Geibel, F. Steglich, C. P\'epin, and P. Coleman,
Nature {\bf 424}, 524 (2003). 


\bibitem{pepin} C. P\'epin, Phys. Rev. Lett.
{\bf 94}, 066402 (2005).

\bibitem{bud} S.L. Bud'ko, E. Morosan, and P.C. Canfield,
Phys. Rev. {\bf 69}, 014415 (2004).

\bibitem{pag}  J. Paglione, M.A. Tanatar, D.G. Hawthorn, E. Boaknin, 
R.W. Hill, F. Ronning, M. Sutherland, and L. Taillefer, Phys. Rev. 
Lett. {\bf 91}, 246405 (2003).

\bibitem{senth1} T. Senthil, M. Vojta, and S. Sachdev,
Phys. Rev. B {\bf 69}, 035111 (2004); T. Senthil, S. Sachdev, and M. Vojta,
Physica B {\bf 359-361}, 9 (2005).

\bibitem{senth3} T. Senthil, A. Vishwanath,
L. Balents, S. Sachdev, and M.P.A. Fisher,
Science {\bf 303}, 1490 (2004).

\bibitem{kuch} R. K\"uchler, P. Gegenwart,
K. Heuser, E.-W. Scheidt, G. R. Stewart, and F. Steglich,
Phys. Rev. Lett. {\bf 93}, 096402 (2004).

\bibitem{movsh} A. Bianchi, R. Movshovich, I. Vekhter, P.G. Pagliuso, and
J.L. Sarrao, Phys. Rev. Lett. {\bf 91}, 257001 (2003).

\bibitem{zhu}  L. Zhu, M. Garst, A. Rosch, and
Q. Si, Phys. Rev. Lett. {\bf 91}, 066404 (2003).

\bibitem{ishida} K. Ishida, K. Okamoto, Y. Kawasaki, Y. Kitaoka,
O. Trovarelli, C. Geibel, and F. Steglich,
Phys. Rev. Lett. {\bf 89}, 107202 (2002).

\bibitem{col1} P. Coleman, {\it 
Lectures on the Physics of Highly Correlated Electron Systems VI}
(Editor F. Mancini, American Institute of Physics, New York, 2002, p. 79).
\bibitem{col2} P. Coleman and A.J. Schofield,
Nature {\bf 433}, 226 (2005).

\bibitem{ckhz} J.W. Clark, V.A. Khodel, and M.V. Zverev,
Phys. Rev B {\bf 71}, 012401 (2005).

\bibitem{lanl2}  E.M. Lifshitz and L.P. Pitaevskii,
{\it Statistical Physics} (Part 1, Butterworth-Heinemann, 2000, p. 168).

\bibitem{shag5}  V.R. Shaginyan,
JETP Lett. {\bf 80}, 263  (2004).

\bibitem{khod2} J.W. Clark, V.A.
Khodel, M.V. Zverev, and V.M. Yakovenko,  Phys. Rep. {\bf 391}, 123 (2004).

\bibitem{zver} M.V. Zverev, V.A. Khodel, and V.R. Shaginyan,
JETP Lett. {\bf 65}, 863 (1997).

\bibitem{noz}  P. Nozi\`eres, J. Phys. I (France) {\bf 2}, 443  (1992).

\bibitem{pash} S. Paschen, T. L\"uhmann, S. Wirth, P. Gegenwart, O. 
Trovarelli, C. Geibel, F. Steglich, P. Coleman and Q. Si, Nature {\bf 
432}, 881 (2004).

\bibitem{spa} V.R. Shaginyan, K.G. Popov, and S.A. Artamonov,
JETP Lett. {\bf 82}, 234 (2005).

\bibitem{geg2} P. Gegenwart, J. Custers, Y. Tokiwa, C. Geibel, and F. 
Steglich, Phys. Rev. Lett. {\bf 94}, 076402 (2005).

\bibitem{khzvya} V.A. Khodel, M.V. Zverev, and V.M. Yakovenko, 
cond-mat/0508275.

\bibitem{oes} N. Oeschler, P. Gegenwart, M. Lang, R. Movshovich, J.L. 
Sarrao, J.D. Thompson, and F. Steglich, Phys. Rev. Lett. {\bf 91}, 
076402 (2003).

\bibitem{bau} E. D. Bauer, J.D. Thompson, J.L. Sarrao, L.A. Morales, 
F. Wastin, J. Rebizant, J.C. Griveau, P. Javorsky, P. Boulet, E. 
Colineau, G. H. Lander, and G. R. Stewart, Phys. Rev. Lett. {\bf 93}, 
147005 (2004).

\bibitem{tun} V.R. Shaginyan, JETP Lett. {\bf 81}, 222 (2005).

\bibitem{zag} A.M. Zagoskin, {\it Quantum Theory of Many-Body Systems}
(Springer-Verlag New York, Inc., 1998).

\bibitem{pan} S.H. Pan, J.P. O'Neal, R.L. Badzey,
C. Chamon, H. Ding, J.R. Engelbrecht,
Z. Wang, H. Eisaki, S. Uchida, A.K. Gupta,
K.-W. Ng, E.W. Hudson, K.M. Lang, and J. C. Davis,
Nature {\bf 413}, 282 (2001).

\end{thebibliography}
\end{document}